\documentclass[referee]{aa} 
\usepackage{natbib}
\usepackage{lscape}
\usepackage{amssymb}
\usepackage{txfonts}
\usepackage[dvips]{graphicx}
\usepackage{rotating}
\usepackage{longtable}
\newcommand{\ecy}{C$_2$H$_5$CN}
\newcommand{\ecys}{C$_2$H$_5$CN }
\newcommand{\vcy}{C$_2$H$_3$CN}
\newcommand{\vcys}{C$_2$H$_3$CN }
\newcommand{\de}{CH$_3$OCH$_3$}
\newcommand{\des}{CH$_3$OCH$_3$ }
\newcommand{\hy}{H$_2$}
\newcommand{\hys}{H$_2$ }
\newcommand{\hco}{HCOOCH$_3$}
\newcommand{\hcos}{HCOOCH$_3$ }
\newcommand{\amm}{NH$_3$}
\newcommand{\amms}{NH$_3$ }
\newcommand{\ace}{CH$_3$C$_2$H}
\newcommand{\asec}{$^{\prime \prime}$}
\def\CIII{\hbox{C$^{17}$O}}
\newcommand{\vlsr}{$V_{\rm LSR}$}
\def\hii{\hbox{H{\sc ii}}}
\def\uchii{\hbox{UC H{\sc ii}}}

\def\kms{\hbox{km~s$^{-1}$}}
\def\cmq{\hbox{cm$^{-2}$}}
\def\cmc{\hbox{cm$^{-3}$}}
\begin{document}
   \title{Comparative study of complex N- and O-bearing molecules in hot molecular cores
\thanks{Based on observations carried out with the IRAM Pico Veleta telescope.
IRAM is supported by INSU/CNRS (France), MPG (Germany) and IGN (Spain).}}
\author{F. Fontani \inst{1}
        \and I. Pascucci \inst{2}  
        \and P. Caselli \inst{3,} \inst{5}
        \and F. Wyrowski \inst{4}
        \and R. Cesaroni \inst{5} \and C.M. Walmsley \inst{5}}
\offprints{F. Fontani, \email{ffontani@ira.inaf.it}}
\institute{INAF, Istituto di Radioastronomia, CNR, Via Gobetti 101, 
           I-40129 Bologna, Italy \and
	   Steward Observatory, The University of Arizona, Tucson, AZ 85721 \and
           Harvard-Smithsonian Center for Astrophysics, 60 Garden Street MS42, Cambridge, MA 02138, USA \and
	   Max-Planck-Insitut fuer Radioastronomie, Auf dem Huegel 69, 53121 Bonn, Germany \and
	   INAF, Osservatorio Astrofisico di Arcetri, Largo E. Fermi 5,
           I-50125 Firenze, Italy
 }
\date{Received date; accepted date}

\titlerunning{Complex molecules in hot cores}
\authorrunning{Fontani et al.}

\abstract
{}
{We have observed several emission lines of two Nitrogen-bearing 
 (\ecys and \vcy) and two Oxygen-bearing (\des and \hco)
molecules towards a sample of well-known hot molecular cores (HMCs) in
order to check whether the chemical differentiation seen in 
the Orion-HMC and W3(H$_2$O) between O- and
N-bearing molecules is a general property of HMCs.}
{With the IRAM-30m telescope we have observed 12 HMCs in 21 bands, 
centered at frequencies from $86250$ to $258280$ MHz.}
{In six sources,
we have detected a number of transitions sufficient to derive their
main physical properties. The rotational temperatures obtained
from \ecy, \vcys and \des range from
$\sim 100$ to $\sim 150$ K in these HMCs. The total column densities of these
molecules are of the order of $\sim 10^{15}-10^{17}$\cmq .
Single Gaussian fits performed to unblended lines show a marginal
difference in the line peak velocities of the \ecys and \des lines, 
indicating a possible spatial separation between the region
traced by the two molecules. On the
other hand, neither the linewidths nor the rotational temperatures 
and column densities confirm such a result.
The average molecular abundances of 
\ecy, \vcys and \des are in the range $\sim 10^{-9}-10^{-10}$,
comparable to those seen in the Orion hot core. In other
HMCs Bisschop et al.~2007 found comparable values
for \ecys but values $\sim 2.5$ times larger for \de.
By comparing the
abundance ratio of the pair \ecy/\vcys with the predictions
of theoretical models, we derive that the age of our cores ranges
between 3.7 and 5.9$\times 10^{4}$ yrs.} 
{The abundances of \ecys and \vcys are
strongly correlated, as expected from theory 
which predicts that \vcys is formed 
through gas phase reactions involving
\ecys. A correlation is also found between the abundances
of \ecys and \de, and \vcys and \de.
In all tracers the fractional abundances increase with the
H$_2$ column density while they are not correlated with
the gas temperature.
On average, the chemical and physical differentiation between
O- and N-bearing molecules seen 
in Orion and W3(H$_2$O) is not revealed by our observations.
We believe that this is partly due to the poor angular resolution
of our data, which allows us to derive only average values over
the sources of the discussed parameters.}


\maketitle
%

\section{Introduction}
\label{Introduction}

The formation process of massive stars ($M\geq 8 M_{\odot}$) is 
still poorly understood. This is mainly due to 
the observational limitations that hinder the study of early-type stars, 
i.e. their shorter evolutionary timescales, larger distances and 
stronger interaction with their environment. 
Despite this, a growing effort has been devoted 
to investigate this process, and important 
results have been obtained both from the observational and the 
theoretical point of view (see e.g. Garay \& Lizano~1999;
Tan \& McKee~2002; Yorke~2004; Beuther et 
al.~2007; Cesaroni et al.~2007).

It is well known that when a high-mass star reaches the ZAMS, 
an ultracompact (UC)
\hii{} region, i.e. an ionized region with diameter
smaller than a few $\sim 0.1$ pc, becomes observable at radio 
wavelengths due to the emission of free--free radiation. 
This has led to using \uchii s as ``signposts'' of
massive star formation (e.g. Wood and Churchwell~1989). 
Various authors have shown that \uchii s are embedded in molecular 
clumps of a few thousand solar masses, diameters $\sim $ 1 pc, and 
average H$_2$ volume density of $\sim 10^5$ \cmc\ 
(Cesaroni et al.~1991; Hofner et al.~2000; Fontani et al.~2002).
Observations at higher angular resolution have also identified
hot ($\geq 100$ K), dense ($\geq10^7$ cm$^{-3}$) and compact 
($\leq0.1$ pc) molecular condensations in these clumps,
called ``hot molecular cores'' (HMCs) from the name of their prototype,
the Orion Hot Core (see Genzel \& Stutzki~1989; 
Walmsley \& Schilke~1992). These HMCs are often 
associated with outflows, water masers, and in some
of them an \uchii\ region has been detected, suggesting that
they represent the environment in which high-mass stars
were recently born (see e.g. Kurtz et al.~2000; 
Cesaroni~2005). 
Their chemical composition is peculiar and very different from
that of the surrounding molecular clump, showing
higher abundances of saturated species (e.g. H$_2$O,
\amm, CH$_3$OH) and complex organic H--rich molecules
(see e.g. Caselli~2005).

Since most of the observed H--rich molecules are difficult to form in
gas-phase reactions, grain-surface chemistry is often invoked
to explain their formation.
One possible scenario is the following: during the collapse of
a pre-stellar core, molecules freeze out onto dust grains, 
react with atoms or other molecules on the grain surface and, 
as the core is heated by the young star, some of them evaporate 
and return to the gas phase. Based on this scenario,
one can distinguish three types of 
molecules (see Charnley~1995):
\begin {itemize}
\item those formed in cold gas through ion-molecule reactions, frozen onto
dust grains and then released into the gas phase when the core is heated up 
by the central star.
Their abundances should hence reflect those of the initial early cold
phase;
\item those formed on the grain surfaces through reactions of
frozen species, and then evaporated during the HMC phase;
\item those formed in hot gas through reactions of evaporated 
molecules.
\end{itemize}

Several studies have been made to investigate the chemistry of HMCs
(Macdonald et al.~1995; Gibb et al.~2000;
Nomura \& Millar~2004; Garrod \& Herbst~2006).
Charnley et al.~(1992) showed that initial
differences in the chemical composition and gas phase 
reactions subsequent to evaporation from dust grains
can explain the existence of many of the complex O- and N-bearing 
species. Interestingly,
Blake et al.~(1987) found that N-bearing species
are more abundant than O-bearing ones in the Orion-HC, while
the opposite is observed in the Compact Ridge. The chemical
model by Caselli et al.~(1993)
explains such a chemical difference as being due to
the different physical properties of the two regions.
The model also predicts a relation between the
core evolutionary stage and the abundance ratio of 
molecular species like \vcys and \ecy.
Therefore, the relative abundance of these species 
can be used as a 'chemical clock' to estimate the HMC age.

In this work we present the results of the first survey of 4
complex molecules, \vcy , \ecy , \des and \hco , 
towards 12 well-known HMCs, all of them 
associated with \uchii\ regions and already observed in 
several molecular tracers (Hatchell et al.~1998;
Hofner et al.~2000; Fontani et al.~2002).
In Sect.~\ref{sect:Observ} we describe the observations and we present 
our sample. The methods used to analyse the data and the immediate
results are presented in Sect.~\ref{sect:results}. 
In Sect.~\ref{discu} we discuss our results,
and give an estimate of the chemical ages of the HMCs. 
Conclusions are drawn in Sect.~\ref{sect:Conclusions}.
       

\section{Observations and Data Reduction}  
\label{sect:Observ}
\subsection{Observations}

\begin{table*}
  \caption[Observed sources]{List of the observed sources. The distances
are taken from Table~1 of Kurtz et al.~(2000) except when
differently specified.}
   \label{tab:source}
\begin{center}
\begin{tabular}{l  c c  c  cc}
\hline
\hline
Source & R.A.(J2000) & Dec.(J2000) & \vlsr{}& $d$ & code\\
    & (h m s)& ($\circ\;\prime\;\prime\prime$) & (km~s$^{-1}$) & (kpc) & \\
\hline
W3(H$_2$O)  &02 27 04.7 &  +61 52 24.5 & $-$45.0& 1.95$^{a}$ & 1\\
G5.89$-$0.39  &18 00 30.4 &  -24 04 00.5 &10.0& 2.0$^{b}$ & 2 \\
G9.62+0.19  &18 06 15.0 &  -20 31 42.2 &4.4& 5.7 & 3\\
G10.47+0.03 &18 08 38.2 &  -19 51 49.7 & 67.8& 5.8 & 4 \\
G10.62$-$0.38 &18 10 28.7 &  -19 55 49.7 & $-$3.1& 4.8$^{c}$ & 5\\
G19.61$-$0.23 &18 27 38.1 &  -11 56 38.5 & 41.6& 12.6$^{d}$ &  6 \\
G29.96$-$0.02 &18 46 04.0 &  -02 39 21.5 &98.0& 7.4 & 7\\
G31.41+0.31 &18 47 34.4 &  -01 12 46.0 & 97.0& 7.9 & 8\\
G34.26+0.15 &18 53 18.5 & +01 14 57.7&58.0& 3.7$^{e}$ & 9\\
G45.47+0.05 &19 14 25.6 &  +11 09 25.9 & 62.0& 8.3 & 10\\
W51D        &19 23 39.9 &  +14 31 08.1 & 60.0& 8.0 & 11\\
IRAS20126+4104  &20 14 26.0 &  +41 13 32.5 &$-$3.5& 1.7 & 12\\
\hline
\multicolumn{5}{c}{ }\\
\multicolumn{5}{l}{R.A.(J2000)= right ascension}\\
\multicolumn{5}{l}{Dec.(J2000)= declination} \\
\multicolumn{5}{l}{\vlsr{}= Local standard of rest velocity}
\end{tabular}
\end{center}
$^{a}$ Xu et al.~(2006) \\
$^{b}$ Acord et al.~(1998) \\
$^{c}$ Fish et al.~(2003) \\
$^{d}$ Kolpak et al.~(2003) \\
$^{e}$ Kuchar \& Bania~(1994) \\
\end{table*}

The observations were carried out with the IRAM 30-m telescope 
during two observing runs: in July 1996, and August 1997.
The observed sources are listed in Table~\ref{tab:source}: in Cols.~2
and 3 we give the equatorial (J2000) coordinates; LSR source 
velocities (\vlsr ) and kinematic distances ($d$)
are given in Cols.~4 and 5, respectively.
The numbers listed in Col.~6 will
be used in the following to identify the source.
We observed simultaneously at 3, 2, and 1.3~mm. The half power beam 
width (HPBW) of the telescope was $\sim$~22\asec, $\sim$~16\asec, and 
$\sim$~12\asec , respectively. We obtained spectra of 
vinyl cyanide (\vcy), ethyl cyanide (\ecy), methyl formate (\hco) 
and dimethyl ether (\de).
The central frequencies of the observed bands and
the sources observed in each band are listed 
in Table~\ref{tab:Frequency setup}. 

System temperatures for both observing runs were 
300--450~K, 500--900~K, and 700--3000~K for the 3~mm, 
2~mm, and 1.3~mm receivers, respectively.
The given temperature ranges depend on weather 
conditions and source elevation. 
The receivers alignment was checked through
continuum cross scans on planets and found to be accurate to within
2\asec\ for the 3~mm, 2~mm, and the first 1.3~mm receiver.  
A misalignment of up to 5\asec\ was observed for the second 1.3~mm
receiver, which however is negligible with respect to the beam size.
Our spectrometers were an autocorrelator with bandwidth $\sim 523$ 
MHz and 0.32~MHz resolution, and two filter-banks with 512
MHz bandwith and 1~MHz resolution.
The wobbling secondary mirror was used with a beam throw of
240\asec\ and a frequency of 0.5~Hz resulting in linear baselines
of the spectra.
The focus was checked at the beginning of each night on Jupiter or
Saturn. Since the two 1.3~mm receivers had somewhat different foci,
we selected in these cases a focus position giving a compromise
between the different receivers more weighted to the 1.3~mm G1
receiver that
had the same focus as the 3 and 2~mm receivers. The resulting gain
loss is smaller than 10\%. Pointing was checked hourly by cross
scans on planets. 
The observations were made in wobbler switching mode, with the wobbling
secondary reflector switching in azimuth at a rate of 0.5 Hz 
between the source position and a reference position offset by 4 arcmin.

The data were reduced and analysed using the software package CLASS.
An advanced version of the package, XCLASS,  was used to fit
the transitions with blending problems (see e.g. Comito et
al.~2005 for details on the XCLASS fitting procedure).


\subsection{Line Identification}

Line identification was done using several catalogues of line 
frequencies: the 
Lovas\footnote{http://physics.nist.gov/PhysRefData/Micro/Html/contents.html} 
and JPL\footnote{http://spec.jpl.nasa.gov} catalogues, as well as 
the compilations of Anderson et al.~(1987,~1988a, 
1988b,~1990a,~1990b,~1992),
Bettens et al.~(1999), Groner et al.~(1998),  
Herbst~(1999), Klisch et al.~(1996), 
Pan et al.~(1998), Pearson et al.~(1991, 1994,
1997), Plummer et al.~(1986, 1987),
Wyrowski et al.~(1999), Yamada et al.~(1986).
Predicted vibrational transitions of \ecys\ were taken from Pearson~(2002).
We have considered as identified those lines in the 
spectra within $\sim$1~MHz from the expected line position 
in the catalogues. We have considered as detected the lines
above the 3$\sigma$ level in the spectra.

 \begin{table}[!]
\caption[Frequency setup]
{\label{tab:Frequency setup} {Observed frequency bands}}
\begin{center}
\begin{tabular}{l lcc }
\hline
\hline
Observing run & Band name & $\nu_c$$^{a}$ & observed sources$^{b}$    \\
    &   &(MHz)  &        \\
\hline
 &BAND3--HCN    &     86250  & 1,4,8\\          
 &BAND3--HNCO   &     87800   & 1,4,7,8\\          
 &BAND3--109    &    109410  & 4,8\\          
 &BAND3--110    &    110095 & 1,4,5,7-9\\          
 &BAND2--NH3    &    140300  &4,8 \\           
July &BAND2--154    &    154230 &1,4,7,8 \\           
1996  &BAND2--155    &    154850 &1,4,5,7-9 \\           
 &BAND1--215    &    215250 & 1,4,5,7-9\\           
 &BAND1--218    &    218970 & 1,4,8\\           
 &BAND1--HNCO   &    219580 & 1,4,7,8 \\           
 &BAND1--228    &    228100  & 4,8\\  
 &BAND1--HCN    &    258280 & 1,4,8\\ 
\hline
 &BAND3MM-1    &98600  &4-9,11,12             \\
 &BAND3MM-2    &104360 &6-9,11                 \\
 &BAND3MM-3    &111600 &1-12                \\
       &BAND2MM-1    &147370 &1-12           \\
August &BAND2MM-2    &161463 &4-9,11,12          \\
1997   &BAND2MM-3    &173150 &6-9,11             \\
 &BAND1MM-1    &209580 &6-9,11             \\
 &BAND1MM-2    &224237 &1-12             \\
 &BAND1MM-3    &237460 &4-9,11,12            \\
\hline
\end{tabular}
\end{center}
$^{a}$ band central frequency \\
$^{b}$ source identification numbers as in Table~\ref{tab:source}\\
\end{table}

\begin{figure*}
 \centering
 \includegraphics[angle=-90,width=13cm]{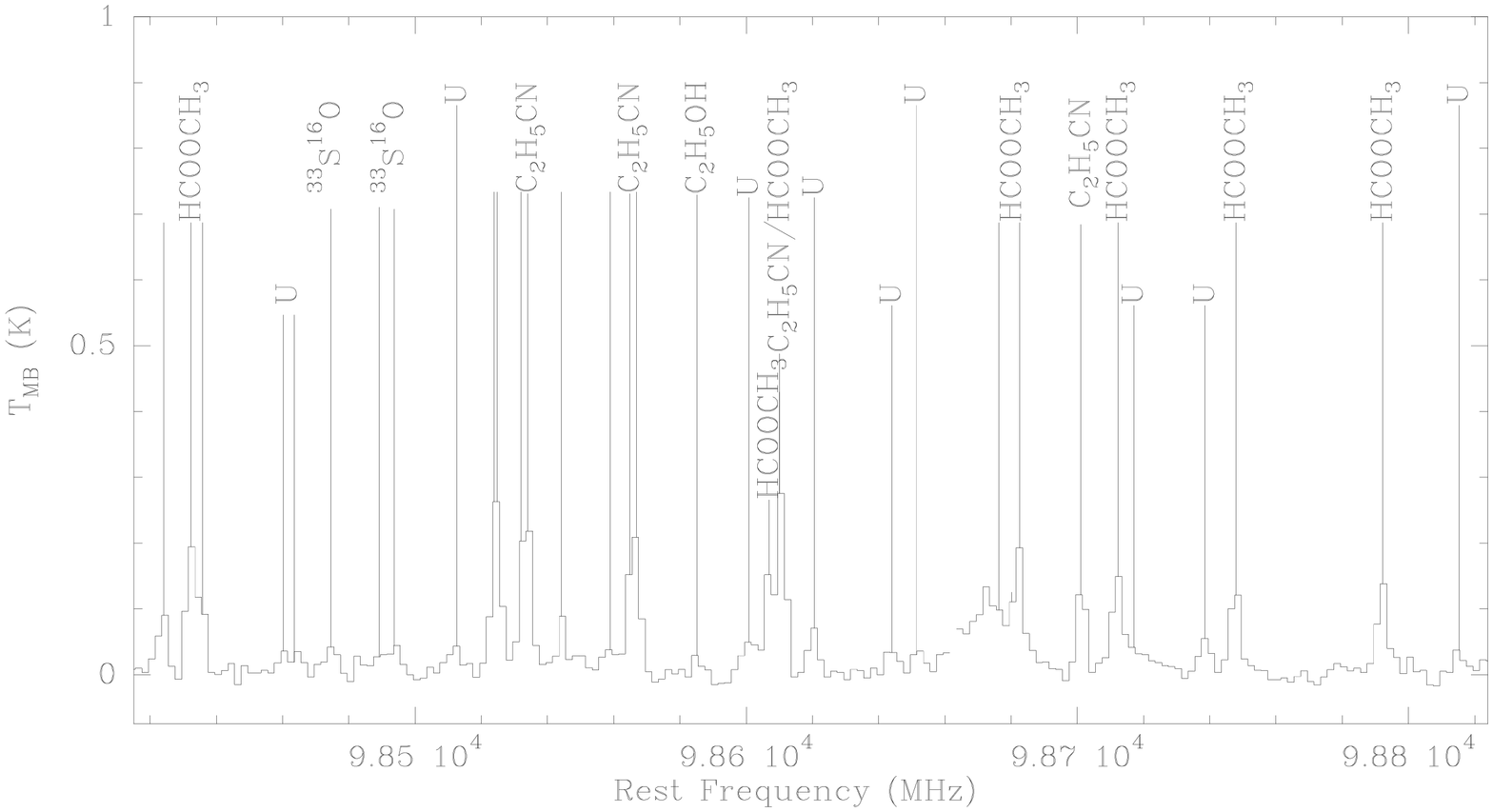}
 \caption{Spectrum of G34.26+0.15 in BAND3MM-1. The identified and
          unidentified (U) lines are marked.}
 \label{fig:iden}
\end{figure*}

In Fig.~\ref{fig:iden} we show a sample spectrum at 3~mm for the source
G34.26+0.15, with identified and unidentified lines. 



\section{Results} 
\label{sect:results}

In this Section we report on and analyse the results for the 6 sources of
our sample which show a number of transitions sufficient to 
derive the physical parameters of our interest (column density,
temperature, abundance). They are listed in Col.~1 of 
Table~\ref{tab:sourcesize}. 
All spectra observed towards these sources 
are shown in Figures A-1 -- A-11 of Appendix A. 
Hereafter, we will use the source name
G10.47 for G10.47+0.03, G10.62 for G10.62--0.38, G19.61 for G19.61--0.23,
G29.96 for G29.96--0.02, G31.41 for G31.41+0.31 and G34.26 for G34.26+0.15.

We will concentrate on the study of \vcy, \ecy, and \de . 
We discuss separately the results of the study of \hcos\
in Sect.~\ref{sect:methyl formate}, because most of the 
transitions of this molecule appear in multiplets with nearly 
the same upper energies, and they are blended with lines of
other molecular species much more than the lines of
\ecy, \vcys and \des. 

\subsection{Masses and H$_2$ Column Densities of the cores} 
\label{sect:sourcesize}


\begin{table}
\begin{center}
\caption[]{Source angular ($\theta_{\rm s}$) and linear ($D$) diameters, 
linewidths at half of the maximum of the \ecy\ transitions ($\Delta v$),
core masses ($M_{\rm gas}$) and H$_2$ total column densities ($N$(\hy )). 
The diameters are derived from mm interferometric data, except when
otherwise specified.
Masses and column densities are computed assuming sources in virial 
equilibrium and with Gaussian distribution of the gas. }
\label{tab:sourcesize}
\begin{tabular}{cccccc}
\hline \hline
Source &   \multicolumn{2}{c}{Diameter}  & $\Delta v$ & $M_{\rm gas}$ & $N$(\hy ) \\
\cline{2-3} 
       &  $\theta_{\rm s}$  & $D$  &  &  &  \\
       &  (\asec)& (pc) & (km s$^{-1}$) & (M$_{\odot}$) & (cm$^{-2}$) \\
\hline
G10.47+0.03& 1.3$^a$& 0.036 & 7.89 & 286 & 1.2$\times 10^{25}$ \\
G10.62$-$0.38& 1.7$^b$& 0.039 & 5.18 & 133 & 4.8$\times 10^{24}$ \\
G19.61$-$0.23& 1.9$^c$& 0.112 & 9.34 & 1239 & 5.4$\times 10^{24}$ \\
G29.96$-$0.02& 1.4$^d$& 0.050 & 6.01 & 228 & 5.1$\times 10^{24}$ \\
G31.41+0.31& 1.1$^e$& 0.042 & 8.52 & 385 & 1.2$\times 10^{25}$ \\
G34.26+0.15& 3.6$^f$& 0.066 & 6.55 & 358 & 4.5$\times 10^{24}$ \\
\hline
\end{tabular}
\end{center}
$^{a}$ Olmi et al.~(1996) \\ 
$^b$ Keto et al.~(1988, from \amms interferometric observations) \\
$^c$ Kurtz et al.~(2000) \\ 
$^d$ Cesaroni et al.~(1998, from \amms interferometric observations)\\
$^e$ Beltr\'an et al.~(2005) \\
$^f$ Akeson \& Carlstrom~(1996, from CH$_3$CN interferometric observations) \\
\end{table}


To derive the abundances of the observed molecules relative
to H$_2$ we need first to estimate the total gas mass, $M_{\rm gas}$.
We have computed $M_{\rm gas}$ and the corresponding H$_2$ total column 
density, $N$(\hy ), assuming the clumps to be 
Gaussian and in virial equilibrium. Under the further 
hypothesis that the lines are not broadened by high optical
depth, one can demonstrate that $M_{\rm gas}$ and $N_{\rm H_2}$ 
are given by:
\begin{equation}
M_{\rm gas}(M_{\odot})=125\, \Delta v^{2} {\rm (\kms )}\,D ({\rm pc})\;\;, 
\label{eq:mass}
\end{equation}
\begin{equation}
N_{\rm H_2} = \frac{4\ln2}{\pi D^2\;m_{H_2}} \, M_{\rm gas}\;\;,
\label{eq_coldens} 
\end{equation}
where $\Delta v$ is the observed full line width at half
maximum, $D=d\theta_{\rm s}$ is the HMC linear diameter
($\theta_{\rm s}$ is the source angular diameter) and
$m_{\rm H_2}$ is the mass of the H$_2$ molecule.

The results are summarised in 
Table~\ref{tab:sourcesize}: HMC angular and linear diameters 
(obtained from the literature) are
listed in Cols.~2 and 3, respectively; line widths, 
core masses and H$_2$ total column dentities are given
in Cols.~4, 5 and 6, respectively. The source linear
diameters have been computed from the distances in 
Table~\ref{tab:source}. The line widths are the average values
of isolated transitions of \ecy . We decided to use
this molecule because it is the species for which we have 
the largest number of unblended lines in each source. 

The angular diameters of the HMCs have been taken from previous 
interferometric observations and require a detailed discussion.
Most of the cores of our sample have been already observed at high 
angular resolution in several molecular tracers, such as
e.g. \amm , CH$_3$CN and CH$_3$OH.
These observations have shown that different molecular
species can probe distinct regions of the core.
Beuther et al.~(2005) have observed at high angular 
resolution the Orion-HC in several molecular tracers, and 
reported significant spatial separation among the regions traced by
O-, N- and
S-bearing species (see also Fig.~2 of Wright et al.~1996).
Notwithstanding this, at present high angular resolution observations 
of complex molecules in the HMCs of our sample are poor: Beltr\'an et 
al.~(2005) have detected one
line of \ecys and \hcos in G31.41, and found that their
integrated emission maps are fairly well overlapping with
that of the 1.4~mm continuum, even though the
two species do not peak exactly at the same position (see their Fig.~26).
Similarly, Mookerjea et al.~(2007) have detected few
lines of \ecy , \hcos and \des in G34.26, showing a separation between 
the emission peaks of O- and N-bearing species of $\sim$0\farcs8, 
but a substantial overlap between their emission maps and that
of the 2.8~mm continuum.
Additionally, Remijan et al.~(2004) have found that
in G19.61, \hcos and \ecys approximately peak at the same position. 
Given these results, as zero-order assumption we will suppose that the 
emission of all the observed molecules arises from the region seen in 
the mm continuum. 

The angular diameters listed in Table~\ref{tab:sourcesize} 
are hence derived from mm continuum interferometric maps with
the exception of three sources: G10.62, G29.96 and G34.26. 
We decided to use the diameter inferred from CH$_3$CN for G34.26 
and from \amms for G29.96, respectively, instead of that from the 
mm~continuum, because the mm continuum position does not agree with that
of the molecular emission (see Akeson \& Carlstrom~1996 and 
Carral \& Welch~1992 for G34.26; Olmi et al.~2003 for G29.96). 
Also, in G34.26 60\% of the mm flux is due to
the free-free continuum emission of an associated \hii\ 
region (Akeson \& Carlstrom~1996).
For G10.62, angular diameters from mm continuum 
interferometric data are not available. Therefore, 
we have assumed the diameter from \amms inversion transitions 
measurements.

The total \hys column density has been derived in our sources also
from C$^{17}$O by Hofner et al.~(2000) and from \ace\ 
by Fontani et al.~(2002). In both works the authors
used Eq.~(\ref{eq:mass}), adopting the diameters and the virial 
masses they derive from the molecular lines that they observed. 
These estimates are $\sim 2$ orders of magnitude lower than those 
obtained from \ecy, consistently with the fact that the
\CIII\ and \ace\ transitions trace the parsec-scale molecular clump 
in which the HMC is embedded.

\subsection{Rotational temperatures, total column densities and
fractional abundances}
\label{sect:temp}

\subsubsection{Derivation of the parameters}

We derive rotational temperatures and total column densities applying 
the population diagram method described by Goldsmith \& Langer 
(1999, hereafter GL99).
The main assumptions in this method are: (i) optically thin lines; 
(ii) local thermodynamic equilibrium (LTE) conditions.
Under assumption (i), one can compute the column 
density $N_{\rm u}$ of the upper level, $u$, of each transition
from its integrated intensity according to the relation:
\begin{equation}
\frac{N_{\rm u}}{g_{\rm u}}=\frac{3k}{8\pi^{3}}\frac{W}{\nu \mu^{2}S}
\label{eq_nu}
\end{equation} 
where $g_{\rm u}$ is the statistic weight of the upper level, $k$ 
the Boltzmann constant, $W$ the integrated line intensity 
(in K \kms ), $\nu$ the line rest 
frequency, $\mu$ the molecule's dipole moment and $S$ the line strength.
Then, under assumption (ii),
we can use Eq.~(21) of GL99 to derive rotational temperatures 
and total column densities of the molecules from least square fits 
to the data. The validity of assumptions (i) and (ii) will be discussed in 
detail in Sect.~\ref{optthin}.
The parameters derived from the XCLASS fitting procedure
for the lines identified and used in the rotation diagrams
are given for each source in Tables~\ref{tab:g1042007} - \ref{tab:g342006} 
of Appendix B.
The rotation diagrams for all sources are shown 
in Figs.~\ref{fig:g10.ps} - \ref{fig:g34.ps} of Appendix C.

The nuclear spin degeneracy and the partition functions used 
in Eq.~(21) of GL99 have been taken from Blake et 
al.~(1987) and Turner et al.~(1991) 
for \ecy , \vcys, and \hco, and from Groner et 
al.~(1998) for \de.
Partition functions are calculated analytically in the 
approximation of high temperatures (i.e. $hA\ll kT$ 
for asymmetric top species, where $A$ is the Einstein coefficient).
For the vibrational transitions, we computed the partition 
function following the 
approximation of Nummelin \& Bergman~(1999). 
In the rotation diagrams, the column density of each level observed 
at 2~mm and 1.3~mm has been normalized to the telescope beam at
3~mm ($\sim 23$\asec ) assuming a source diameter negligible with
respect to the beam size, thus obtaining {\it beam-averaged} 
total column densities. Then, {\it source-averaged} total column 
densities have been inferred by multiplying the beam-averaged 
values by $(23/ \theta_{\rm s})^{2}$.

The molecule's fractional abundance, $X$, is given by the 
source-averaged total column density devided by the H$_2$ 
total column density listed in Col.~5 of Table~\ref{tab:sourcesize}: 
$X=N({\rm molecule})/N({\rm H_{2}})$.

\subsubsection{Temperatures and column densities}

\begin{table*}
  \caption[Results]{Rotational temperatures ($T$), beam and source 
averaged total column densities ($N_{\rm b}$ and $N_{\rm s}$) and 
abundances ($X$) with respect to H$_2$ of \ecy, \vcys and \de.}
   \label{tab:Results} 
\begin{center}
\begin{tabular}{l l  c  c c c}
\hline
\hline
Molecule      & Source & $T$          & $N_{\rm b}$ & $N_{\rm_s}$  & $X$ \\
	      &	       &(K)           &  (cm$^{-2})$ & (cm$^{-2})$ & \\
\hline
\ecy          &G10.47&103$\pm$12 &3.6$\pm$0.9$\times 10^{14}$ & 1.1$\pm$0.3$\times 10^{17}$ &9.6$\times 10^{-9}$ \\
	      &G10.62&89$\pm$13  &2.5$\pm$0.9$\times 10^{13}$ & 4.6$\pm$1.6$\times 10^{15}$&9.5$\times 10^{-10}$ \\
	      &G19.61&116$\pm$12 &1.5$\pm$0.2$\times 10^{14}$ & 2.2$\pm$0.3$\times 10^{16}$&4.1$\times 10^{-9}$ \\
	      &G29.96&121$\pm$17 &5.4$\pm$1.3$\times 10^{13}$& 1.5$\pm$0.3$\times 10^{16}$&3.0$\times 10^{-9}$  \\
	      &G31.41&118$\pm$13 &2.3$\pm$0.5$\times 10^{14}$ & 1.0$\pm$0.2$\times 10^{17}$&8.3$\times 10^{-9}$ \\ 
	      &G34.26&130$\pm$13 &1.9$\pm$0.3$\times 10^{14}$ & 8.4$\pm$1.3$\times 10^{15}$&1.9$\times 10^{-9}$ \\     
\hline
\vcy          &G10.47&176$\pm$35&2.0$\pm$0.6$\times 10^{14}$ & 6.1$\pm$1.7$\times 10^{16}$&5.1$\times 10^{-9}$  \\
              &G10.62& --       & $\leq$1.7$\times 10^{13}$ & $\leq$3.2$\times 10^{15}$ &$\leq$6.8$\times 10^{-10}$ \\ 
              &G19.61&123$\pm$24&6.5$\pm$2.1$\times 10^{13}$ & 9.7$\pm$3.2$\times 10^{15}$&1.8$\times 10^{-9}$ \\
              &G29.96&fixed     &2.7$\times 10^{13}$ & 7.3$\times 10^{15}$ &1.4$\times 10^{-9}$ \\
	      &G31.41&111$\pm$12&8.4$\pm$1.8$\times 10^{13}$ & 3.7$\pm$1.0$\times 10^{16}$&3.1$\times 10^{-9}$ \\
	      &G34.26&67$\pm$15 &5.6$\pm$2.5$\times 10^{13}$ & 2.4$\pm$1.0$\times 10^{15}$&5.3$\times 10^{-10}$ \\
\hline
\de           &G10.47&156$\pm$37&1.2$\pm$0.3$\times 10^{14}$ & 3.9$\pm$1.0$\times 10^{16}$&3.3$\times 10^{-9}$  \\       
              &G10.62&16$\pm$1  &1.4$\pm$0.2$\times 10^{13}$ & 2.6$\pm$0.4$\times 10^{15}$&5.4$\times10^{-10}$ \\
	      &G19.61&158$\pm$17&1.3$\pm$0.1$\times 10^{13}$ & 2.0$\pm$0.2$\times 10^{15}$&3.7$\times10^{-10}$ \\
	      &G29.96&141$\pm$26&2.3$\pm$0.4$\times 10^{13}$ & 6.2$\pm$1.0$\times 10^{15}$&1.2$\times 10^{-9}$ \\
	      &G31.41&127$\pm$25&7.9$\pm$1.9$\times 10^{13}$ & 3.5$\pm$1.0$\times 10^{16}$&2.9$\times 10^{-9}$  \\
	      &G34.26&116$\pm$25&7.6$\pm$2.3$\times 10^{13}$ & 3.3$\pm$1.0$\times 10^{15}$&7.3$\times 10^{-10}$ \\
\hline
\multicolumn{5}{l}{ }\\
\end{tabular}
\end{center}
\end{table*}

Rotational temperatures ($T$), beam- and source-averaged total column 
densities ($N_{\rm b}$ and $N_{\rm s}$)
are given in Cols.~3, 4 and 5 of Tab.~\ref{tab:Results}, respectively.

The rotational temperatures typically range from $\sim 100$ to
$\sim 150$ K in all tracers. In each source, the estimates obtained 
from \ecy , \vcys and \des are in good agreement among them,
with the exception of G34.26 and G10.62. 
For G34.26, the estimate from \vcys is a factor of 2 lower 
than those derived from \des and \ecy , but this is likely due to
the fact that in the rotation diagram there are no points
above $\sim 200$ K. In G10.62, the temperature from \des is much 
lower ($\sim 16$ K) than the other two estimates, but it has been 
obtained from only few line detections and is thus less accurate. 

The temperatures derived from \ecys and \des are in agreement 
both with those obtained by Bisschop et al.~(2007) 
in other HMCs,
and with those derived from other HMC tracers
by Kurtz et al.~(2000, see their Table~1),
further attesting they are tracing the hot gas of the cores.
The column densities of \ecys are in good agreement with the values 
of Bisschop et al.~(2007) in similar objects,
while for \des we find column densities one order of magnitude 
smaller.
In computing source-averaged column densities Bisschop et 
al.~(2007) assume a source diameter corresponding to
that of the region where the temperature is higher than
$\sim 100$ K, that is comparable to those of our
objects ($\sim 1$\asec ). Since masses and luminosities of the
cores are comparable as well, the observed difference might
reflect a different \des chemistry in the HMCs of our sample and 
those of Bisschop et al.~(2007).

Only three transitions of \vcys\ have been identified towards G29.96.
In this case the column densities have been estimated fixing the rotation 
temperature to that found from \ecy .
The assumption of equal rotational temperature for 
\vcys\ and \ecys\ is justified both by observations 
(Schilke et al.~1997), which have shown that
these molecules trace a very similar region,
and by chemical models (Charnley et al.~1992;
Caselli et al.~1993), which predict that \vcys\ forms 
in gas phase reactions involving \ecy .

\subsubsection{Molecular fractional abundances}

Molecular fractional abundances are listed in Col.~6 of
Table~\ref{tab:Results}.
The average abundances of \ecys, \vcys and \des are 
$\sim 4.6\times 10^{-9}$, $2.4\times 10^{-9}$ and 
$\sim 1.5\times 10^{-9}$, respectively. These values
are comparable to those found in the Orion Hot Core
(Sutton et al.~1995) and in Sgr B2
(Nummelin et al.~2000). \ecys abundances
are also comparable to the values obtained by
Bisschop et al.~(2007) in a similar sample of HMCs,
while for \des we find a discrepancy of $\sim 2$ orders of magnitude
with respect to the estimates of Bisschop et al.~(2007).
Again, we have to consider that 
in deriving their H$_2$ column densities Bisschop et 
al.~(2007) use a model with a power-law density
profile in the HMC of the type $n\propto r^{-1.5}$. Since
we derive H$_2$ total column densities directly from observational
measurements, the observed difference might be due to
some assumptions in the model used by Bisschop et 
al.~(2007) which cannot be applied to our sources.


\subsubsection{LTE conditions and optical depth of the lines}
\label{optthin}

Our results have been derived assuming LTE conditions. The critical
densities of the observed transitions are of the 
order of $10^{6}-10^{7}$ cm$^{-3}$ (see Genzel 1991 for the
collisional coefficients), and always $\leq 10^{8}$ cm$^{-3}$.
Assuming spherical and homogeneous cores,
from the linear diameters and gas column densities
listed in Cols.~3 and 6 of Tab.~\ref{tab:sourcesize},
respectively, we find that all of our HMCs have H$_2$ volume densities
larger than $10^7$~cm$^{-3}$. This makes us confident that the LTE 
approximation is correct. 

The other fundamental assumption of the rotation diagram method
is that of optically thin lines.
If the lines are optically thick, the expression for the column 
density of the upper levels given in Eq.~(\ref{eq_coldens}) has 
to be multiplied by the factor $\tau / (1-{\rm exp}(-\tau))$
(see e.g. Eq.~16 of GL99), 
where $\tau$ is the line opacity. Such a correction 
modifies the results of the rotation diagrams: large optical 
depths are expected to affect mostly the low-excitation
transitions, thus causing a flattening of the plots
at lower energies. In this case, the temperatures estimated
assuming optically thin lines have to be taken as upper limits.

Line opacities can be derived from the line intensity ratio of 
two isotopologues of the same species. With the help of the 
JPL catalogue, we checked whether rotational transitions of
isotopologues of the molecules of interest fall in the
observed frequency bands. We have not
found lines of the \vcys isotopologues: this is likely due 
to the fact that 
not all of the line frequencies have been determined yet in the laboratory.
On the other hand, the observed bandwiths cover 7 lines of 
\ecys isotopologues (5 CH$_3$C$^{13}$H$_2$CN lines and 2 
CH$_3$CH$_2$C$^{13}$N lines), but they are all undetected. 
Therefore, we can derive only upper limits on their optical depth:
assuming a $^{12}$C/$^{13}$C ratio of 10, the minimum upper limit
turns out to be $\tau = 1.7$ which is not sufficient to
conclude that the \ecys lines are optically thin.
Neverthless, the good agreement between the data and
the linear fits indicates that the assumption of optically
thin lines is reasonable in our sources.

\subsection{\hcos}
\label{sect:methyl formate}

\begin{figure*}[t!]
 \centering
 \resizebox{\textwidth}{!}{\includegraphics[angle=-90]{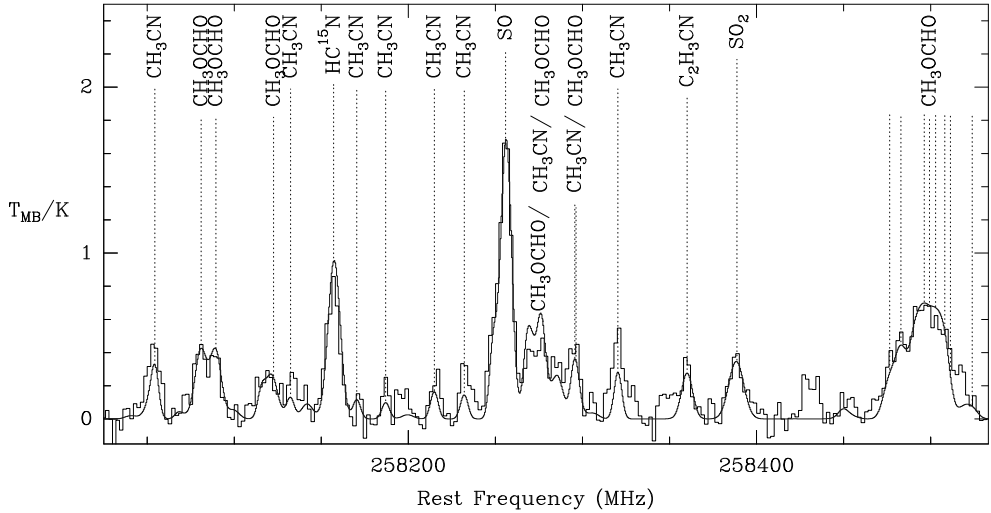}}
 \caption[HCOOCH3 for G10.47]
 {\label{fig:g10-258}{Sample spectrum of \hcos in BAND1-HCN towards the source G10.47. We used 
the XCLASS software to solve the blending problem in this frequency band.}}
\end{figure*}

As pointed out in Sec.~\ref{sect:results}, many \hcos transitions
are blended with other lines. Therefore, we used the software 
XCLASS to solve this problem (see Sect.~\ref{sect:Observ}).
The spectra in BAND1-HCN offer the only opportunity in our data sets to fit 
simultaneously many \hcos lines with energy of the upper level 
from about 150 to 300 K. 
In Fig.~\ref{fig:g10-258} we show an example spectrum at 258500\,MHz
with highly blended lines fitted by XCLASS. 
The rotation diagrams of only two sources, G10.47 and G31.41, 
give reliable fit results.
Although the temperature is not well determined by the fit, 
the total column density is quite accurate since it accounts 
mainly for the total intensity of the blended feature. 
We derive source averaged total column densities of 
3.7$\times 10^{17}$\,cm$^{-2}$ for G10.47 and 
5.3$\times 10^{17}$\,cm$^{-2}$ for G31.41. 
The \hcos fractional abundances are therefore $\sim 3.1\times 10^{-8}$
and $\sim 4.4\times 10^{-8}$ for G10.47 and G31.41,
respectively. Both column densities and abundances
are consistent with the values found by Bisschop et 
al.~(2007) in similar HMCs.

\subsection{Vibrationally excited \vcys in G10.47}
\label{sect:vcyvibr} 

\begin{figure*}[t!]
 \centering
 \resizebox{\textwidth}{!}{\includegraphics[angle=-90]{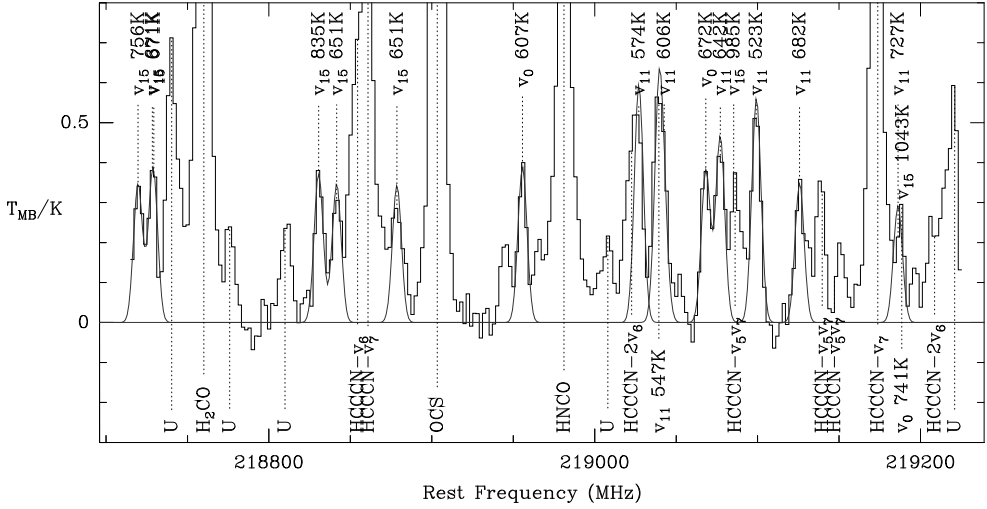}}
 \caption[vibrational lines of G10.47]
 {\label{fig:c2h3cn_vibr}{
 Vibrationally excited vinyl cyanide in G10.47 at 218 GHz.
 The lines are marked with their vibrational mode and upper energies are given.}}
\end{figure*}

Several lines of vibrationally excited \vcys have been identified
 at 218\,GHz (see Fig.~\ref{fig:c2h3cn_vibr}) towards the source G10.47.
The main line parameters are given in Table~\ref{vcyvib}.
These lines have been identified for the first time by Nummelin et 
al.~(1998) in the Sagittarius B2 molecular cloud.
The frequencies for the in-plane CCN bend $v_{11}$ at 342 K and the CCN 
out-of-plane bend $v_{15}$ at 486 K are taken from the JPL catalogue 
(Poynter \& Pickett~1985, Pickett et al.~1996).

\begin{table}
  \caption[Results]{Rest frequency ($\nu$), energy of the upper level 
($E_{\rm u}$) and integrated intensity ($\int T_{\rm MB}{\rm d}v$) of
9 unblended vibrational lines of \vcys detected towards G10.47.}
   \label{vcyvib} 
\begin{center}
\begin{tabular}{c  c c}
\hline
\hline
$\nu$     & $E_{\rm u}$ & $\int T_{\rm MB}{\rm d}v$ \\
(MHz)     & (K)    &  (K \kms ) \\
\hline
 218719.6 &   756 &    3.23 \\ 
 218728.9 &   671 &    4.15 \\ 
 218830.5 &   835 &     3.57 \\ 
 218841.6 &   651 &    3.31 \\ 
 218878.6 &   651 &    3.28 \\ 
 219027.1 &   574 &      5.68 \\ 
 219077.1 &   642 &     4.71 \\ 
 219099.1 &   523 &     5.81 \\ 
 219125.8 &   682 &    3.46 \\ 
\hline
\end{tabular}
\end{center}
\end{table}

\begin{table}
  \caption[Results]{Rest frequency ($\nu$), energy of the upper level 
($E_{\rm u}$) and integrated intensity ($\int T_{\rm MB}{\rm d}v$) of
12 unblended vibrational lines of \vcys detected towards 
G10.47 and G31.41.}
   \label{ecyvib} 
\begin{center}
\begin{tabular}{c  c c c}
\hline
\hline
Source & $\nu$     & $E_{\rm u}$ & $\int T_{\rm MB}{\rm d}v$ \\
       & (MHz)     & (K)    &  (K \kms ) \\
\hline
G10.47 & 98464 &   410 &  2.0466 \\
       & 98539 &   409 & 0.85978 \\
       & 98619 &   361 & 5.3485 \\
       & 98672 &   377 & 1.9452 \\
       & 161338 &  422 & 3.4767 \\
       & 224380 &  342 & 2.9466 \\
       & 224396 &  616 & 2.1549  \\
       & 224409 &  542 & 2.8934 \\
       & 224480 &  522 & 2.7522 \\
G31.41 &  98464 & 410  &  0.64 \\
       & 98619  & 361 & 1.83 \\
       & 98672  & 377 &  1.27 \\
       & 98737  & 342 &  1.90 \\
       & 161338 & 422 &   2.35 \\
       & 224378 & 577 &   4.28 \\
       & 224408 &  542 &   3.63 \\
       & 224479 & 522  &   2.06 \\ 
\hline
\end{tabular}
\end{center}
\end{table}

We also have found two "weak" \vcys transitions (i.e. transitions with low 
$\mu^2$S, at $E_{\rm u}<$\,100 K, see Fig.~\ref{fig:g10.ps}).
Both vibrational and ``weak'' lines are unlikely to be optically thick.
Therefore, we combine them to derive both temperature and column density.
We obtain a rotational temperature of 135 $\pm$ 13 K and a source averaged 
total column density of 1.3$\pm$0.6$\times 10^{18}$ cm$^{-2}$. 
The temperature is equal, within the error bar, to that given in 
Tab.~\ref{tab:Results} but the density is about a factor of 20 higher. 
Note that it is also a factor of 10 higher than \ecys column density 
(see Tab.~\ref{tab:Results}).
The presence of both "weak" rotational and vibrational lines suggest
that some "strong" rotational transitions can suffer from high optical 
depth. Therefore, correcting the data following the method 
described in Sect.~\ref{optthin}, we obtain a best fit for the following 
parameters: $N_{\rm s}\,=\,(2.3\,\pm\,0.4)\,\times\,10^{18}$\,cm$^{-2}$,
$T\,=\,133\,\pm\,3$\,K, and $\theta\,\sim\,1.6$\,\asec .

\subsection{Vibrationally excited \ecys in G10.47 and G31.41}\label{sect:ecyvibr} 

\begin{figure*}
   \centering
   \includegraphics[width=14cm]{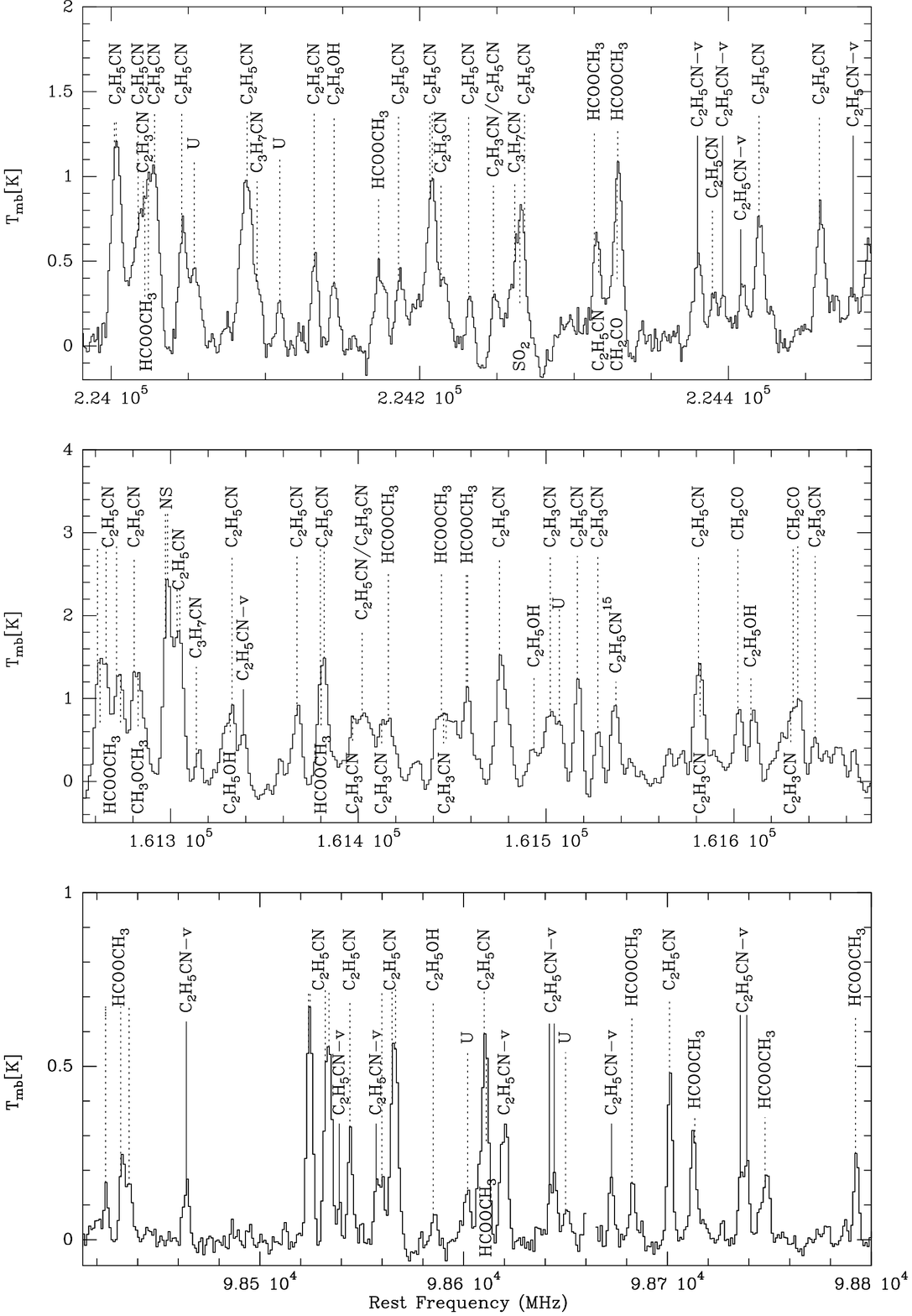}
      \caption{Bands in which vibrationally excited \ecys transitions
      		have been identified (solid line) towards the source G10.47.
              }
         \label{fig:c2h5cnvibrg10}
   \end{figure*}

\begin{figure*}
   \centering
   \includegraphics[width=14cm]{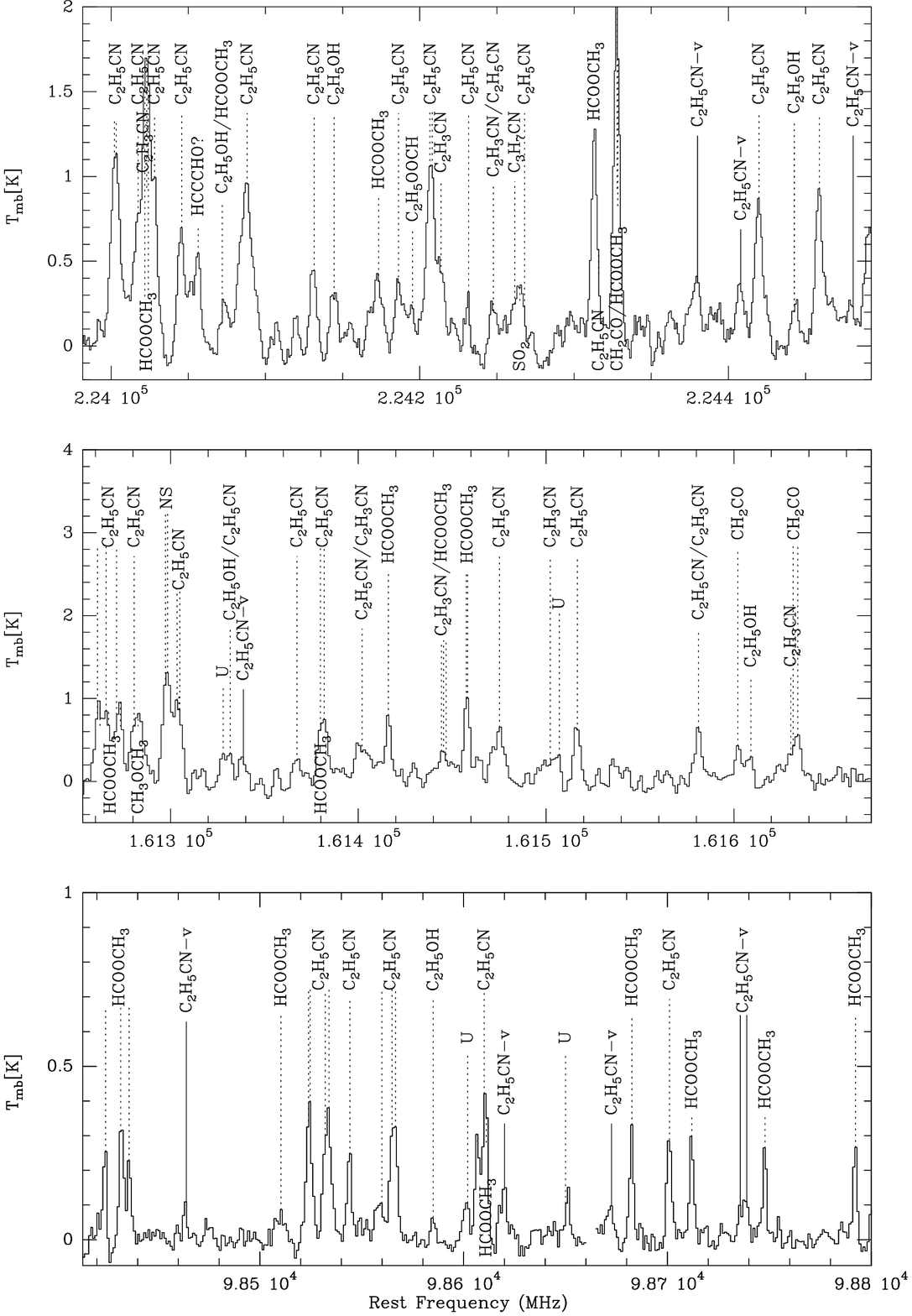}
      \caption{Bands in which vibrationally excited \ecys transitions
      		have been identified (solid line) towards the source G31.41.
              }
         \label{fig:c2h5cnvibrg31}
   \end{figure*}

Using the compilation of Pearson~(2002), we have searched 
for vibrationally excited \ecys transitions.
We have identified 9 lines in G10.47 and 8 in G31.41.
In Fig.~\ref{fig:c2h5cnvibrg10} and Fig.~\ref{fig:c2h5cnvibrg31}, we
show the 3 bands in which vibrationally excited lines
of \ecys have been identified. The main parameters of the 
identified lines are listed in Table~\ref{ecyvib}.
Vibrationally excited \ecys was firstly identified by Mehriger et
al.~(2004) in Sgr B2.

\section{Discussion}
\label{discu}

\subsection{Comparison between N-bearing and O-bearing molecules}
\label{comparison}

A different spatial distribution between N-bearing and 
O-bearing molecules has been observed in some HMCs
(see Blake et al.~1987, Wright et al.~1992,
1996, and Beuther et al.~2005 for Orion; 
Wyrowski et al.~1999 for W3(H$_2$O)). 
Caselli et al.~(1993) interpreted the
difference seen in Orion as being due to the fact that
N-bearing and O-bearing species trace regions of the HMCs with 
different physical properties. Charnley et al.~(1992)
simply use different grain mantle compositions for the Orion-HC and 
Compact Ridge to reproduce the observed differentiation.

In order to see if such a differentiation is observable
in the HMCs of our sample, we compare in each source the
main physical properties (velocity, 
column density, temperature and abundance) obtained from 
N- and O-bearing species. 

\subsubsection{Line velocities and widths at half maximum}
\label{sec:velo}

Peak velocities, $V_{\rm LSR}$, and line widths at half maximum,
$\Delta V$, of the unblended lines detected in each source
are listed in Cols.~3 and 4, respectively, of 
Table~\ref{tab:velocities}. 
Both \vlsr{} and $\Delta V$ have been
obtained from single Gaussian fits to the lines.
In Cols.~2 and 3 of Table~\ref{tab:averages} we report the average 
values of \vlsr{} for the unblended lines of \ecy , $V_{\rm N}$, 
and \de , $V_{\rm O}$, respectively, and their difference
($V_{\rm N}-V_{\rm O}$) is given in Col.~4.
In Cols.~5, 6 and 7 of the same Table we also list the
average value of the linewidths for \ecy , $\Delta V_{\rm N}$, and
\de , $\Delta V_{\rm O}$, and their ratio 
($\Delta V_{\rm N}/\Delta V_{\rm O}$), respectively.

\begin{figure}
 \begin{center}
 \resizebox{\hsize}{!}{\includegraphics{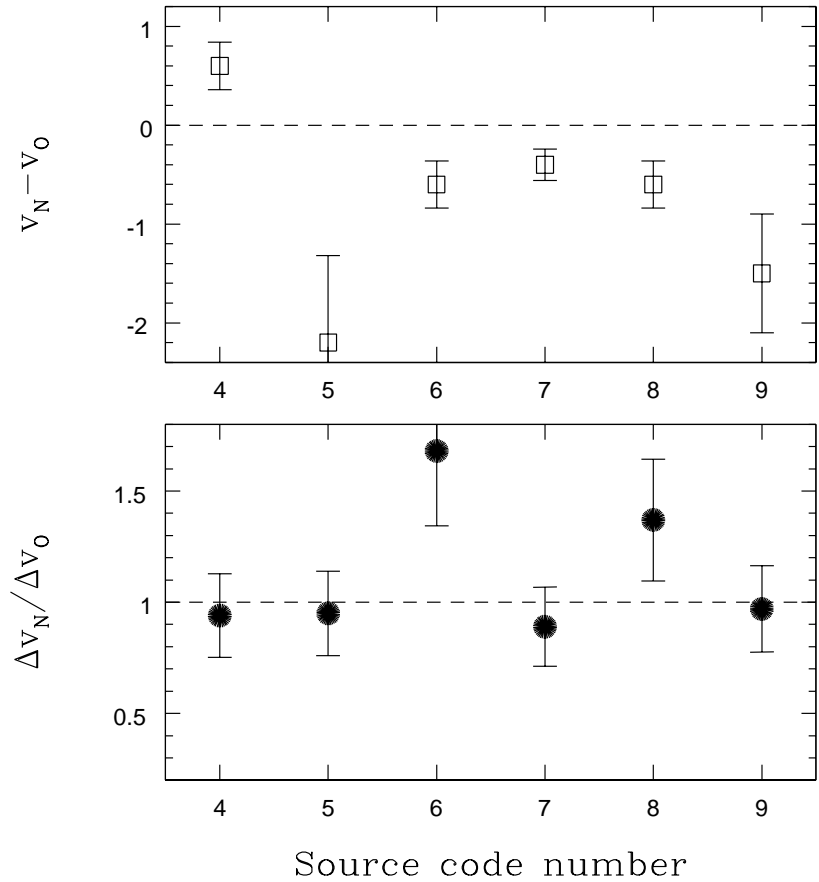}} 
 \end{center}
 \caption[Differences from N-bearing / O-bearing molecules]
 {\label{fig:del_velfw}
Top panel: difference between the average velocities of unblended transitions of
\ecys ($V_{\rm N}$) and \des ($V_{\rm O}$). Bottom panel: ratio between the average line widths at
half maximum of \ecys ($\Delta V_{\rm N}$) and \des ($\Delta V_{\rm O}$).
Source code number is taken from Table~1.}
\end{figure}

The velocity differences $V_{\rm N}-V_{\rm O}$ and 
line width ratios $\Delta V_{\rm N}/\Delta V_{\rm O}$ found
in each source are then plotted in Fig.~\ref{fig:del_velfw}.
$V_{\rm N}-V_{\rm O}$ shows that on average the \ecys lines
have different velocities from the \des lines, in
agreement with the idea that they are tracing different regions.
The largest difference is for G10.62, for which however 
we stress that $V_{\rm N}$ has been derived from 3 lines only,
and one of these (the line at 224045 MHz) has a very uncertain
velocity.
On the other hand, no systematic differences are present in the
line widths, except for G19.61 for which the \ecys lines are
more than $\sim 1.5$ times larger than the \des lines.
We point out that the values plotted in  Fig.~\ref{fig:del_velfw}
are {\it average} values between lines observed at 
different wavelenghts, and this can
influence a lot the discussion of these quantities. In fact, the
higher excitation lines, which trace the denser and hotter
portion of the HMC, are also expected to have highest linewidths.
Therefore, the number of unblended lines detected in each band, which
are different for each source, may influence the average values
both of velocities and linewidths. 
Also, each rotational transition of \de\ is split into 
four components coming from torsional motions of the molecule. 
The separation of these components in velocity is much smaller than 
the spectral resolution of our data, so that they are not resolved. 
Even though the software can derive the intrinsic $\Delta V$ 
and \vlsr{} from the theoretical separation of the different components,
we believe that $\Delta V$ and \vlsr{} from these lines are less 
accurate than those derived from \ecy .
For these reasons, we stress that the quantities
plotted in Fig.~\ref{fig:del_velfw} have to be interpreted carefully.

\subsubsection{Abundance correlations}
\label{mol_abb}

\begin{figure*}[t!]
 \begin{center}
 \resizebox{\textwidth}{!}{\includegraphics[angle=-90]{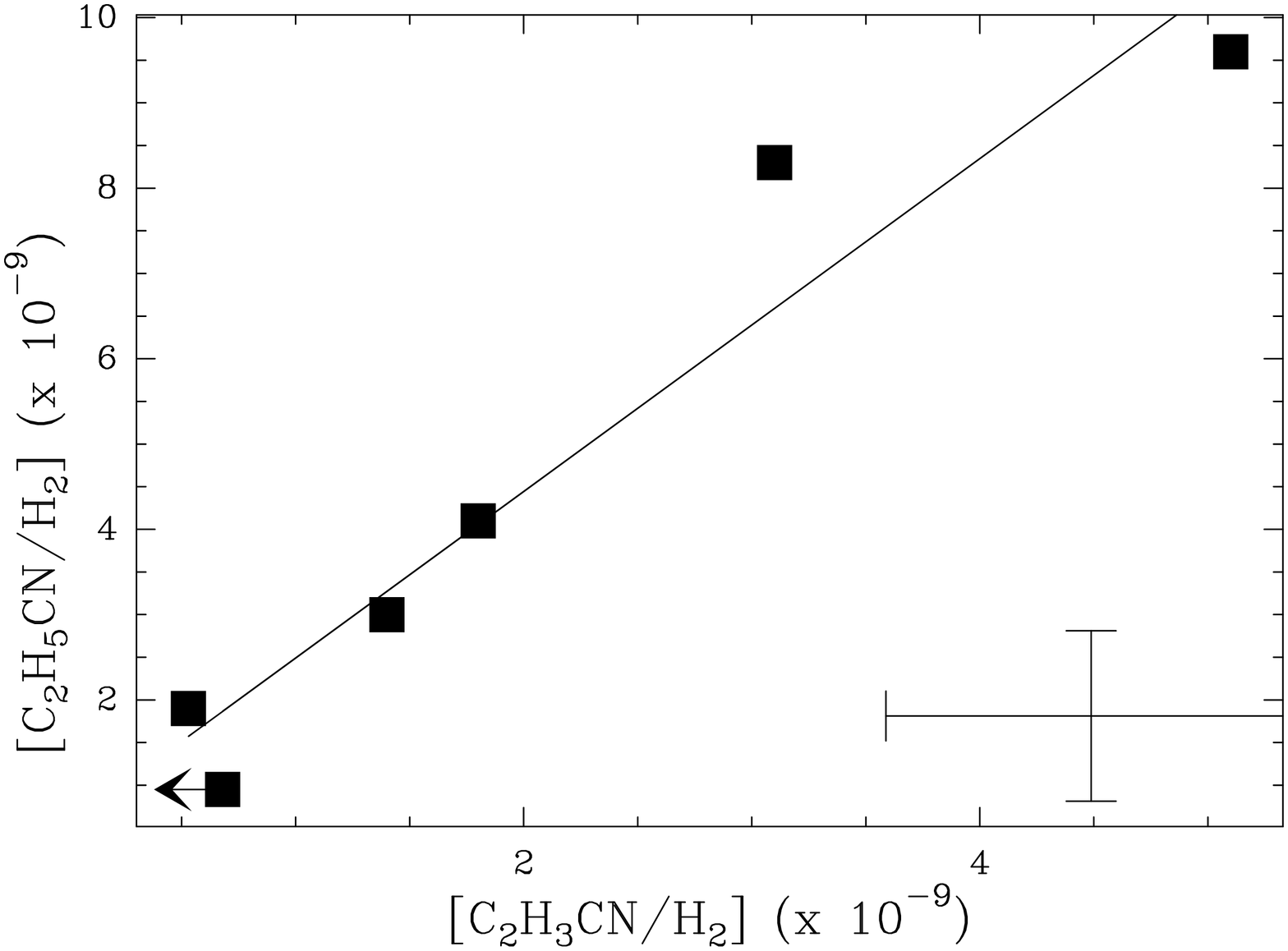}%
 			   \includegraphics[angle=-90]{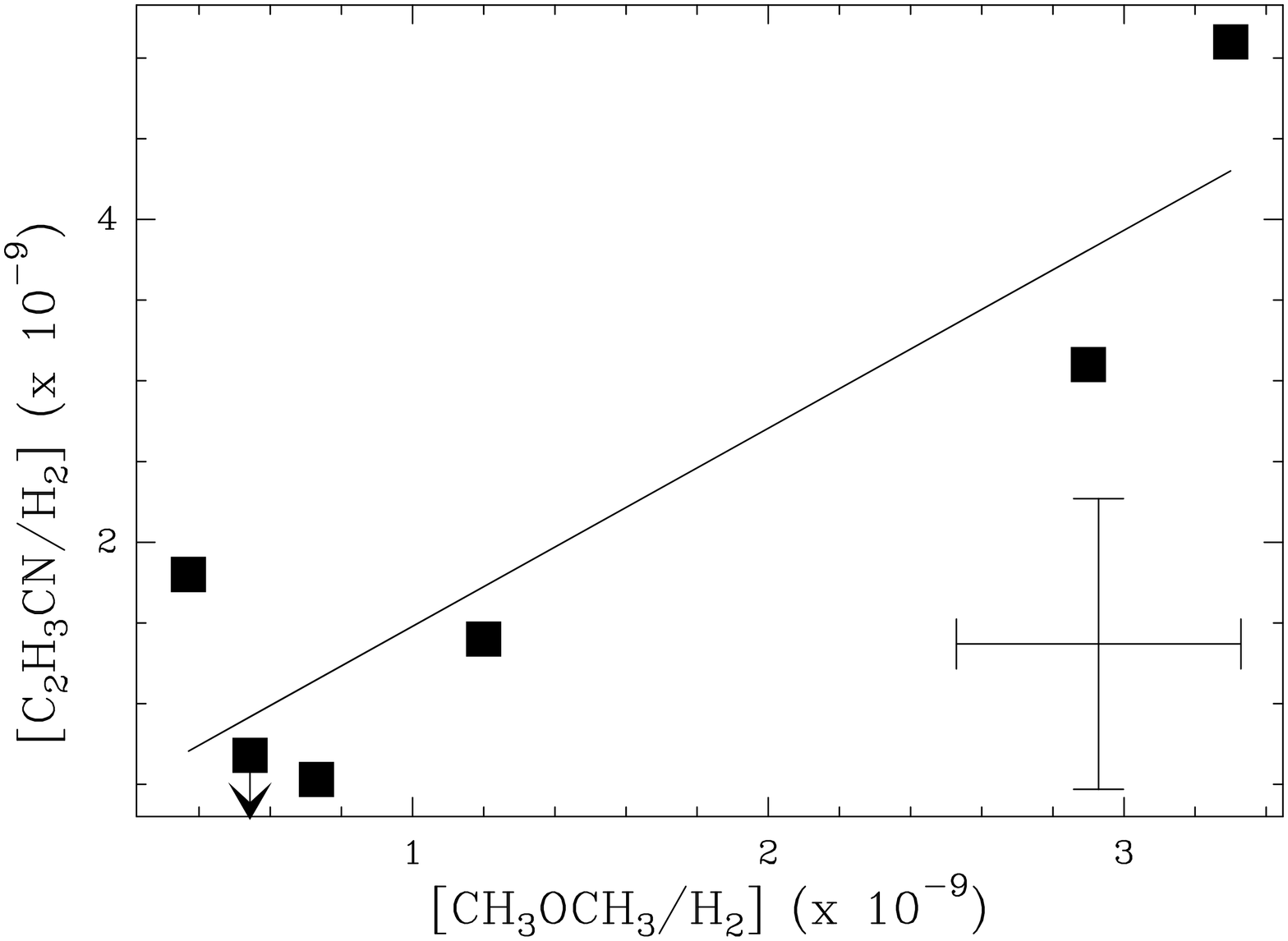}%
			   \includegraphics[angle=-90]{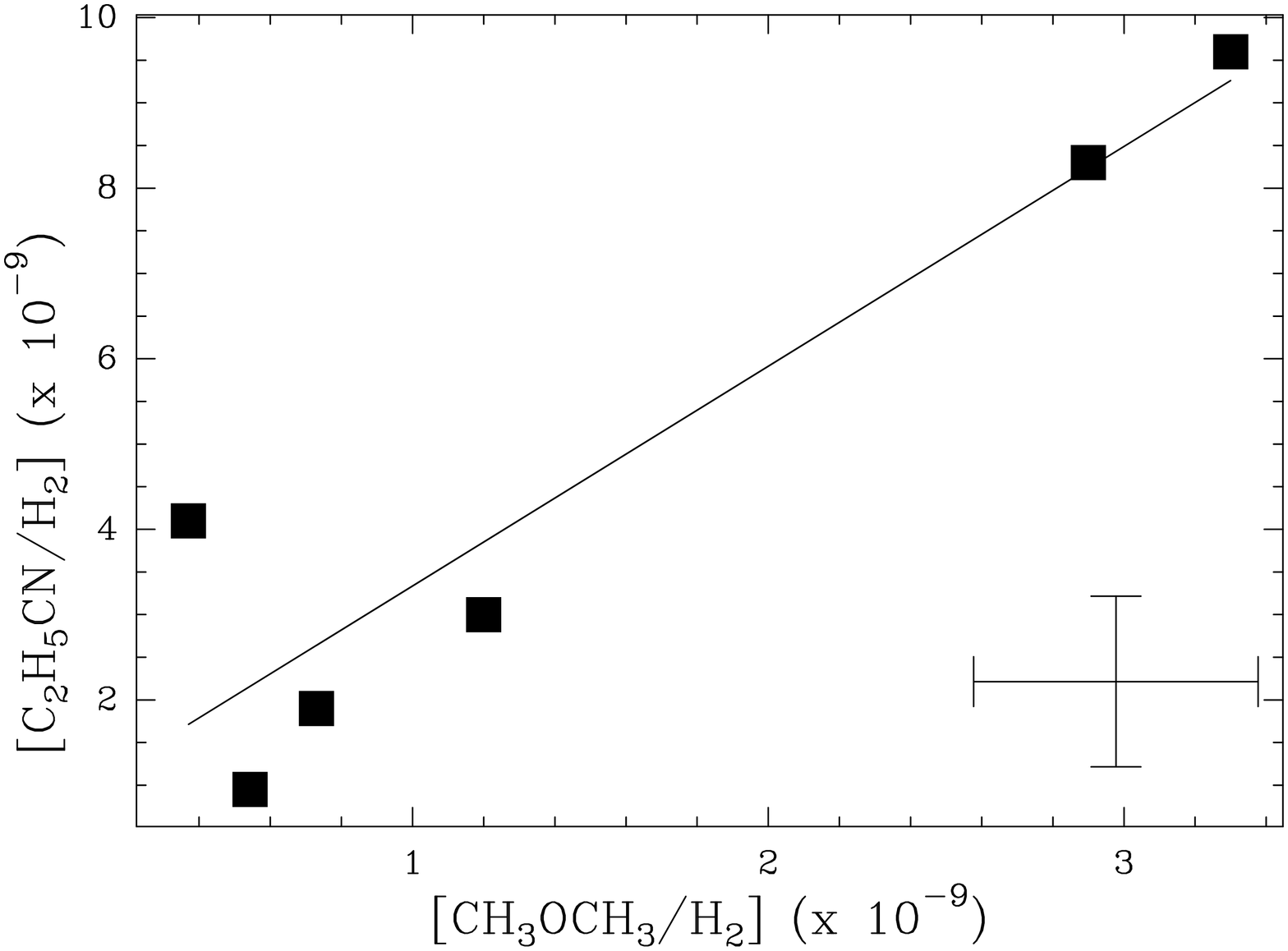}} 
 \end{center}
 \caption
 {\label{fig:abb_correl}{Comparison between the molecular abundances of
\ecys and \vcys (left panel), \vcys and \des (central panel) and
\ecys and \des (right panel). All abundances are relative to H$_2$.
The solid lines correspond to the best-fits. The upper limit for
the \vcys abundance in G10.62 is indicated by an arrow. 
Errorbars representing the
typical uncertainties are shown in the lower right corners.}}
\end{figure*}

The abundances relative to H$_2$ of \ecy, \vcys and \des
are compared in Figure~\ref{fig:abb_correl}. For each pair
of tracers, we have performed a least square fit to the data to check
a possible correlation between the different abundances. 
The left panel of Figure~\ref{fig:abb_correl} shows that the
abundances of the two N-bearing molecules, \ecys and \vcy, are strongly 
correlated (correlation coefficient $r\sim 0.96$). This result
agrees with the theoretical prediction that \vcys is
producted through gas-phase reactions involving \ecys
(Caselli et al.~1993). On the other hand, central
and right panels of Figure~\ref{fig:abb_correl} show that there is 
correlation also between the abundances of \ecys and \vcy,
and that of \des (correlation coefficients $\sim 0.90$ and
$\sim 0.86$, respectively). \des is thought to form both
through reactions on grain surfaces (CH$_3$ + CH$_3$O $\rightarrow$ \de,
Allen \& Robinson~1977), and in the gas phase
through reactions between methanol and protonated methanol
(Garrod \& Herbst~2006), but both formation processes are
not expected to involve species containing  nitrogen.
Furthermore, Bisschop et al.~(2007) have indeed shown 
that in their HMCs the \des abundance does not correlate with 
that of other 
N-bearing species (see their Table~8). Neverthless, based on
considerations about abundance ratios and temperatures, they suggest
that the O- and N-bearing molecules observed in their work
may have experienced a common formation scheme.

\begin{figure*}[t!]
 \begin{center}
 \resizebox{\textwidth}{!}{\includegraphics[angle=-90]{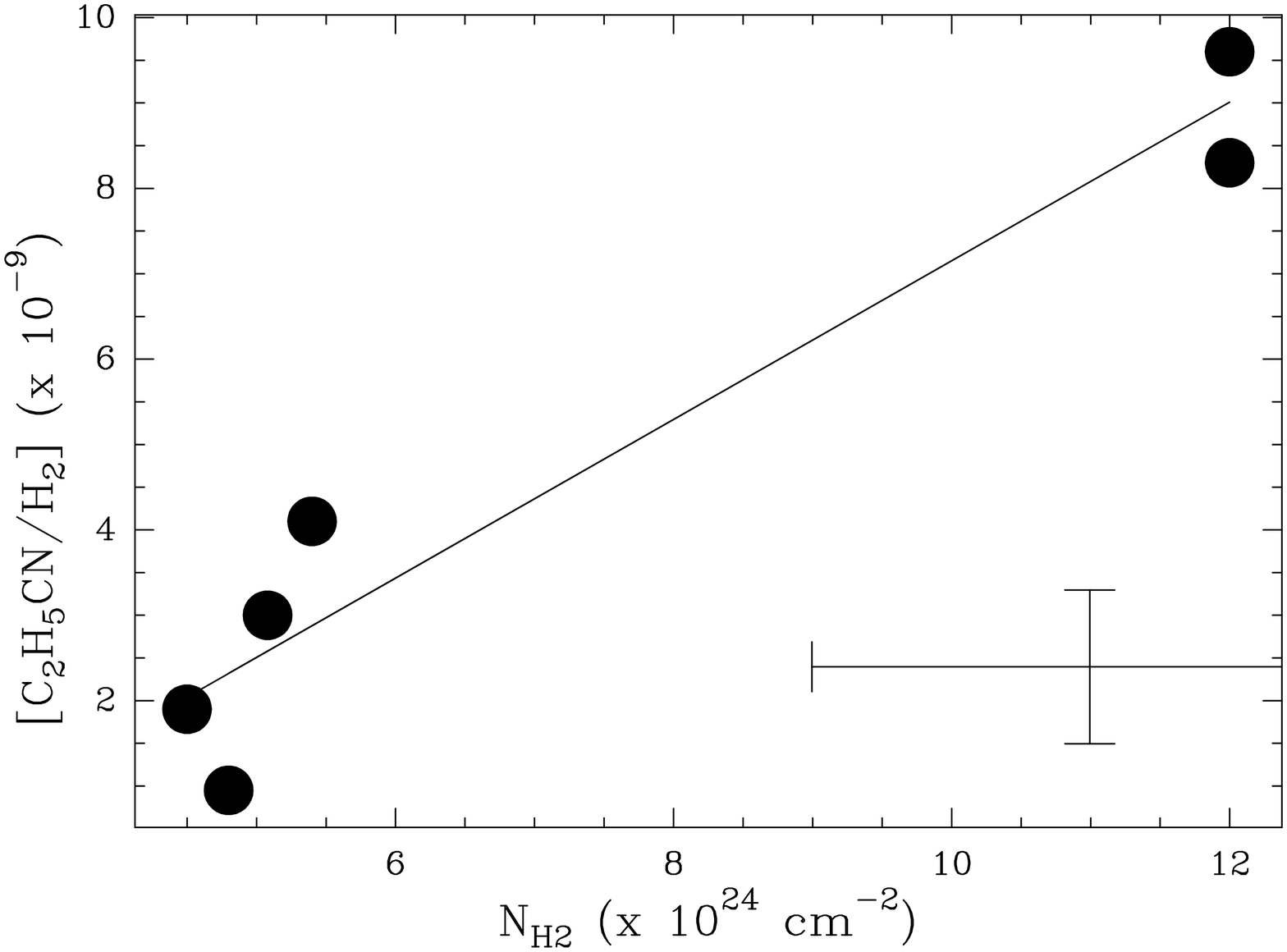}%
 			   \includegraphics[angle=-90]{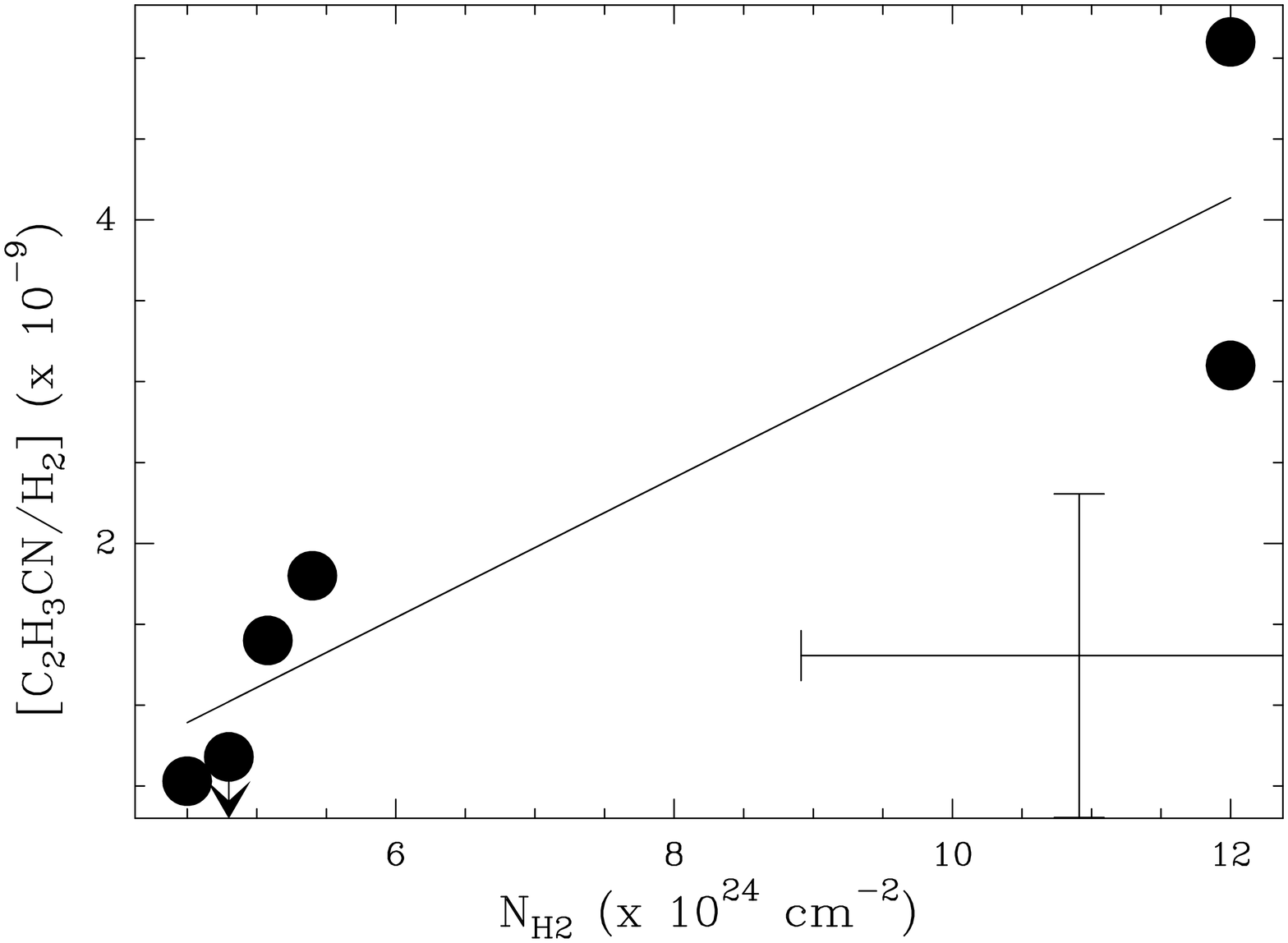}%
			   \includegraphics[angle=-90]{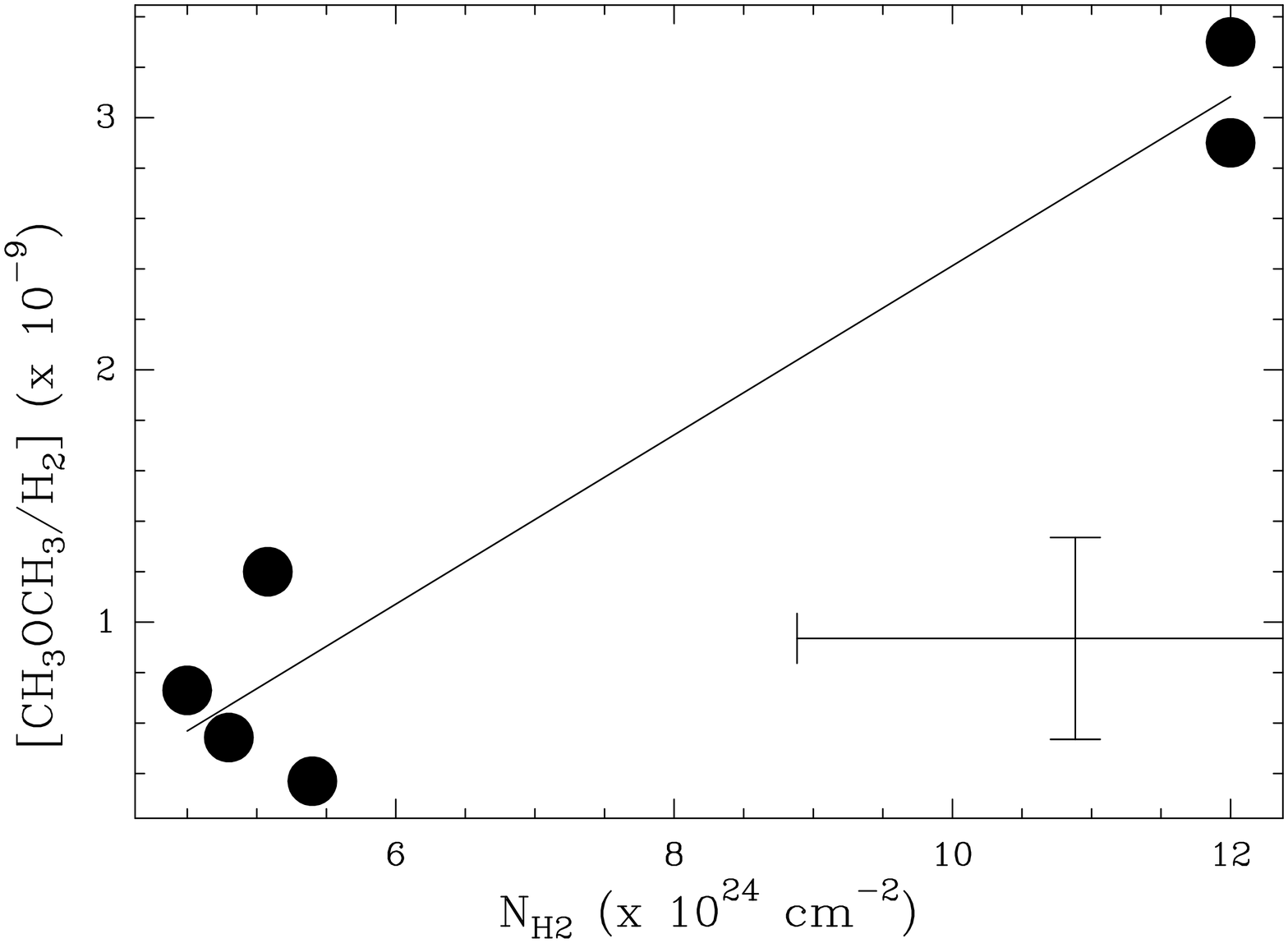}} 
 \end{center}
 \caption
 {\label{fig:abb_coldens} Comparison between the H$_2$ total column density
and the molecular abundances of \ecys (left panel), \vcys (central panel)
and \des (right panel). The upper limit for
the \vcys abundance in G10.62 is indicated by an arrow. Errorbars representing the
typical uncertainties are shown in the lower right corners.}
\end{figure*}

In Figure~\ref{fig:abb_coldens}
we plot the molecular abundances against the total
H$_2$ column densities. Even though the uncertainties on
these quantities are large, Figure~\ref{fig:abb_coldens} shows
that on average cores with higher H$_2$ column density have also
larger column densities of \ecy, \vcys and \de, without 
any substantial difference between O- and N-bearing species. 
We have also checked if a correlation is present between the observed 
abundances and the gas temperature: none of the sources show 
a correlation between these two parameters.
In summary, we do not see any difference between \des and the two
N-bearing species when comparing their abundances both
to the gas temperature and density of the cores.

\subsubsection{Beam-averaged column densities and rotational temperatures}
\label{sec:temp_col}

\begin{figure}
 \begin{center}
 \resizebox{\hsize}{!}{\includegraphics{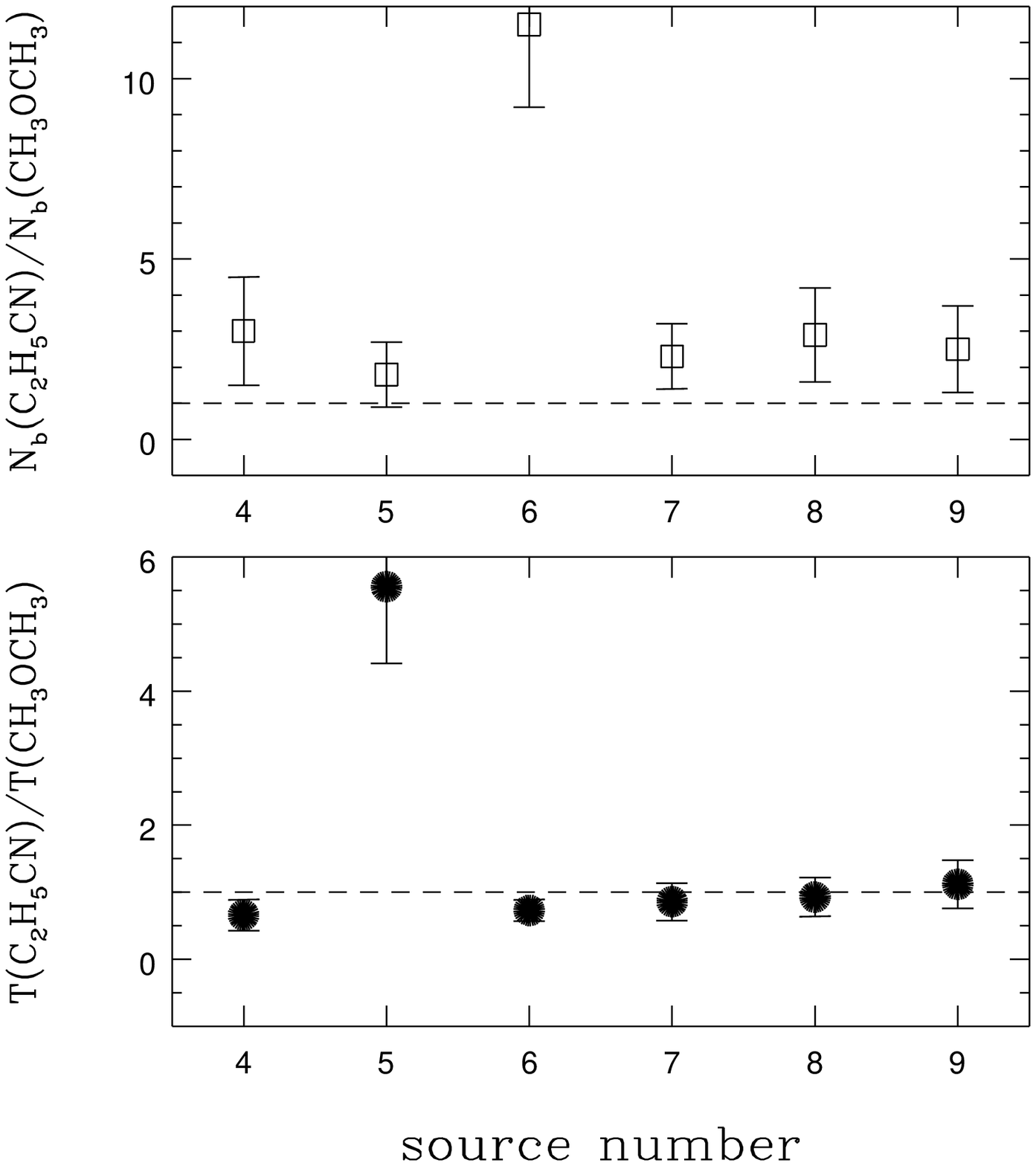}} 
 \end{center}
 \caption[Differences from N-bearing / O-bearing molecules]
 {\label{fig:rat_tn}{Top panel: ratio between the beam averaged 
total column densities (N$_{\rm b}$ in Tab.~\ref{tab:Results}) derived
from \ecys and \de. Bottom panel: ratio between the rotational 
temperatures derived from \ecys and \de.
The numbers on the x-axis indicate the identification number of each source 
as in Tab.~\ref{tab:source}. The dashed line represents equal
column density or rotational temperature from \ecys and \de.}}
\end{figure}

Figure~\ref{fig:rat_tn} shows the ratios between the beam averaged 
column densities (top panel) and
those between the rotational temperatures derived from \ecys
and \des (see Tab.~\ref{tab:Results}). The plot indicates that 
the column densities obtained from \ecys are on average marginally higher 
(a factor of $\sim 2.5$) than those derived from \des , suggesting 
that \ecys may indeed trace a region marginally denser than that
traced by \de. We note a significant (a factor of $\sim 11$)
difference in G19.61, which is the source in which there is also the 
largest difference in linewidths between \ecys and \de, suggesting that
\ecys is actually associated with a denser and more turbulent
sub-region of the HMC. On the other hand, 
no systematic difference is found in the rotational temperatures, 
with the significant exception of G10.62, for which we find 
a value more than 5 times higher from \ecy. However, for this 
source the temperature from \des is less accurate because it is derived 
from three transitions only.

Figs.~\ref{fig:del_velfw}, ~\ref{fig:abb_correl} and \ref{fig:rat_tn} 
do not allow to conclude that the O- and N-bearing species observed in this work
trace different HMC regions. However, when discussing these
results one has to bear in mind that the angular resolution of our 
data is much larger than 
the size of the HMC. High angular resolution observations
have demonstrated that the environment of some of our HMCs, 
and in general of massive YSOs, has a complex structure, and
may be fragmented in objects with different 
masses and in different evolutionary stages (see e.g. 
Garay et al.~1998 for G19.61; Avalos et 
al.~2006 and Mookerjea et al.~2007 for G34.26), whose angular
separation is comparable to or even smaller than the angular
resolution of the observations. The observed emission is hence 
affected both by the emission of other molecular condensations 
close to the main one and, less likely, by the emission of the
cooler molecular gas of the envelope surrounding the HMC. 
Therefore, it is difficult to reveal a
clear differentiation between O- and N-bearing species since 
the observed line parameters represent {\it average} values in 
the whole source.

  \begin{longtable}{cccc}
  \caption{Line velocities ($V_{\rm LSR}$) and widths 
at half maximum ($\Delta V$) derived from
single Gaussian fits to unblended lines for the sources G10.47,
G10.62 and G19.61. The errors of the Gaussian fits to the lines
are given between brackets.}
   \label{tab:velocities} 
\\
\hline
\hline
Molecule      & frequency & $V_{\rm LSR}$ & $\Delta V$ \\
              & (MHz)     & (\kms ) & (\kms ) \\
\hline
            & \multicolumn{3}{c}{G10.47} \\
\ecys & 98701.1090  &    67.249 (0.15) &    8.411 (0.36) \\
      & 161581.2030 &    66.782 (0.48) &   13.001 (1.37)  \\
      & 161367.344  &    67.564 (0.37) &    9.842 (0.86) \\
      & 224458.8590 &   66.546 (0.26) &   10.180  (0.73) \\
      & 224419.8120 &   66.952 (0.35) &   11.136  (1.12) \\
\vcys & 147561.719  &   66.866 (0.49)  &   9.277 (1.20) \\
\des  & 111783.241  &   66.745 (0.14)  &   9.939 (0.36) \\
      & 111782.596  &   66.754 (0.14)  &   9.555 (0.22) \\
      & 147456.863  &   66.940 (0.54)  &  13.100 (1.19) \\
      & 237620.371  &   65.184 (0.43)  &  11.751 (1.07) \\
\hline
      &    \multicolumn{3}{c}{G10.62} \\
\ecys & 161581.203  & -1.667 (0.82)  &   5.934 (1.51) \\
      & 224045.750  & 0.395 (0.35)  &   4.703 (0.78) \\
      & 224419.812  & -1.409 (0.40) &   6.208 (0.83) \\
\vcys & -- & -- & -- \\
\des  & 111783.241 & -3.440 (0.28)  &   7.852 (0.58) \\
      & 111782.596 & -3.429 (0.28)  &   7.357 (0.38) \\
      & 147210.732 & -2.726 (0.27)  &   2.730 (0.57) \\
      & 237620.371 & -2.615 (0.54)  &   5.690 (1.01) \\
\hline
     &    \multicolumn{3}{c}{G19.61} \\
\ecys & 98701.109 &  39.665 (0.40) &   10.119 (0.95) \\ 
      & 161581.203 &  40.198 (0.52) &   12.758 (1.57) \\
      & 161367.344 &  43.908 (0.72) &    9.451 (2.02) \\ 
      & 224045.750 &  39.835 (0.43) &    6.861 (1.06) \\
      & 224419.812 &  40.334 (0.39) &   7.634 (1.25)  \\
\vcys & -- & -- & -- \\
\des  & 111783.241 &  41.429 (0.53) &    5.955 (0.99) \\
      & 111782.596 &  41.427 (0.53) &    5.264 (0.66) \\
\hline
             &    \multicolumn{3}{c}{G29.96} \\
\ecys  & 98701.109 &  97.792 (0.29) &    4.721 (0.91)  \\
       & 161581.203 &   97.413 (0.73) &   7.728 (2.19)  \\
       & 161367.344 &   98.110 (1.47) &   11.032 (4.69) \\
       & 224045.750 &   98.124 (0.73) &    6.617 (1.53) \\
       & 224419.812 &   97.169 (0.24) &    2.854 (0.59) \\ 
       & 224458.859 &   96.595 (0.60) &    5.633 (1.39) \\ 
\vcys  &  -- & -- & -- \\
\des   & 111783.241 & 97.929 (0.34) &    5.675 (0.88) \\
       &  111782.596 & 97.976 (0.32) &    4.919 (0.58) \\
       &  147210.732 & 105.190 (1.13) &   10.947 (3.52) \\
\hline  
      &    \multicolumn{3}{c}{G31.41} \\
\ecys  & 98701.109 &  97.711 (0.26) &    8.289 (0.58) \\ 
       & 161581.203 &   97.179 (0.38) &   8.365 (1.07) \\
       & 161367.344 &   97.006 (1.02) &   8.245 (2.25) \\
       & 224458.859 &   96.673 (0.50) &   9.998 (1.67) \\ 
       & 224419.812 &   97.140 (0.36) &   9.065 (0.93) \\
\vcys  & 147561.719 & 97.316 (0.46)  &   9.643 (1.10) \\
\des   & 111783.241 & 98.104 (0.06)  &   7.181 (0.15) \\
       & 111782.596 & 98.125 (0.06)  &   6.617 (0.10) \\
       & 147210.732 &  97.580 (0.18)  &   5.502 (0.49) \\
       & 173293.064 & 98.255 (0.36) &   5.641 (0.83) \\
       & 237620.371 & 96.334 (0.20) &   7.338 (0.46) \\
\hline
       &    \multicolumn{3}{c}{G34.26} \\
\ecys  & 98701.109  &  57.333 (0.26) &   7.449 (0.66) \\ 
       & 161581.203 &  57.511 (0.34) &   8.542 (0.93) \\ 
       & 161367.344 &  57.103 (0.53) &   8.368 (1.46) \\ 
       & 224045.750 &  57.610 (0.39) &   3.691 (0.92) \\
       & 224419.812 &  57.478 (0.34) &   6.052 (0.81) \\
\vcys  & -- & --  & -- \\
\des  & 111783.241 & 59.091 (0.11) &  7.629 (0.25) \\
      & 111782.596 & 59.107 (0.10) &  7.117 (0.16) \\
      & 147210.732 & 59.653 (0.35) &   6.181 (0.84) \\
      & 173293.064 & 58.985 (0.27) &   6.262 (0.72) \\
      & 237620.371 & 57.523 (0.17) &   7.674 (0.38) \\
\hline
\end{longtable}

\begin{table*}[htbp]
  \caption[Results]{Average line velocities and widths computed from the 
unblended lines of \ecys ($V_{\rm N}$ and $\Delta V_{\rm N}$,
respectively) and \des ($V_{\rm O}$ and $\Delta V_{\rm O}$, respectively). 
Velocity differences ($V_{\rm N}-V_{\rm O}$) and linewidth ratios 
($\Delta V_{\rm N}/\Delta V_{\rm O}$) are also given.}
\label{tab:averages} 
\begin{center}
\begin{tabular}{ccccccc}
\hline
\hline
Source  & $V_{\rm N}$ & $V_{\rm O}$ & $V_{\rm N}-V_{\rm O}$ & $\Delta V_{\rm N}$& $\Delta V_{\rm O}$ & $\Delta V_{\rm N}/\Delta V_{\rm O}$  \\
\cline{2-6}
        & \multicolumn{5}{c}{(\kms )} & \\
\hline
G10.47 & 67.0 & 66.4 & 0.6 & 10.5 & 11.1 & 0.94 \\
G10.62 & --0.9 & --3.1 & --2.2 & 5.6 & 5.9 &0.95 \\
G19.61 & 40.8 & 41.4 & --0.6 & 9.4 & 5.6 & 1.68 \\
G29.96 & 97.5 & 97.9 & --0.4 & 6.4 & 7.2 & 0.89 \\
G31.41 & 97.1 & 97.7 & --0.6 & 8.9 &  6.5 & 1.37 \\
G34.26 & 57.4 & 58.9 & --1.5 & 6.8 & 7.0 & 0.97 \\
\hline
\end{tabular}
\end{center}
\end{table*}


\subsection{Hot core chemical ages}\label{sect:ages}
 
When comparing the observed abundance of different molecular
species (see Sect.~\ref{sect:temp}) from predictions of
theoretical models, we can get some insights into the evolutionary
stage of HMCs.  

H-rich species are thought to form in icy mantles of dust grains and 
then released into the gas phase when the central massive 
(proto)star begins to heat up the surroundings. A following gas
phase chemistry transforms many of the grain mantle products 
into other species.
In particular, \ecy, formed onto dust grains and then
evaporated, is expected to 
form \vcys through gas-phase reactions (Caselli et al.~1993).
The occurrence of these reactions, as well as those which destroy
\vcy , depends on the core physical parameters, 
which are thought to change with the core evolution.
Therefore, the abundance ratio of \ecys and \vcys is
correlated to the core evolutionary stage, and can be
used as a 'chemical clock' for the HMC.

\begin{figure}
 \begin{center}
 \resizebox{\hsize}{!}{\includegraphics[angle=-90]{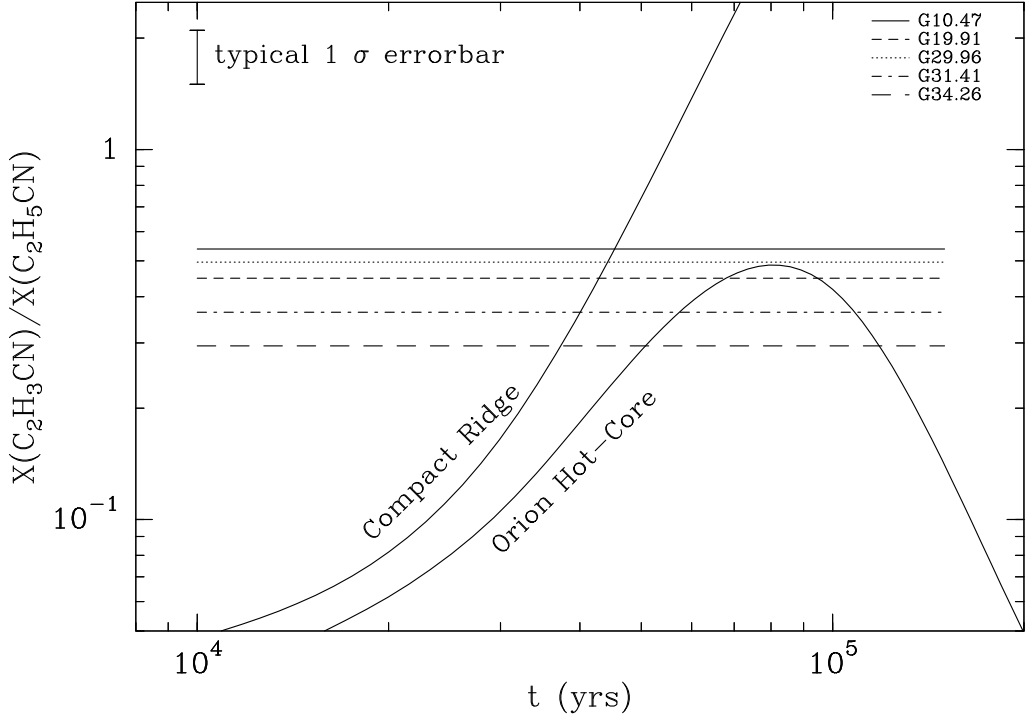}}
 \end{center}
 \caption[Chemical ages]
 {\label{fig:age}{Comparison between the observed abundance 
ratio of the daughter/parent pair \vcys/\ecys
and the predictions of the chemical models of Caselli et al~(1993).
The observed abundance ratio of each source is plotted as a 
horizontal line (see the top right corner for the source 
identification). 
The curves correspond to the predictions of
the models by Caselli et al.~(1993) for the
Orion-HC (corresponding to $T=200$ K and 
$n_{\rm H_2}=10^{7}$\cmc) and Compact Ridge (corresponding 
to $T=100$ K and $n_{\rm H_2}=10^{6}$\cmc).}}
\end{figure}

In Col.~2 of Tab.~\ref{tab:age} we list the observational values 
for the abundance ratio $R = X$(\vcy )/$X$(\ecy ), derived 
from the molecule's abundances given in 
Col.~6 of Table~\ref{tab:Results}.
We also plot these ratios in Fig.~\ref{fig:age} together
with the predictions of the chemical models
of Caselli et al.~(1993) for the 
Orion-HC and Compact Ridge. The physical and chemical properties
of these two regions can be considered as 'extrema' for our HMCs, 
so that we will use them to constrain the chemical ages of 
our sample.
Chemical models predict a sharp decrease in the abundances of 
many complex molecules, among which are \vcys and \ecy , in HMCs 
after $\sim 10^5$ yrs (see Fig.~\ref{fig:age}). Therefore, the
detection of several lines of both \vcys and \ecys suggest 
that all of our HMCs are likely younger than $\sim 10^5$ yrs.
In fact, we find ages in the range from $\sim 3.7 \times 10^{4}$ 
to $\sim 5.9 \times 10^{4}$ yrs, with the HMC
in G34.26 being the youngest and that in G29.96 the oldest. 
We point out that this procedure does not take into account 
the errors in the model predictions. 

\begin{figure}
 \begin{center}
 \resizebox{\hsize}{!}{\includegraphics[angle=-90]{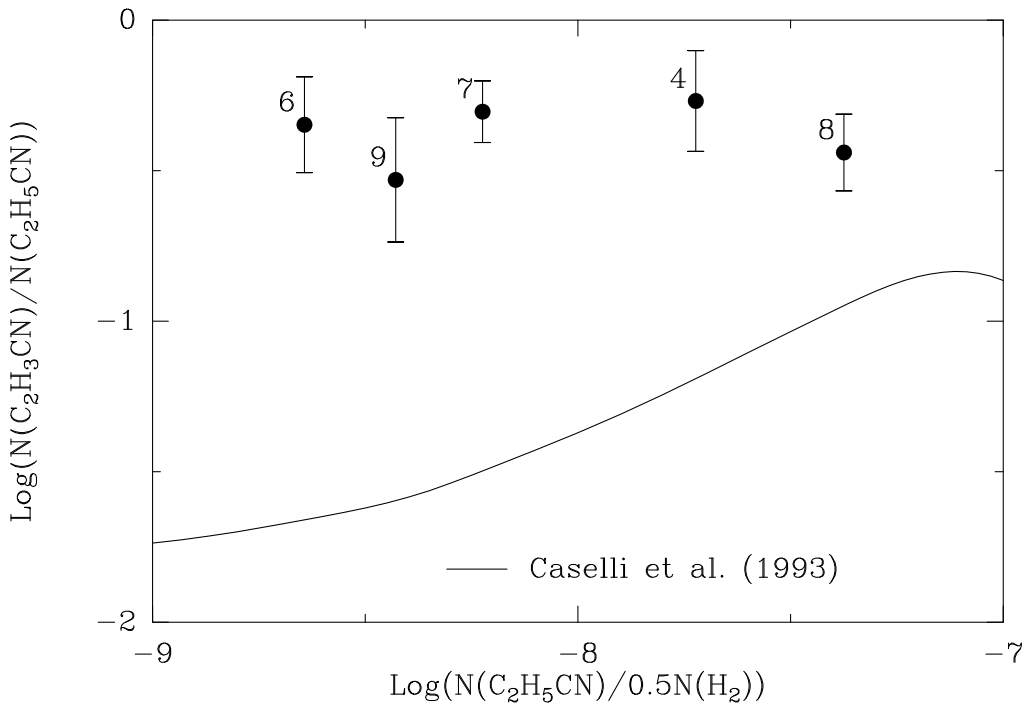}}
 \end{center}
 \caption[R against X(C2H5CN)]
 {\label{fig:r_abbfrac}{Observed abundance ratio $R = X$(\vcy )/$X$(\ecy ) 
against \ecys fractional abundance. The solid line represents the
prediction of the model by Caselli et al.~(1993) for
the Orion-HC ($T=200$ K, $n=10^{7}$ \cmc). The numbers identify the
sources as in Table~\ref{tab:source}.}}
\end{figure}

In Fig.~\ref{fig:r_abbfrac} we also plot $R$ against the
\ecys fractional abundance and we compare 
it with the theoretical curve predicted for the Orion HC. 
From Fig.~\ref{fig:r_abbfrac} we deduce that the model 
underestimates the observed ratio by a factor greater than 10. 
This could be due to a missing reaction in the chemical network 
between \ecys and H$_3$O$^+$. In fact,
the hydronium ion is quite abundant in the HMC phase and could destroy 
some of the \ecys molecules, thus increasing the 
observed \vcy /\ecys abundance ratio (see e.g. Caselli et
al.~1993).
Another possibility is that additional \vcys  forms in gas 
phase reactions as suggested by Charnley et al. 
(1992) and Rodgers \& Charnley (2001).

\begin{table}
\caption[chemical ages]{Hot cores chemical ages}
\label{tab:age}
\begin{center}
\begin{tabular}{c c c c}
\hline
\hline
Source & $R$    &    t$_{min}$& t$_{max}$\\
       &        &\multicolumn{2}{c}{($\times$10$^4$ yr)}\\
\hline
G10.47& 0.5$\pm$0.2 &5.0  &5.4 \\
G19.61& 0.4$\pm$0.2 &4.7  &5.0 \\
G29.96& 0.5$\pm$0.1 &4.7  &5.9 \\
G31.41& 0.4$\pm$0.1 &4.3  &4.7 \\
G34.26& 0.3$\pm$0.1 &3.7  &4.2 \\
\hline
\multicolumn{4}{c}{}\\
\multicolumn{4}{l}{$R = X$(\vcy )/$X$(\ecy ) }\\
\end{tabular}
\end{center}
\end{table} 

\subsection{Comments on individual sources} \label{sect:SbyS}

In this Section we describe the main properties of each
source on the basis of the results obtained both in this
work and from previous observations.


\paragraph{G10.47+0.03}

In this region three \uchii s are associated with a HMC
(see e.g. Gibb et al.~2004).
Observations of CH$_3$CN (13--12) and (19--18) 
(Hatchell et al.~1998) show that the temperature increases 
from $\sim$ 87 K at a distance of $\sim$ 2.3\asec\ from
the centre, to $\sim$ 134 K at 0.8\asec\ from the centre, 
indicating a temperature gradient. 
This is confirmed by observations of the NH$_3$ (4,4) inversion 
transition (Cesaroni et al.~1998), that show 
temperature and velocity gradients, which can be interpreted
in terms of a rotating structure with temperature increasing 
towards the centre. Our temperature measurements
agree with the estimates obtained by Hatchell et al.~(1998)
at $\sim 1$\asec\ from the core centre.

Fig.~\ref{fig:del_velfw} indicates a small difference in 
the line velocities of O- and N-bearing molecules,
suggesting that they are likely arising from marginally different
regions. 
This result agrees with that of Olmi et al.~(1996), who found a 
displacement between \hcos and CH$_3$CN of $\sim$0.5\asec .  
The most surprising property of this source is the high \vcys column 
density. The presence of both
vibrationally excited (see Sect.~\ref{sect:vcyvibr}) and 
"weak" b-type lines (see Sect.~\ref{optthin})
suggest that \vcys is about 10 times more abundant than \ecy . 
If we use in the rotation diagram all our data ("strong", "weak" and
vibrationally excited lines) and we correct for the optical depth 
(see Wyrowski et al.~1999 for the procedure)
we obtain the following best fit parameters:
N$_{\rm s} \sim$ 1.7 $\times$ 10$^{18}$ cm$^{-2}$, 
T$_{\rm rot} \sim$ 140 K and $\theta \sim$ 1\asec .
A previous study of vibrationally excited HC$_3$N also showed 
anomalous high abundance for this molecule,
a column density of $\sim 4\times 10^{18}$ cm$^{-2}$, comparable with that
from \vcy. This result is in agreement
with the model predictions of Caselli et al.~(1993), 
who showed that the abundances of HC$_3$N and \vcy, after
evaporation from dust grains, vary in a very similar way
both in the Orion-HC for $t > 100$ yrs (see 
their Fig.~4) and in the Compact Ridge for $t\geq 10^4$ yrs (see 
their Fig.~5), while those of \ecys and HC$_3$N are very different
between them. In fact, the destruction of \ecys by molecular
ions (in particular H$_3^+$) leads to the production of
both \vcys and HC$_3$N via dissociative recombination of
C$_3$H$_4$N$^+$ (formed in the reaction with H$_3^+$).


\paragraph{G10.62--0.38}

This star forming region consists of a cluster of OB stars
embedded in a dense, collapsing molecular cloud
(Ho \& Haschick~1986, Keto et al.~1987, 1988, 
Sollins et al.~2005).
Keto et al.~(1987) find an average temperature 
of 95 K for the absorption components
at -3.0 and -0.5 \kms\ and 140 K at 1.9 \kms, while the average 
temperature for the gas seen in emission, away from the \uchii{}, 
is 54 K: they suggest a 
temperature decreasing from the HMC center with a power-law of 
the type R$^{-0.5}$.
The temperature derived from \ecys (see Tab.~\ref{tab:Results})
indicates that this molecule traces a region at 
about 0.1~pc from the centre of the core, 
which also coincides with the centre of the continuum emission of 
the \uchii{} region. The \ecys and \des line velocities in
Fig.~\ref{fig:del_velfw} show that these molecules are 
spatially separated, with the latter more distant from the \uchii.
This is also suggested by the temperature derived from \de , even though 
the statistics are very poor because only three lines at E$_u<$ 100 K have
been identified.  
We find only one blended transition of \vcys in all our spectra so 
we cannot estimate rotational temperature and column density for 
this molecule.


\paragraph{G19.61--0.23}

This is a complex region with an irregular
structure both of the ionized gas (Wood \& Churchwell~1990),
and of the molecular emission from the associated molecular clumps.
Garay et al. (1998) found five distinct \uchii{} regions 
excited by individual stars from VLA observations of the ionized  
and molecular gas (NH$_3$(2,2) inversion line): the 
cometary-like and most compact of them is associated with the densest 
ammonia clump, called Middle clump to distinguish it from the Northern 
and the Southwestern clumps.
The Middle clump exhibits very broad line widths ($\sim$ 9.5 \kms) and 
is characterized by a \vlsr{} of $39.3\pm0.1$ \kms\ in the main 
NH$_3$(2,2) line: these values are similar to those we find for \ecy .
The Southwestern clump shows lower $\Delta V$ and higher \vlsr{} equal 
to those for \de . This seems to suggest that N- and O-bearing molecular 
emission originates from separate regions. 
This is confirmed by the column density ratio between \ecys and \de,
much higher than those of the other sources (see Fig.~\ref{fig:rat_tn}),
suggesting that \ecys indeed traces a region denser than that
traced by O-bearing species.
In spite of this, high angular resolution observations of two 
rotational lines
of \ecys and \hco , performed with BIMA (Remijan et al.~2004), 
indicate that these molecules seem to trace the same region.


\paragraph{G29.96--0.02}

G29.96 is an example of \uchii{} cometary region.
High angular resolution VLA observations in the ammonia (4,4) inversion 
transition (Cesaroni et al.~1998) show that the HMC has
a nearly symmetric profile, and 
is not affected by the presence of the embedded \uchii{} region.
Our rotational temperatures agree with those obtained by
Hatchell et al.~(1998) from CH$_3$CN (13--12) and (19--18) 
and are higher than the estimate that they derive from CH$_3$OH 
($T_{\rm rot}\sim 48$ K). 
This could indicate that O-bearing molecules map colder and extended gas. 
Nevertheless, Fig.~\ref{fig:del_velfw} does not show any displacement 
between the location of O- and N-bearing molecules: 
both their $\Delta V$ and \vlsr{} are, in fact, equal within the error bars.
This is supported by high angular resolution observations 
(Olmi et al.~2003)
which have shown that some lines of \hcos and the N-bearing
molecule CH$_3$CN trace approximately the same region. 
Our \vlsr{} also agrees with those of C$^{17}$O (2-1) and (3-2) 
measured by Hofner et al.~(2000).

\paragraph{G31.41+0.31}

The region is made of a 
core-halo \uchii\ region, offset by $\sim 5$\asec\ from a hot core in
which Beltr\'an et al.~(2005) have revealed a rotating
toroidal structure. Also,
as already mentioned in Sect.~\ref{sect:sourcesize}, they
serendipitously detected the \ecys ($25_{2,24}-24_{2,23}$)
and \hco -A (25$_{11,15}-26_{9,18}$) lines,
and found from their high-angular resolution maps that,
even though the two tracers peak at slightly different positions, their
integrated emissions and that of the 1.4~mm continuum
are fairly well overlapping. Interferometric observations
of CH$_3$CN lines (Olmi et al.~1996; Beltr\'an et al.~2005) 
and of the NH$_3$(4,4) main and satellite inversion lines 
(Cesaroni et al.~1998)
show that both the column density and the temperature increase 
towards the HMC center. 
This hot and dense gas has also been detected through HC$_3$N (17-16) in 
its v$_7$ and v$_6$ vibrationally excited states 
(Wyrowski et al.~1999). The ratio between the intensity 
of these two lines point to temperatures of $\sim$ 250 K.

Lower angular resolution maps of the CH$_3$CN (19-18) and (13-12) 
transitions (Hatchell et al.~1998) 
give a temperature in agreement with the estimate
that we derive from our molecules (see Tab.~\ref{tab:Results}). 
The $\Delta V$ of the CH$_3$CN lines are
comparable to those measured by us from \ecy . \des and \hcos have 
lower $\Delta V$, but the difference in \vlsr{} 
between the O- and N-bearing molecules is small.


\paragraph{G34.26+0.15}

This is a complex massive star formation region located
at a distance of 3.7 kpc (Kuchar \& Bania~1994)
extensively studied both in radio continuum
(Wood \& Churchwell~1989; Avalos et al.~2006) 
and in molecular lines (e.g. Garay \& Rodr\'{\i}guez~1990;
Hatchell et al.~1998; Sewilo et al.~2004).
VLA observations of the \amms (2,2) and (3,3)
inversion transitions (Garay \& Rodr\'{\i}guez~1990) reveal 
the presence of three distinct regions of the molecular gas:
\begin{itemize}
\item a low density (n(H$_2$)$\simeq 10^4$ cm$^{-3}$), low-temperature
(T$_{\rm rot}\simeq 18$ K) gas in front of a bright cometary \uchii{} 
region.
\item a warm gas (T$_{\rm rot}\simeq 70$ K) in a molecular disk-like structure
$\simeq 7.3''\times 2.8''$ in size, mapped by the main hyperfine 
component of the (3,3) transition, seen in absorption.
\item an ultracompact region $\simeq 1.6$\asec\ in size, 
$\simeq 2$\asec\ to the east
of the bright cometary \uchii{} region, traced by the satellite 
hyperfine lines of the (2,2) and (3,3) transitions. It is characterized by a 
rotational temperature of $185^{+75}_{-45}$ K and by a H$_2$ 
density of $\simeq 7\times 10^7$ cm$^{-3}$.
\end{itemize}
The H$_2$ density and temperatures estimated from the tracers 
studied in this paper agree with those given by 
Garay \& Rodr\'{\i}guez~(1990).
Mookerjea et al.~(2007) have observed with $\sim 1$\asec\ 
resolution two transitions of \ecy , and one transition of \des and \hco,
but all of them fall out of the bands observed in this work.  
From these maps and those of other molecular species, 
Mookerjea et al.~(2007) have concluded that 
O- and N-bearing species peak at different
positions, separated by $\sim 0$\farcs8, with a spatial
separation similar to that observed in Orion.
Given the low angular resolution of our
data and the complexity of the region, such a
differentiation is not shown from our observations.
Hatchell et al.~(1998) found evidence for 
temperature and density gradient in this source: 
the high excitation CH$_3$CN (19-18) transitions trace a smaller, 
hotter and denser region than that traced by CH$_3$CN (13-12). 
Our \ecys and \des temperatures agree with those  
of the higher excitation lines, indicating that they trace
the same hot gas.
 


\section{Conclusions} \label{sect:Conclusions}

We have surveyed rotational transitions of 4 complex O- and
N-bearing molecules, \vcy, \ecy , \des and \hco , with the 
IRAM-30m telescope towards a selected sample of 12 well-known HCs,
all of them associated with \uchii\ regions. 
For 6 sources of our sample, we have detected a sufficient 
number of transitions to derive the main physical properties of 
the cores. We focus our analysis on the 3 species
\vcy, \ecys and \des only, because the \hcos lines are
usually blended. 
The main results of our study are summarised as follows:
\begin{itemize}
\item From the rotational lines of \vcy, \ecys and \de ,
we have derived rotational temperatures from $\sim 100$ to 
$\sim 150$ K, and source averaged total column densities 
of order of $10^{15}-10^{17}$\cmq . 
Temperatures and column densities derived from the three 
tracers are typically in good agreement among them
in each source, indicating that they are tracing approximately 
the same dense and hot gas of the cores. 
\item The abundances relative to H$_2$ 
are of the order of $\sim 10^{-9}-10^{-10}$ for all species,
and are comparable to the values found in the
Orion-hot core and in Sgr B2. The \ecys abundances are also comparable
to those derived in other HMCs by Bisschop et al.~(2007),
while for \des we find values a factor of $\sim 2.5$ lower.
There is a strong correlation between the abundances of \ecys
and \vcy, as expected from theory that predicts that the latter
is formed through gas phase reactions involving \ecy. Interestingly,
a correlation is present also between the two N-bearing species
and \de, in disagreement with the results found by
Bisschop et al.~(2007) in similar sources.
\item Single Gaussian fits to unblended lines reveal a small
difference between the average peak velocity of \ecys and \des lines,
suggesting a possible spatial separation of the two tracers,
as seen in Orion and W3(H$_2$O). On the other
hand, no systematic differences are found in the linewidths.
We find a clear difference only in G19.61. 
We believe that this partly is due to the poor angular
resolution of our observations, which allows us to derive
only average values over the sources, which show a complex
morphology. 
\item We have compared the abundance ratio of the 
daughter/parent pair \ecy /\vcys with the predictions 
of chemical models to constrain the HCs' chemical ages.
We find ages between 3.7-5.9$\times 10^{4}$ yrs. 
We also compared this ratio against \ecys\ fractional abundance, finding 
that chemical models understimate even by a factor greater than 
10 our observational results. 
\end{itemize}

The chemical and physical differentiation between
O- and N-bearing molecules seen in the Orion HC and
W3(H$_2$O) is not revealed by our observations of other
HMCs. We stress 
however that our data have beam size much larger than the 
diameters of the HMCs, providing only {\it average} values
of the physical parameters of interest in the whole molecular
clump hosting the cores, and may be affected by other
objects embedded inside the clumps themselves. Follow-up observations
with millimeter and sub-millimeter interferometers 
(in particular ALMA, available in the near
future) will be fundamental to provide maps with resolution comparable
or even smaller than the core sizes, thus allowing us
to derive with better spatial accuracy the physical parameters
of the HMCs. 

\begin{acknowledgements}
We are grateful to the IRAM-30m telescope staff for their 
assistance.
We thank Peter Schilke for providing us with the XCLASS program.
Ilaria Pascucci would like to thank Thomas Henning for helpful discussions.
Many thanks to Geoff Macdonald for his corrections and suggestions.
\end{acknowledgements}
\newpage
\begin{appendix}

\section*{Appendix A: Observed spectra} 

\begin{figure*}
 \begin{center}
 \resizebox{\hsize}{!}{\includegraphics[angle=0]{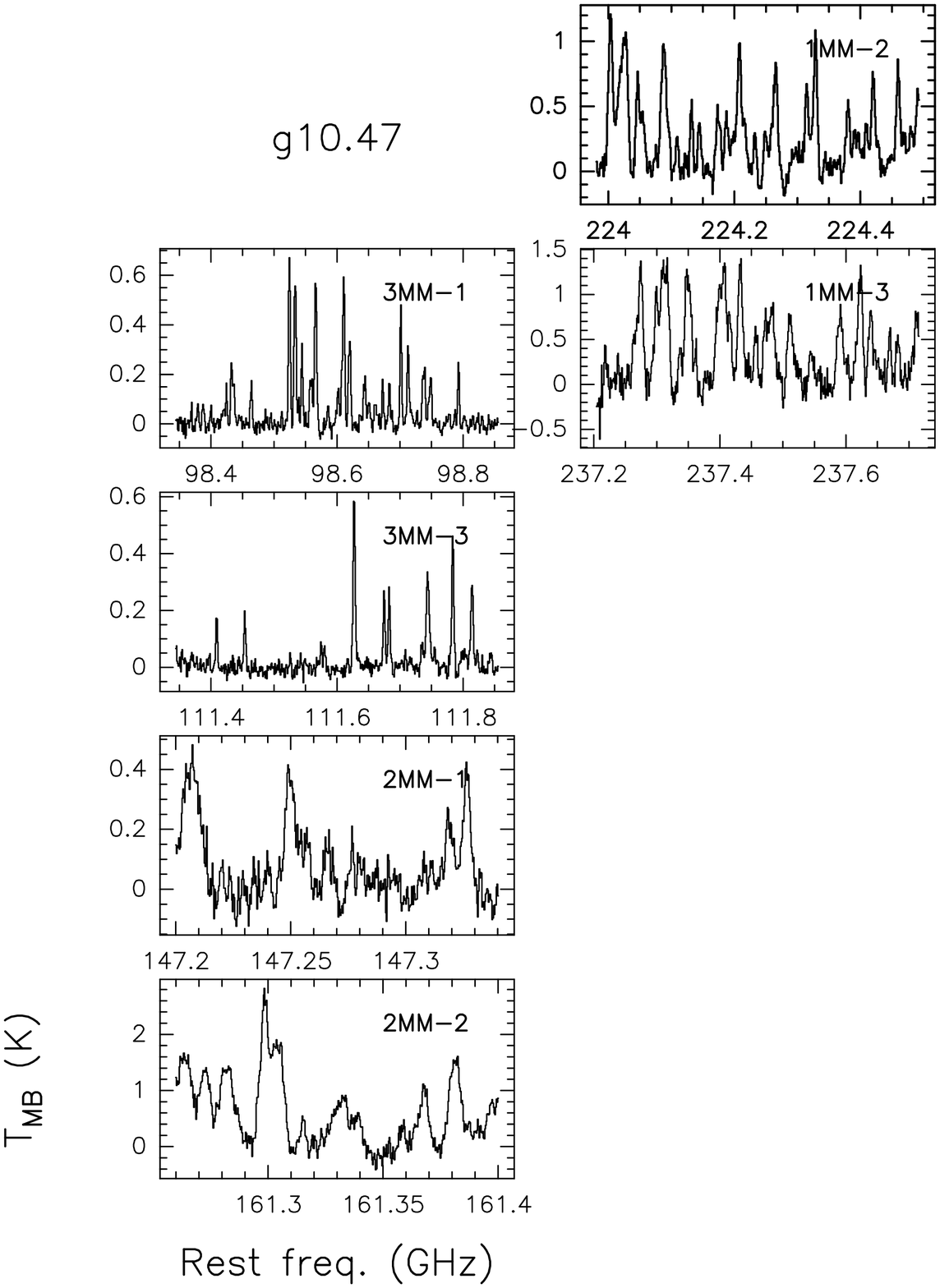}}
 \caption[] 
{\label{spectrag10.47}{Spectra obtained towards G10.47
with the IRAM-30m Telescope during the 1997 observing run.}}
\end{center}
\end{figure*}
\begin{figure*}
 \begin{center}
 \resizebox{\hsize}{!}{\includegraphics[angle=0]{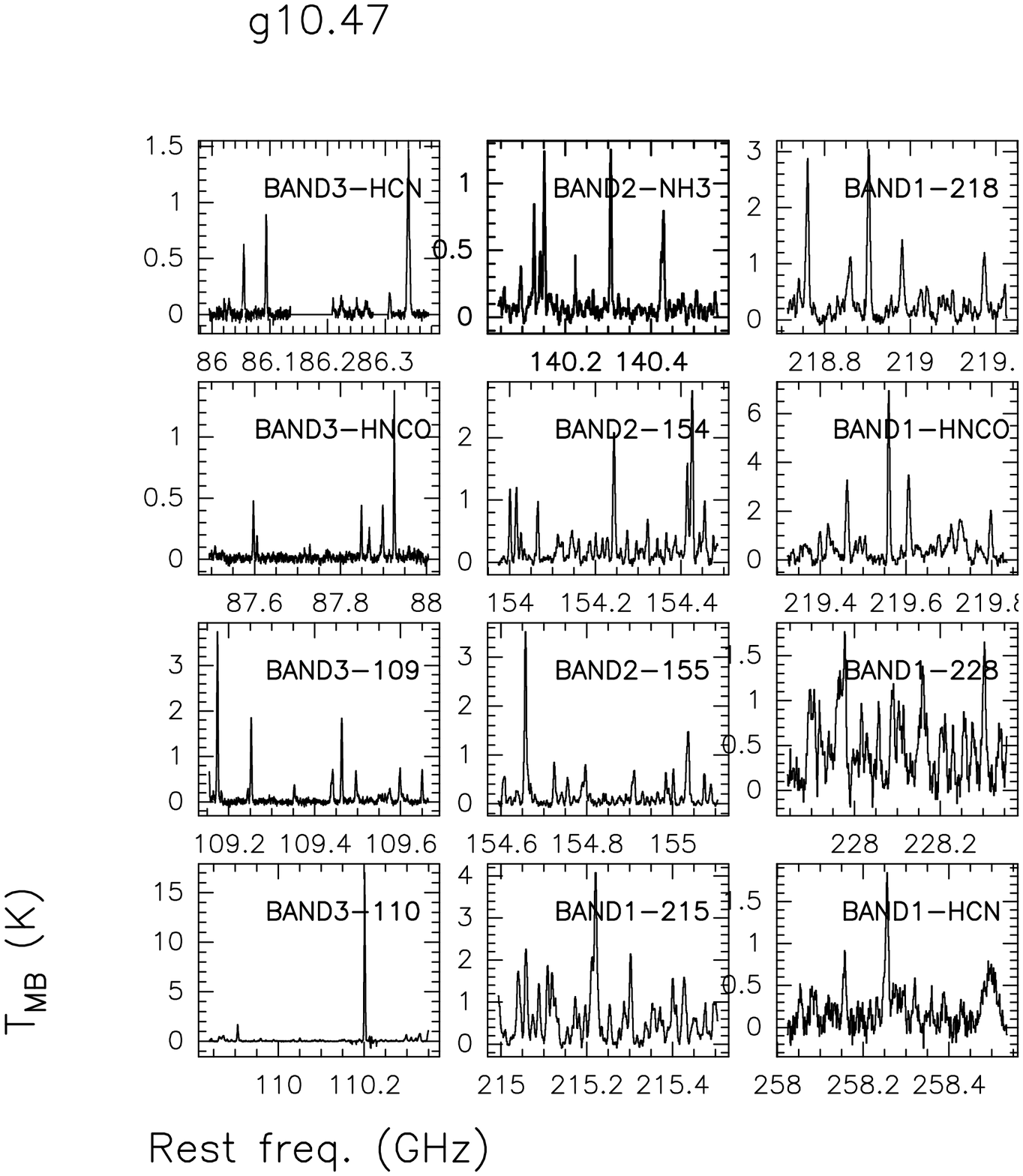}}
 \caption[] 
{\label{spectrawyrg10.47}{Spectra obtained towards G10.47
with the IRAM-30m Telescope during the 1996 observing run.}}
\end{center}
\end{figure*}

\begin{figure*}
 \begin{center}
 \resizebox{\hsize}{!}{\includegraphics[angle=0]{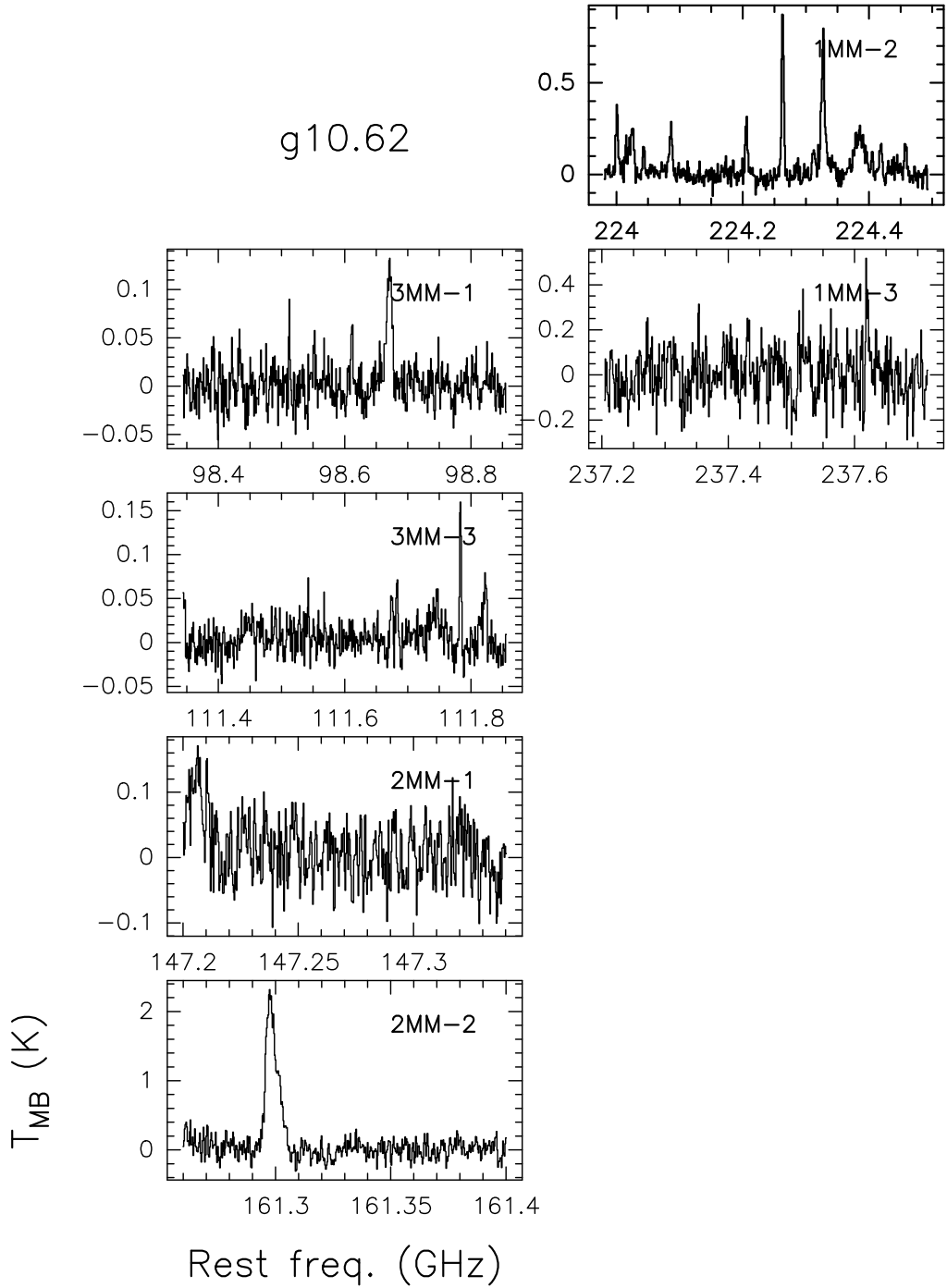}}
 \caption[]
 {\label{spectrag10.62}{Same as Fig.~\ref{spectrag10.47}
for G10.62.}}
\end{center}
\end{figure*}

\begin{figure}
\resizebox{\hsize}{!}{\includegraphics[angle=0]{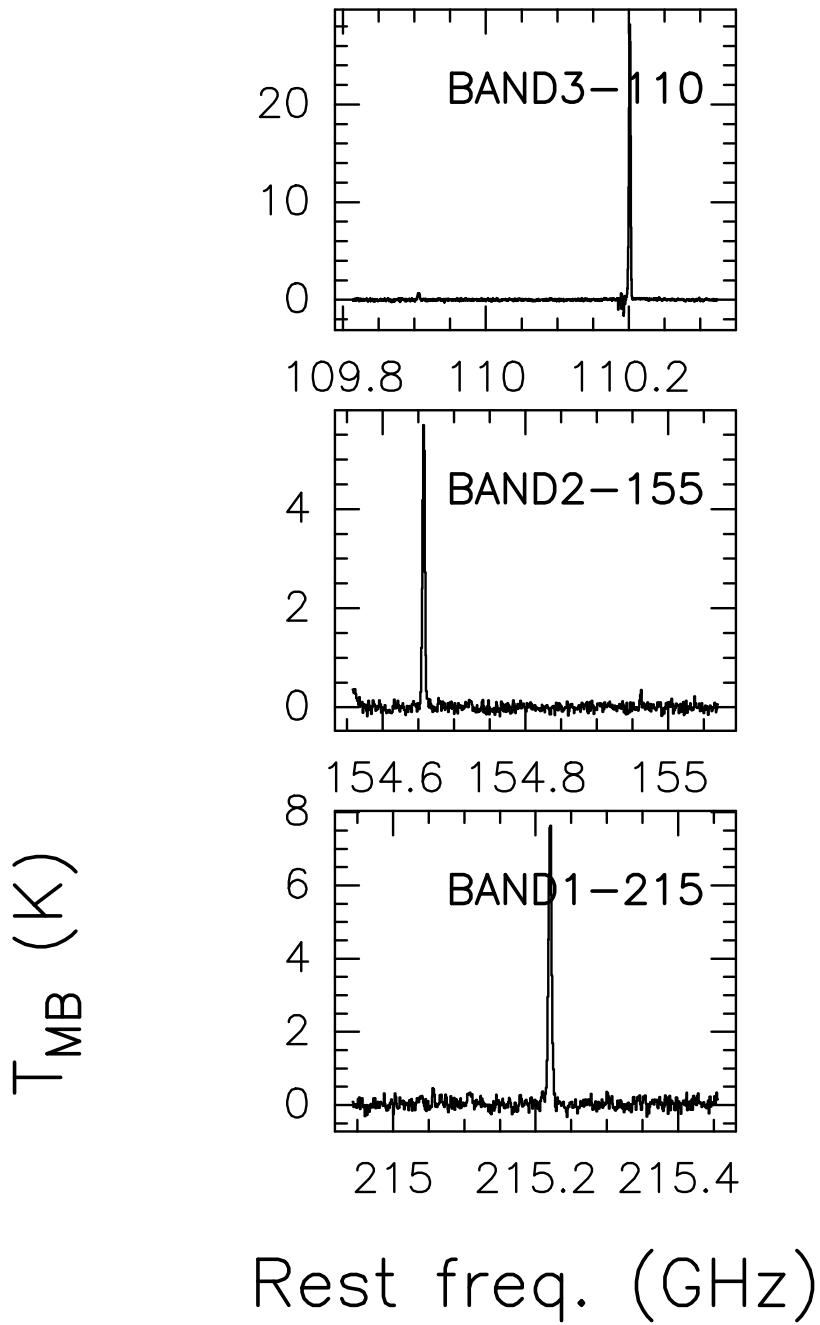}}
\caption[] 
{\label{spectrawyrg10.62}{Same as Fig.~\ref{spectrawyrg10.47} for G10.62.}}
\end{figure}

\begin{figure*}
 \begin{center}
 \resizebox{\hsize}{!}{\includegraphics[angle=0]{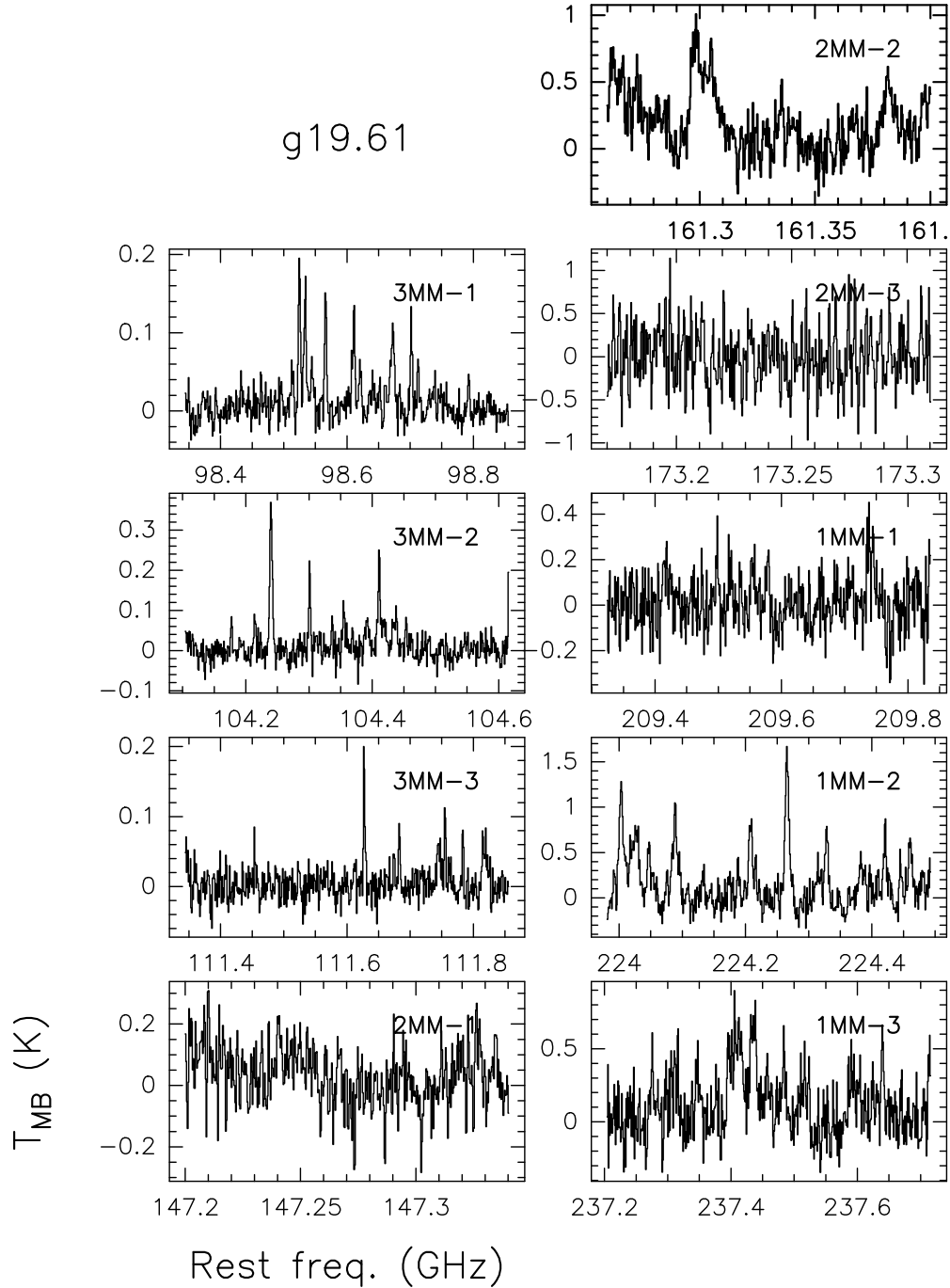}}
 \caption[]
 {\label{spectrag19.61}{Same as Fig.~\ref{spectrag10.47}
for G19.61.}}
\end{center}
\end{figure*}

\begin{figure*}
 \resizebox{\hsize}{!}{\includegraphics[angle=0]{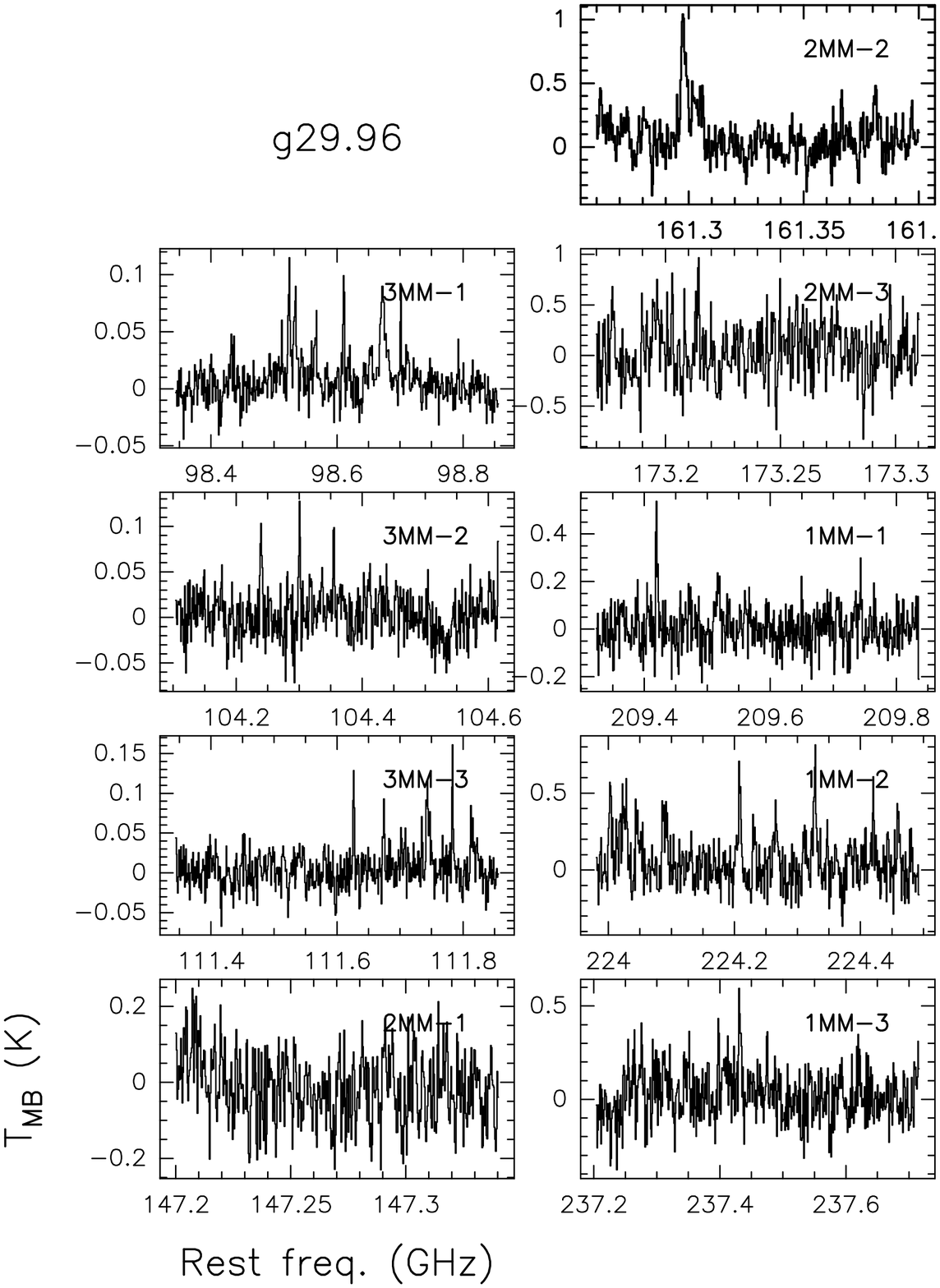}}
 \caption[] 
{\label{spectrag29.96}{Same as Fig.~\ref{spectrag10.47}
for G29.96.}}
\end{figure*}

\begin{figure*}
 \resizebox{\hsize}{!}{\includegraphics[angle=0]{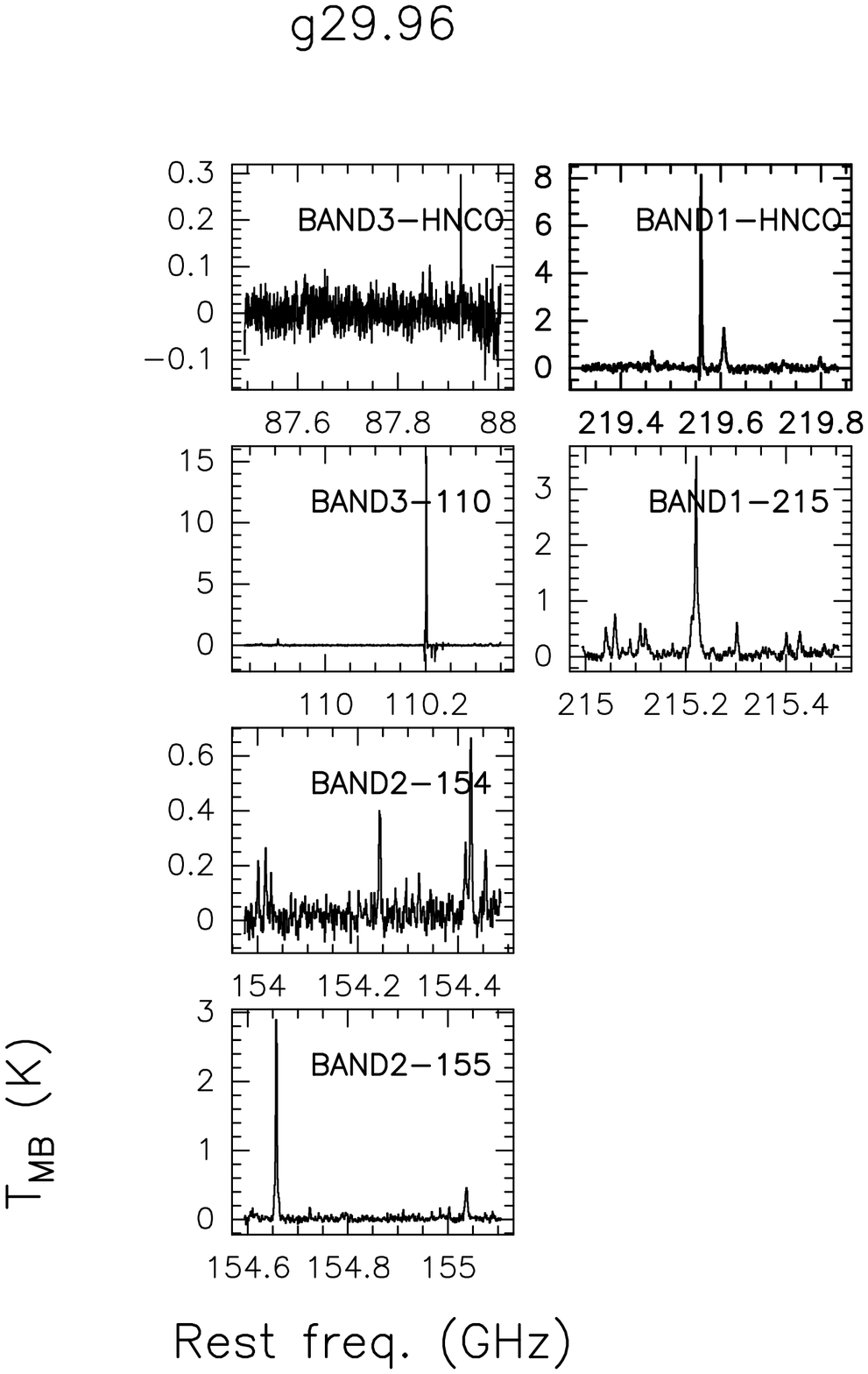}}
 \caption[] 
{\label{spectrawyrg29.96}{Same as Fig.~\ref{spectrawyrg10.47} for G29.96.}}
\end{figure*}

\begin{figure*}
\begin{center}
 \resizebox{\hsize}{!}{\includegraphics[angle=0]{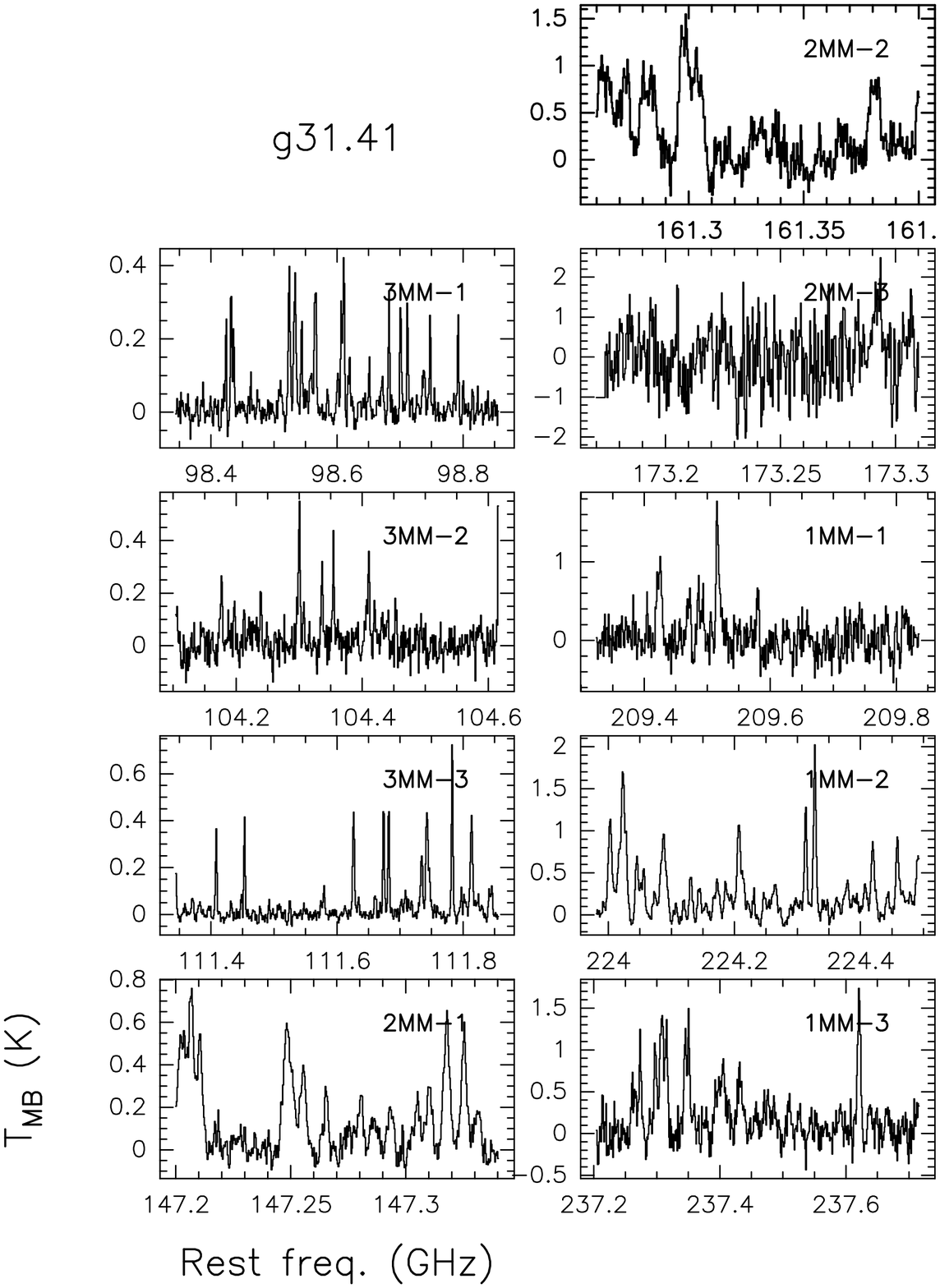}}
 \caption[]
 {\label{spectrag31.41}{Same as Fig.~\ref{spectrag10.47}
for G31.41.}}
 \end{center}
\end{figure*}

\begin{figure*}
 \begin{center}
 \resizebox{\hsize}{!}{\includegraphics[angle=0]{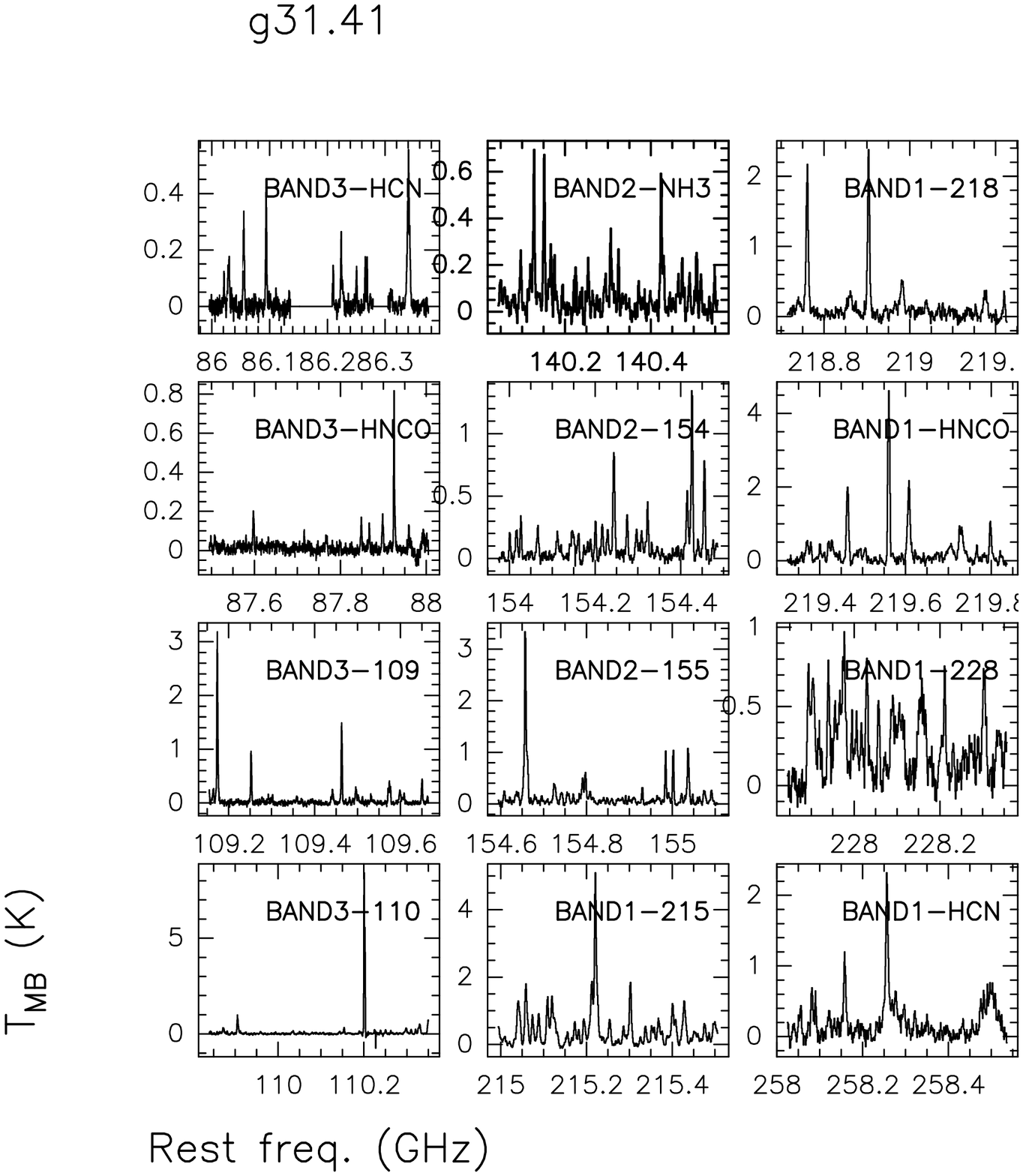}}
 \caption[] 
{\label{spectrawyrg31.41}{Same as Fig.~\ref{spectrawyrg10.47} for G31.41.}}
\end{center}
\end{figure*}

\begin{figure*}
 \begin{center}
 \resizebox{\hsize}{!}{\includegraphics[angle=0]{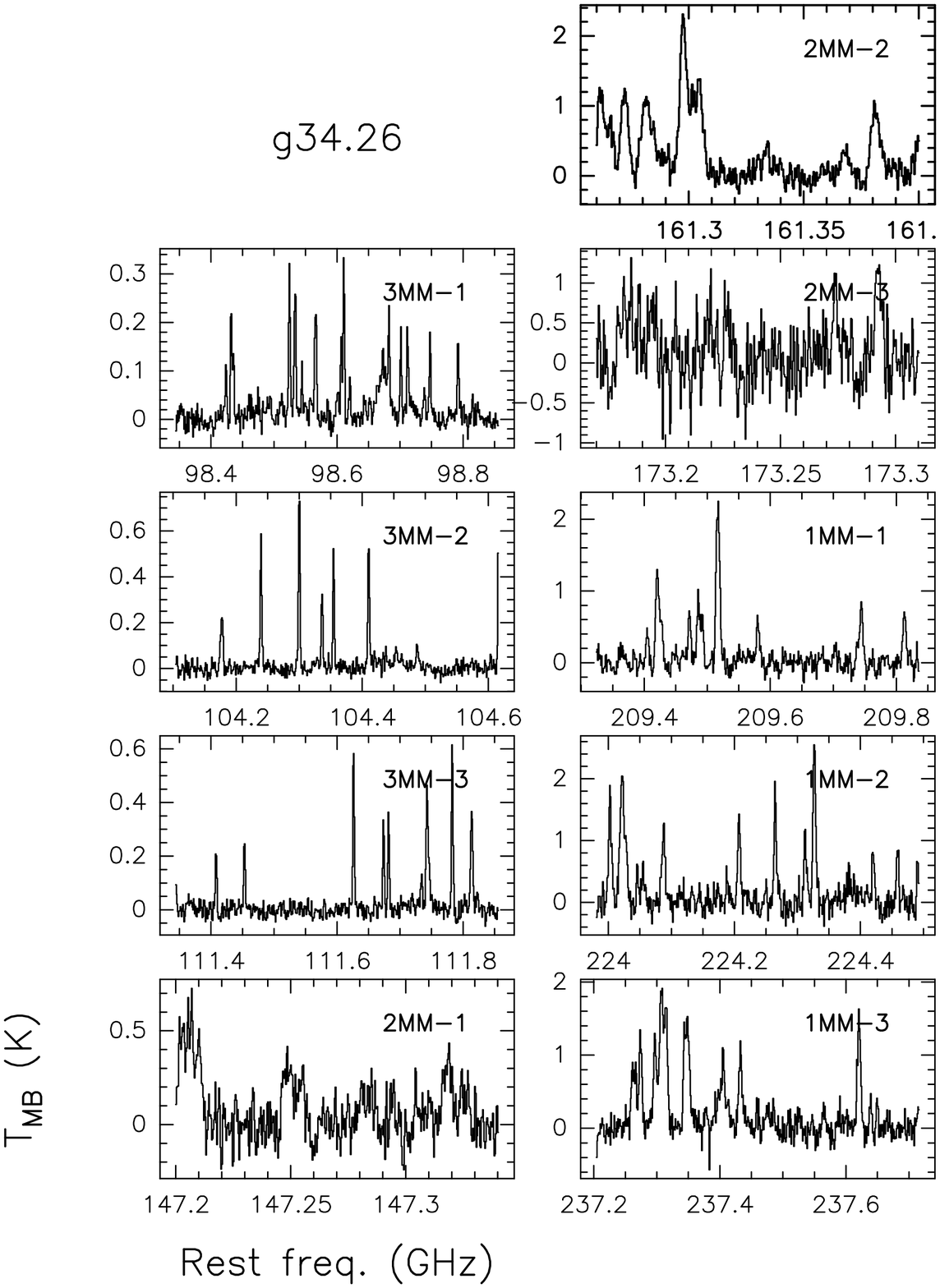}}
 \caption[]
 {\label{spectrag34.26}{Same as Fig.~\ref{spectrag10.47}
for G34.26.}}
 \end{center}
\end{figure*}

\begin{figure}
 \resizebox{\hsize}{!}{\includegraphics[angle=0]{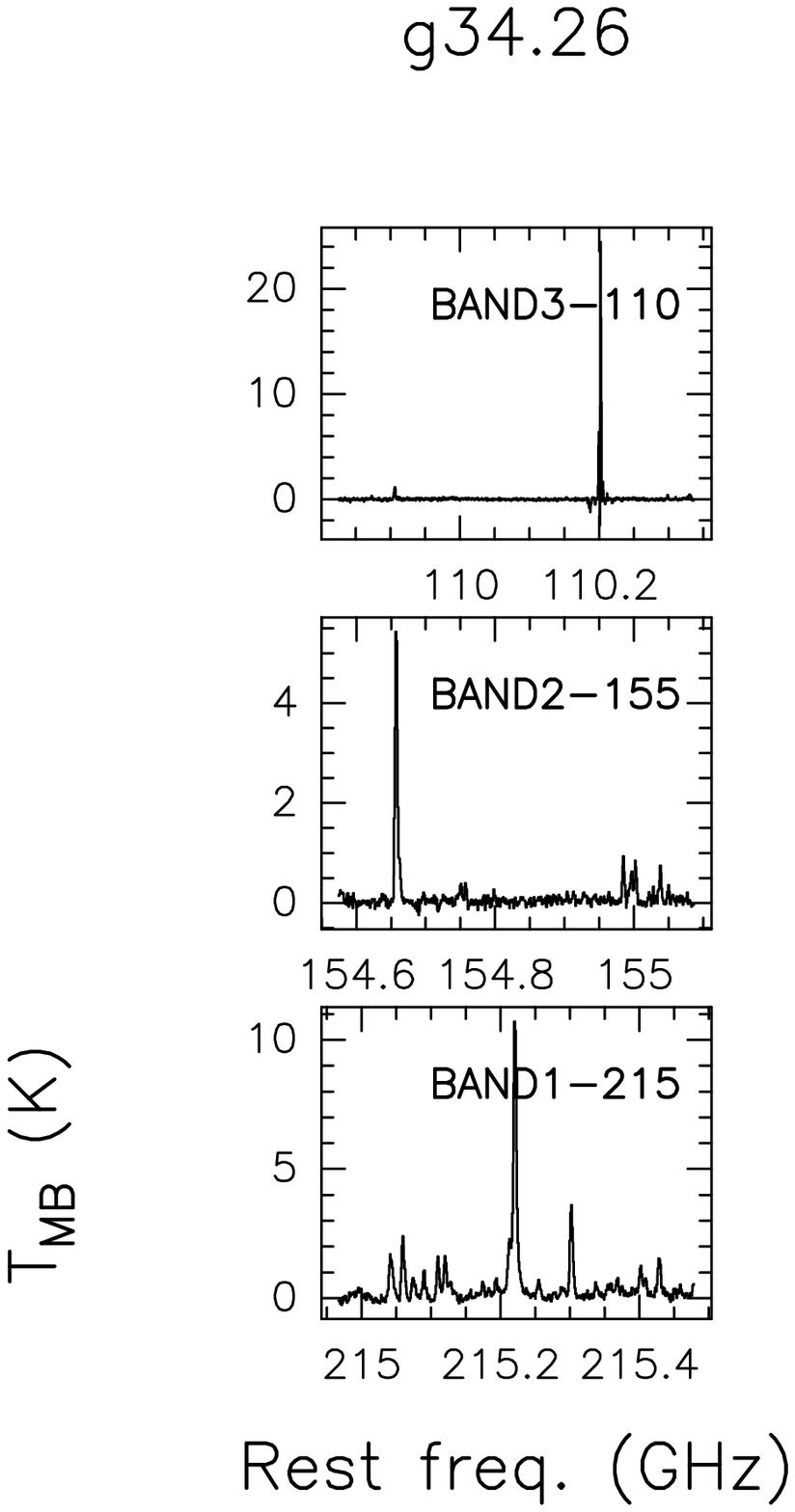}}
 \caption[] 
{\label{spectrawyrg34.26}{Same as Fig.~\ref{spectrawyrg10.47} for G34.26.}}
\end{figure}

\end{appendix}
\begin{appendix}
\section*{Appendix B: Tables} 
Tables~\ref{tab:g1042007}--\ref{tab:g342006}
list the molecular transitions used to derive the
rotational temperatures and total column densities of the
sources in Table~\ref{tab:sourcesize}.
For each line we give: rest frequency, $\nu$, in MHz (Col.~1,
indicated with "b" if it is a b-type transition); the molecular species
(Col.~2); the energy of the lower level, 
$E_{\rm l}$, in cm$^{-1}$ (Col.~3);  
the line integrated emission, $\int T_{MB}\:dv$, in K km s$^{-1}$;
(Col.~4); the quantity $\mu^{2}S$, where $\mu$ (in Debye) is the 
molecule's dipole moment and $S$ is the line strength (Col.~5).
The \des lines are separated in four components not resolved in our 
spectra (see also Sect.~\ref{comparison}). For this reason,
for the \des transitions we give in Col.~4 the total integrated 
emission of the four components
divided by the sum of the statistical weights of each 
component (for the statistical weights see Groner et al.~1998). 
\vskip 7cm
\begin{longtable}{ccccc}
\caption
{\label{tab:g1042007}  Transitions observed towards G10.47 during the
1997 observing run.}
\\
\hline
\hline
$\nu$ & Molecule & $E_{\rm l}$ & $\int{T_{MB}}\:dv$ & $\mu^{2}S$\\
(MHz) &             &  (cm$^{-1}$)      &   (K \kms)  &  $(D^2)$\\
\hline
b98399.6170 & C$_2$H$_3$CN &  66.0140 & 0.498   & 4.682 \\
 98523.883 & C$_2$H$_5$CN &  44.251 &  3.33 & 113.939\\   
 98524.664 & C$_2$H$_5$CN &  54.287 &  3.05 & 96.525\\   
 98532.070 & C$_2$H$_5$CN &  65.860 &  3.19 & 76.412\\   
 98533.9845 & C$_2$H$_5$CN &  35.754 &  4.16 & 128.685\\   
 98544.148 & C$_2$H$_5$CN &  78.967 &  2.78 & 53.615\\   
 98559.867 & C$_2$H$_5$CN &  93.604 &  1.64 & 28.149\\   
 98564.836 & C$_2$H$_5$CN &  28.801 &  4.11 & 140.761\\   
 98566.797 & C$_2$H$_5$CN &  28.801 &  3.15 & 140.761\\
 98701.109 & C$_2$H$_5$CN &  23.398 &  4.38 & 150.140\\
 b111574.6250  &  C$_2$H$_5$CN &  73.7330 &  0.74     &  10.807  \\
 b111783.241 & \de\ & 17.550   & 0.1439  & 27.2197 \\
 b111802.549 & \de\ & 42.142   & 0.0287  & 4.43093  \\	
 b111813.417 & \de\ & 118.013  & 0.1204  & 96.5637 \\
 b147205.839 & \de\ & 22.082   & 0.2271  & 19.52177 \\
 b147456.863 & \de\ & 257.141  & 0.1097  & 145.57157 \\
 147561.719 & C$_2$H$_3$CN &  38.4910 & 4.70    & 232.515\\ 
 161261.141 & C$_2$H$_5$CN &  95.144 &  13.0   & 213.000\\
 161265.734 & C$_2$H$_5$CN & 108.253 &  12.5  & 199.066\\ 
 b161281.953 & \de\ & 75.305   & 0.3944  & 49.41313   \\	
  161289.996 & \de\ & 333.668  & 0.1341    & 162.13361  \\
 161303.484 & C$_2$H$_5$CN & 139.060 &  9.28 & 166.301\\  
 161304.922 & C$_2$H$_5$CN & 73.536 &  11.3  & 235.930\\  
 161367.344 & C$_2$H$_5$CN & 175.957 &  9.38 & 126.974\\   
 161380.828  &  C$_2$H$_5$CN &  101.33  &  15.13  &  246.2506 \\
 161397.047 & C$_2$H$_3$CN &  97.1140 & 6.92      & 217.164\\
 161441.2660 & C$_2$H$_3$CN  & 139.0100 & 5.87      & 386.26 \\
 161445.3910 & C$_2$H$_3$CN  &  67.1130 & 7.04      & 234.339 \\
 161450.3590 & C$_2$H$_3$CN  &  67.1130 & 7.04      & 234.339 \\
 161475.172 & C$_2$H$_5$CN & 52.712 &  19.70 & 258.004\\
 161502.266 & C$_2$H$_3$CN &  56.6050 & 9.00       & 240.337\\
 161516.7190  &  C$_2$H$_5$CN &  58.1110  &  9.38   &  252.312 \\
 161527.500 & C$_2$H$_3$CN & 192.6770 & 5.77      & 162.238\\ 
 161581.203 & C$_2$H$_5$CN & 58.115 &  14.3 &    252.312\\
 161582.047 & C$_2$H$_3$CN & 223.8800 & 3.37  & 144.202\\
 161643.250 & C$_2$H$_3$CN & 257.9690 & 4.50  & 124.459\\
 224002.109 & C$_2$H$_5$CN &  166.828 &  4.40 & 309.672\\  
 224003.438 & C$_2$H$_5$CN &  152.185 &  6.00 & 320.879\\
 224045.750 & C$_2$H$_5$CN &  200.696 &  5.73 & 283.720\\  
 224084.281 & C$_2$H$_5$CN &  219.912 &  5.20 & 268.960\\ 
 224088.2035 & C$_2$H$_5$CN &  127.510 &  6.00 & 339.753\\  
 224131.516 & C$_2$H$_5$CN &  240.640 &  3.63 & 253.034\\  
 224186.344 & C$_2$H$_5$CN &  262.876 &  3.27 & 235.930\\  
 224206.609 & C$_2$H$_5$CN &  117.492 &  4.09 & 347.406\\  
 224208.078 & C$_2$H$_5$CN &  117.492 &  4.09 & 347.406\\ 
 b224231.6880 &  C$_2$H$_5$CN &  94.4160  &  2.05  &   33.527 \\
 224248.016 & C$_2$H$_5$CN &  286.614 &  2.21 & 217.645\\  
 224389.719 & C$_2$H$_5$CN &  338.567 &  2.25 & 177.537\\  
 224419.812 & C$_2$H$_5$CN &  109.033 &  6.34 & 353.894\\  
 224458.859 & C$_2$H$_5$CN &  109.036 &  8.10 & 353.894\\
b237360.8910 &  C$_2$H$_5$CN &  114.9850  &  2.33  &  24.209 \\
 237396.9765 & C$_2$H$_3$CN & 168.5260 & 5.15      & 336.209\\
 237405.1880  &  C$_2$H$_5$CN &  101.8870  &  13.71  &  380.894 \\  
 237456.2500 & C$_2$H$_3$CN & 216.3420 & 5.07     & 635.06 \\
 237482.766 & C$_2$H$_3$CN & 132.5690 & 4.06      & 350.218\\
 237485.016 & C$_2$H$_3$CN & 132.5700 & 3.61      & 350.218\\
  237548.541 & \de\ & 447.259  & 0.200              & 34.69413 \\
 237585.484 & C$_2$H$_3$CN & 275.8720 & 3.58      & 294.183\\
 237591.391 & C$_2$H$_3$CN & 108.5880 & 6.11      & 359.528\\
 b237620.371 & \de\ & 32.291   & 0.551   & 23.17836 \\
 237638.0160 & C$_2$H$_3$CN &  119.0860 &  6.74     & 355.471 \\
 237668.766 & C$_2$H$_3$CN & 309.9800 & 4.29      & 280.758\\ 
\hline
\end{longtable} 
\begin{longtable}{ccccc}
\caption
{\label{tab:g1042006} Transitions observed towards G10.47 during the 1996
observing run.}
\\
\hline
\hline
$\nu$ & Molecule & $E_{\rm l}$ & $\int{T_{MB}}\:dv$ & $\mu^{2}S$\\
(MHz) &             &  (cm$^{-1}$)      &   (K \kms)     &  $(D^2)$\\
\hline
 b140163.0   & \des &308.85  &0.1086            &23.138   \\
 b140226.172& \des&159.81  &0.0911            &9.896  \\
  a140429.484 &  \vcys &54.24   &8.45    &         202.872  \\
  b154027.062 &  \vcys &10.02   &4.68    &          2.344 \\
  a154724.531 &  \vcys &65.42   &9.77    &         215.731 \\ 
 b154311.469& \des &185.5   &0.0667            &12.855  \\ 
 b154456.5  & \des & 62.92   &0.4117            &21.739  \\
 215040.8     &  \ecys & 229.5   &     24.1   &        1199.4 \\     
 215088.203   & \ecys &  288.76  &     15.5   &        533.6  \\
 215109.047   & \ecys &  183.46  &     19.7   &        651.1  \\
 215119.203   & \ecys &  135.84  &     16.5   &        369.7 \\
 215126.703   & \ecys &  316.4   &     10.0   &        502.8  \\
 215400.797   &  \ecys & 156.87  &     17.6   &       340.3  \\
 215427.984   &  \ecys & 156.88  &     19.0   &        340.332 \\
 227897.625 &  \vcys &  242.47  &13.12    &        594.744  \\
 227906.719 &   \vcys & 214.46  &10.29    &        609.457  \\
 227918.656 &   \vcys &    274.75  &9.18     &     577.798 \\
 227960.234 &  \vcys &     311.27  &7.23     &     558.56 \\
 227967.594 &  \vcys &     190.74  &19.11    &     621.8  \\
 228017.375 &  \vcys &     352.    &10.23    &     537.313 \\
 228087.297 &  \vcys &     396.91  &6.95     &     513.431 \\
 228090.531 &   \vcys &    156.23  &8.88     &     319.903 \\
 228104.609 &   \vcys &    171.34  &11.91    &     315.959 \\
 228160.297 &  \vcys &     171.35  &16.7     &     631.917 \\
  \hline
\end{longtable}

\begin{longtable}{ccccc}
\caption
{\label{tab:g1062007} Transitions observed towards G10.62--0.38 during the
1997 observing run.}
\\
\hline
\hline
$\nu$ & Molecule & $E_{\rm l}$ & $\int{T_{MB}}\:dv$ & $\mu^{2}S$\\
(MHz) &             &  (cm$^{-1}$)      &   (K \kms)     &  $(D^2)$\\
\hline
 b111783.241 & \de\ & 17.550   & 0.0403  & 27.2197 \\
  111802.17  & \de\ & 42.141   & 0.016              & 2.53886 \\
  111802.93  & \de\ & 42.142   & 0.013              & 2.53886 \\
  111813.417 & \de\ & 118.013  & 0.016              & 96.5637 \\
 b147205.839 & \de\ & 22.082   & 0.0478  & 19.52177 \\       
  147456.863 & \de\ & 257.141  & 0.0196     & 145.57157 \\
 161261.141 & C$_2$H$_5$CN & 95.144 & 1.82 & 213.000\\  
 161265.734 & C$_2$H$_5$CN &  108.253 & 0.32 & 199.066\\ 
  161281.953 & \de\ & 75.305   & 0.0572             & 49.41313 \\
  161289.996 & \de\ & 333.668  & 0.0572             & 162.13361 \\
 161475.172 & C$_2$H$_5$CN &  52.712  & 0.86  & 258.004\\ 
 161581.203 & C$_2$H$_5$CN &  58.115  & 1.07  & 252.312\\ 
 224002.109 & C$_2$H$_5$CN &  166.828  & 0.78  & 309.672\\ 
 224003.438 & C$_2$H$_5$CN &  152.185  & 0.97  & 320.879\\
 224045.750 & C$_2$H$_5$CN &  200.696  & 0.65  & 283.720\\ 
 224084.281 & C$_2$H$_5$CN &  219.912  & 0.39  & 268.960\\ 
 224088.2035 & C$_2$H$_5$CN &  127.510 &  1.16  & 339.753\\ 
 224206.609 & C$_2$H$_5$CN &  117.492  & 1.05  & 347.406\\ 
 224208.078 & C$_2$H$_5$CN &  117.492  & 0.67  & 347.406\\ 
 224419.812 & C$_2$H$_5$CN &  109.033  & 1.20 & 353.894\\  
 224458.859 & C$_2$H$_5$CN &  109.036  & 1.05 & 353.894\\
  237261.970 & \de\ & 214.188  & 0.117              & 66.88355  \\  
  237548.541 & \de\ & 447.259  & 0.117              & 34.69413 \\
 b237620.371 & \de\ & 32.291   & 0.0875  & 23.17836  \\
\hline
\end{longtable}  

\begin{longtable}{ccccc}
\caption
{\label{tab:g1062006} Transitions observed towards G10.62--0.38 during the
1996 July observing run.}
\\
\hline
\hline
$\nu$ & Molecule & $E_{\rm l}$ & $\int{T_{MB}}\:dv$ & $\mu^{2}S$\\
(MHz) &             &  (cm$^{-1}$)      &   (K \kms)    &  $(D^2)$\\
\hline
  215040.8   &  \ecys & 229.5    & 2.49    &     1199.4    \\
  215109.047  &  \ecys &   183.46   &  3.25    &   651.1   \\
  215427.984   &  \ecys &  156.88   &  1.67    &  340.332 \\      
\hline
\end{longtable} 
\begin{longtable}{ccccc}
\caption
{\label{tab:g192007} Transitions observed towards G19.61--0.23 during the 1997 observing run.}
\\
\hline
\hline
$\nu$ & Molecule & $E_{\rm l}$ & $\int{T_{MB}\:dv}$ & $\mu^{2}S$\\
(MHz) &             &  (cm$^{-1}$)      &   (K \kms)     &  $(D^2)$\\
\hline
 98523.883 & C$_2$H$_5$CN & 44.251 &  1.24 & 113.939\\
 98524.664 & C$_2$H$_5$CN &  54.287 &  0.87 & 96.525\\
 98532.070 & C$_2$H$_5$CN &  65.860 &  0.75 & 76.412\\
 98533.9845 & C$_2$H$_5$CN &  35.754 &  1.49 & 128.685\\
 98544.148 & C$_2$H$_5$CN & 78.967 &  0.69 & 53.615\\
 98564.836 & C$_2$H$_5$CN & 28.801 &  1.21 & 140.761\\
 98566.797 & C$_2$H$_5$CN & 28.801 &  0.72 & 140.761\\
 98701.109 & C$_2$H$_5$CN &  23.398 &  1.24 & 150.140\\
 b104177.000 & \de\ & 102.625  & 0.185  & 79.56847 \\
104212.656 & C$_2$H$_3$CN &  23.4110 & 0.80     & 155.205\\
104419.312 & C$_2$H$_3$CN &  71.4640 & 0.80     & 112.653\\
104432.797 & C$_2$H$_3$CN &  30.9450 & 0.72     & 148.58\\
104437.516 & C$_2$H$_3$CN &  90.9290 & 0.94     & 95.507\\
104453.930 &  C$_2$H$_3$CN &  30.9460 & 0.72     & 148.58\\
  104490.168 & \de\ & 585.600  & 0.026              & 4.48006 \\
 b111783.241 & \de\ & 17.550   & 0.017   & 27.2197 \\
 b111813.417 & \de\ & 118.013  & 0.024   & 96.5637  \\
  147205.839 & \de\ & 22.082   & 0.0376             & 19.52177  \\                   
  147456.863 & \de\ & 257.141  & 0.0376             & 145.57157  \\
  147561.719 & C$_2$H$_3$CN &  38.4910 & 2.46      & 232.515\\
 161261.141 & C$_2$H$_5$CN &  95.144 & 5.56 & 213.\\  
 161265.734 & C$_2$H$_5$CN & 108.253 & 4.72 & 199.066\\
 b161281.953 & \de\ & 75.305   & 0.0483  & 49.41313   \\  	
  161289.996 & \de\ & 333.668  & 0.0512  & 162.13361  \\
 161303.484 & C$_2$H$_5$CN & 139.060 & 3.33 & 166.301\\  
 161304.922 & C$_2$H$_5$CN & 73.536 &  3.89 & 235.93\\  
 161367.344 & C$_2$H$_5$CN & 175.957 &  2.78 & 126.974\\ 
 161379.844 & C$_2$H$_5$CN & 65.047 &  2.79 & 244.939\\ 
 161397.0470 &  C$_2$H$_3$CN &  97.1140 & 2.84   & 434.328 \\
 161381.875 & C$_2$H$_5$CN & 65.047 &  2.79  & 244.939\\
161441.266 & C$_2$H$_3$CN & 139.0100 & 2.21   & 193.130\\
161445.391 & C$_2$H$_3$CN &  67.1130 & 2.94   & 234.339\\ 
 161475.172 & C$_2$H$_5$CN & 52.712 &  5.04  & 258.004\\
161480.219 & C$_2$H$_3$CN & 164.3800 & 3.68   & 178.538\\
161527.500 & C$_2$H$_3$CN & 192.6770 & 1.10   &  162.238\\ 
 161581.203 & C$_2$H$_5$CN & 58.115 &  4.78  & 252.312\\
  209422.190 & \de\ & 136.101  & 0.1111             & 71.12975  \\
  209515.532 & \de\ & 41.221   & 0.1111             & 54.9886 \\
  209735.828 & C$_2$H$_3$CN &  86.7860 & 1.14      & 315.05\\
  209812.329 & \de\ & 168.885  & 0.0864             & 64.85627  \\
 224002.109 & C$_2$H$_5$CN &  166.828 &  4.21      & 309.672\\
 224003.438 & C$_2$H$_5$CN &  152.185 &  7.58      & 320.879\\
 224045.750 & C$_2$H$_5$CN &  200.696 &  4.21 & 283.72\\
 224084.281 & C$_2$H$_5$CN &  219.912 &  2.95 & 268.96\\
 224088.203 & C$_2$H$_5$CN &  127.510 &  8.01 & 339.753\\
 224186.344 & C$_2$H$_5$CN &  262.876 &  2.63 & 235.93\\
 224206.609 & C$_2$H$_5$CN &  117.492 &  4.01 & 347.406\\
 224208.078 & C$_2$H$_5$CN &  117.492 &  4.01 & 347.406\\
 224419.812 & C$_2$H$_5$CN &  109.033 &  6.01 & 353.894\\
 224458.859 & C$_2$H$_5$CN &  109.036 &  6.01 & 353.894\\
  237261.970 & \de\ & 214.188  & 0.1533             & 66.88355 \\
237396.976 & C$_2$H$_3$CN & 168.5260 & 3.14      & 336.209\\
237411.906 & C$_2$H$_3$CN & 149.0560 & 3.14      & 343.797\\
237415.359 & C$_2$H$_3$CN & 190.9600 & 2.62      & 327.453\\
237456.250 & C$_2$H$_3$CN & 216.3420 & 2.32      & 317.531\\
237483.891  &  C$_2$H$_3$CN & 132.5690 & 2.98    & 700.436 \\
237514.016 & C$_2$H$_3$CN & 244.6530 & 1.85      & 306.440\\
  237548.541 & \de\ & 447.259  & 0.1533             & 34.69413  \\
 237585.4840 &  C$_2$H$_3$CN & 275.8720 & 2.04      & 588.366 \\
 237591.3910 &  C$_2$H$_3$CN & 108.5880 & 2.38      & 359.528 \\
  237620.371 & \de\ & 32.291   & 0.1533             & 23.17836 \\
  237638.016 & C$_2$H$_3$CN & 119.0860 & 3.49      & 355.471\\
\hline
\end{longtable}
\begin{longtable}{ccccc}
\caption
{\label{tab:g292007} Transitions observed towards G29.96--0.02 during the
1997 observing run.}
\\
\hline
\hline
$\nu$ & Molecule & $E_{\rm l}$ & $\int{T_{MB}\:dv}$ & $\mu^{2}S$\\
(MHz) &             &  (cm$^{-1}$)      &   (K \kms)     &  $(D^2)$\\
\hline
 98523.883 & C$_2$H$_5$CN & 44.251 & 0.41 &  113.939\\
 98524.664 & C$_2$H$_5$CN & 54.287 & 0.52 &   96.525\\
 98532.070 & C$_2$H$_5$CN & 65.860 & 0.34 &  76.412\\
 98533.9845 & C$_2$H$_5$CN & 35.754 & 0.55 & 128.685\\
 98564.836 & C$_2$H$_5$CN & 28.801 & 0.19 &  140.761\\ 
 98566.797 & C$_2$H$_5$CN & 28.801 & 0.40 &  140.761\\
 98701.109 & C$_2$H$_5$CN & 23.398 &  0.69 & 150.140\\
  104177.000 & \de\ & 102.625  & 0.021            & 79.56847 \\
  104490.168 & \de\ & 585.600  & 0.027              & 4.48006 \\
 104437.5160 &  C$_{2}$H$_{3}$CN &  90.9290 & 0.353     & 191.014 \\
  111802.17  & \de\ & 42.141   & 0.024        & 2.53886  \\
  111802.93  & \de\ & 42.142   & 0.0183         & 2.53886 \\  	  	
 b111813.417 & \de\ & 118.013  & 0.0368  & 96.5637 \\
 b147205.839 & \de\ & 22.082   & 0.0458  & 19.52177 \\
  147456.863 & \de\ & 257.141  & 0.040      & 145.57157 \\
 161261.141 & C$_2$H$_5$CN & 95.144 &  2.14 &  213.0\\
 161265.734 & C$_2$H$_5$CN & 108.253 &  1.16 &  199.066\\
  161289.996 & \de\ & 333.668  & 0.0467             & 162.13361  \\
 161303.484 & C$_2$H$_5$CN & 139.060 &  2.14 &  166.301\\
 161304.922 & C$_2$H$_5$CN & 73.536  & 1.43  &  235.930\\
 161367.344 & C$_2$H$_5$CN & 175.957 &  1.79 & 126.974\\
 161379.844 & C$_2$H$_5$CN & 65.047 &  1.34 &  244.939\\
 161381.875 & C$_2$H$_5$CN & 65.047 &  1.34 &  244.939\\
 161397.047 & C$_2$H$_3$CN &  97.1140 & 0.89  & 217.164\\
 161475.172 & C$_2$H$_5$CN & 52.712 &  1.63 &  258.004\\
 161480.219 & C$_2$H$_3$CN & 164.3800 & 0.71  & 178.538\\
 161581.203 & C$_2$H$_5$CN & 58.115 &  1.59   &  252.312\\
  173083.162 & \de\ & 110.480  & 0.1542     & 34.27361 \\
  173094.261 & \de\ & 149.025  & 0.1542      & 85.5836 \\
  173293.064 & \de\ & 34.233   & 0.1542    & 46.73276   \\
 209735.8280 & C$_2$H$_3$CN  &  86.7860  &  0.73  & 315.05 \\
  209422.190 & \de\ & 136.101  & 0.094    & 71.12975 \\
  209515.532 & \de\ & 41.221   & 0.094          & 54.9886 \\
  209812.329 & \de\ & 168.885  & 0.0733           & 64.85627 \\   
 224002.109 & C$_2$H$_5$CN &  166.828 &  1.86 &  309.672\\
 224003.438 & C$_2$H$_5$CN &  152.185 &  1.72 &  320.879\\
 224017.5310 & C$_2$H$_5$CN &  183.0000  &  1.21  &     594.572 \\
 224028.1410  & C$_2$H$_5$CN &  139.0770  &  1.08  &    661.812 \\
 224045.750 & C$_2$H$_5$CN &  200.696 &  2.19 &   283.720\\
 224084.281 & C$_2$H$_5$CN &  219.912 &  2.09 &   268.960\\
 224088.2035 & C$_2$H$_5$CN &  127.510 &  2.00 &   339.753\\
 224206.609 & C$_2$H$_5$CN &  117.492 &  2.36 &   347.406\\
 224208.078 & C$_2$H$_5$CN &  117.492 &  1.68 &   347.406\\
 224419.812 & C$_2$H$_5$CN &  109.033 &  2.86 &   353.894\\
 224458.859 & C$_2$H$_5$CN &  109.036 &  2.36 &   353.894\\
  237261.970 & \de\ & 214.188  & 0.146    &       66.88355 \\
  237396.976 & C$_2$H$_3$CN & 168.526 & 1.22      & 336.209\\
  237548.541 & \de\ & 447.259  & 0.146            & 34.69413 \\            
  237620.371 & \de\ & 32.291   & 0.146           & 23.17836 \\
\hline
\end{longtable}
\begin{longtable}{ccccc}
\caption
{\label{tab:g292006} Transitions observed towards G29.96-0.02 during the
1996 observing run.}
\\
\hline
\hline
$\nu$ & Molecule & $E_{\rm l}$ & $\int{T_{MB}\:dv}$ & $\mu^{2}S$\\
(MHz) &             &  (cm$^{-1}$)      &   (K \kms)     &  $(D^2)$\\
\hline
  b154311.469& \des &185.5   &0.0114            & 12.859  \\
  b154456.5  & \des &62.92   &0.063             & 21.739  \\
 215040.8    & \ecys &     229.5   &  5.22     & 1199.4  \\  
 215088.203  & \ecys &     288.76  &  2.31     & 533.6 \\
 215109.047  & \ecys &     183.46  &  4.82     & 651.1 \\
 215119.203  & \ecys &     135.84  &  4.22     & 369.666 \\
 215126.703  & \ecys &     316.4   &  1.66     & 502.8 \\
 215173.234  & \ecys &     346.23  &  1.5      & 469.4 \\
 215400.797  & \ecys &     156.87  &  3.44     & 340.336 \\
 215427.984  & \ecys &     156.88  &  3.88     & 340.332  \\   
\hline 
\end{longtable}
\begin{longtable}{ccccc}
\caption
{\label{tab:g312007} Transitions observed towards G31.41+0.31 during the
1997 August observing run.}
\\
\hline
\hline
$\nu$ & Molecule & $E_{\rm l}$ & $\int{T_{MB}\:dv}$ & $\mu^{2}S$\\
(MHz) &             &  (cm$^{-1}$)      &   (K \kms)     &  $(D^2)$\\
\hline
 98523.883 & C$_2$H$_5$CN &  44.251 &  1.60 & 113.939\\ 
 98524.664 & C$_2$H$_5$CN &  54.287 &  2.00 & 96.525\\
 98532.070 & C$_2$H$_5$CN &  65.860 &  1.80 & 76.412\\   
 98533.9845 & C$_2$H$_5$CN &  35.754 &  2.79 & 128.685\\   
 98544.148 & C$_2$H$_5$CN &  78.967  & 2.22 & 53.615\\   
 98559.867 & C$_2$H$_5$CN &  93.604  & 0.59 & 28.149\\   
 98564.836 & C$_2$H$_5$CN &  28.801  & 1.87 & 140.761\\   
 98566.797 & C$_2$H$_5$CN &  28.801  & 2.42 & 140.761\\
 98701.109 & C$_2$H$_5$CN &  23.398  & 2.60 & 150.140\\
 b104177.000 & \de\ & 102.625  & 0.0843  & 79.56847  \\
104212.656 & C$_2$H$_3$CN &  23.411 & 1.18       & 155.205\\
104419.312 & C$_2$H$_3$CN &  71.464 & 1.48           & 112.653\\
104432.7970 & C$_2$H$_3$CN &  30.9450 & 1.13            & 148.58 \\
104453.930 & C$_2$H$_3$CN &  30.9460 & 1.62           & 148.580\\
  104490.168 & \de\ & 585.600  & 0.0424             & 4.48006 \\
b 111574.6250 & C$_2$H$_5$CN &  73.7330  &  0.38    &  10.807 \\ 
 b111742.636 & \de\ & 130.299  & 0.1514  & 103.90668 \\
 b111783.241 & \de\ & 17.550   & 0.1647  & 27.2197 \\
 b111801.945 & \de\ & 42.142   & 0.029   & 4.43093 \\
 b111813.417 & \de\ & 118.013  & 0.1694  & 96.5637 \\
 b147205.839 & \de\ & 22.082   & 0.3239  & 19.52177  \\
 b147456.863 & \de\ & 257.141  & 0.0816  & 145.57157 \\
 147561.719 & C$_2$H$_3$CN &  38.4910 & 2.66   & 232.515\\
 161261.141 & C$_2$H$_5$CN &  95.144 & 7.89 & 213.\\
 161265.734 & C$_2$H$_5$CN & 108.253 & 6.91  & 199.066\\
 b161281.953 & \de\ & 75.305   & 0.1839   & 49.41313  \\
  161289.996 & \de\ & 333.668  & 0.0667    & 162.13361  \\     
 161303.484 & C$_2$H$_5$CN & 139.060 & 5.67  & 66.301\\
 161304.922 & C$_2$H$_5$CN & 73.536 & 3.70  & 235.93\\
 161367.344 & C$_2$H$_5$CN & 175.957 & 2.47  & 126.974\\  
 161379.844 & C$_2$H$_5$CN & 65.047 & 3.86   & 244.939\\
 161381.875 & C$_2$H$_5$CN & 65.047 & 4.12 & 244.939\\
 161397.047 & C$_2$H$_3$CN &  97.114 & 3.62      & 217.164\\
 161447.8750 & C$_2$H$_3$CN  &  67.1130 & 5.4     & 468.678 \\
 161475.172 & C$_2$H$_5$CN & 52.712 & 8.24 & 258.004\\
161502.266 & C$_2$H$_3$CN &  56.605 & 2.16 & 240.337\\
 161581.203 & C$_2$H$_5$CN & 58.115 & 5.92 & 252.312\\
 172998.766 & C$_2$H$_5$CN & 55.578 & 9.50 & 293.998\\
 173092.859 & C$_2$H$_3$CN &  54.7720 & 11.2  & 259.424\\
  173083.162 & \de\ & 110.480  & 0.3097             & 34.27361 \\
  173094.261 & \de\ & 149.025  & 0.1542             & 85.5836  \\
 b173293.064 & \de\ & 34.233   & 0.4182   & 46.73276 \\
 b209515.532 & \de\ & 41.221   & 0.479    & 54.9886  \\
  209812.329 & \de\ & 168.885  & 0.1334             & 64.85627  \\
 224002.109 & C$_2$H$_5$CN & 166.828 & 5.36 & 309.672\\  
 224003.438 & C$_2$H$_5$CN & 152.185 & 6.13 & 320.879\\
 224045.75 & C$_2$H$_5$CN & 200.696 & 6.72 & 283.720\\ 
 224084.281 & C$_2$H$_5$CN & 219.912 & 4.60 & 268.960\\  
 224088.2035 & C$_2$H$_5$CN & 127.510 & 6.89 & 339.753\\  
 224131.516 & C$_2$H$_5$CN & 240.640 & 3.82 & 253.034\\
 224186.344 & C$_2$H$_5$CN & 262.876 & 3.73 & 235.930\\  
 224206.609 & C$_2$H$_5$CN & 117.492 & 5.22 & 347.406\\  
 224208.078 & C$_2$H$_5$CN & 117.492 & 5.22 & 347.406\\  
 224248.016 & C$_2$H$_5$CN & 286.614 & 1.91 & 217.645\\  
 224419.812 & C$_2$H$_5$CN & 109.033 & 7.66 & 353.894\\  
 224458.859 & C$_2$H$_5$CN & 109.036 & 7.63 & 353.894\\
 237396.9765 & C$_2$H$_3$CN & 168.526 & 2.50 & 336.209\\
237405.1880   &  C$_2$H$_5$CN &  101.8870  &  8.49 &  380.894   \\
b237476.0470   &  C$_2$H$_5$CN &  91.4820  &  3.77 &  19.864   \\     
237456.250 & C$_2$H$_3$CN & 216.3420 & 2.10       & 317.531\\
237485.016 & C$_2$H$_3$CN & 132.5700 & 3.12       & 350.218\\
  237548.541 & \de\ & 447.259  & 0.1445             & 34.69413 \\
  237591.3910 &  C$_2$H$_3$CN & 108.5880 & 2.17      & 359.528 \\
 b237620.371 & \de\ & 32.291   & 0.452    & 23.17836 \\
237638.0160 &  C$_2$H$_3$CN & 119.0860 & 2.17      & 355.471 \\
237668.7660 &  C$_2$H$_3$CN & 309.9800 & 2.03      & 561.516 \\
\hline
\end{longtable} 
\begin{longtable}{ccccc}
\caption
{\label{tab:g312006} Transitions observed towards G31.41+0.31 during the
1996 July observing run.}
\\
\hline
\hline
$\nu$ & Molecule & $E_{\rm l}$ & $\int{T_{MB}}\:dv$ & $\mu^{2}S$\\
(MHz) &             &  (cm$^{-1}$)      &   (K \kms)     &  $(D^2)$\\
\hline
  b140226.172& \des &159.81  &0.0686            &9.896 \\
  a140429.484 &    \ecys   &54.24   &2.2               &202.872 \\
  b154027.062 &    \ecys   &10.02   &2.64              &2.344 \\
  b154311.469& \des &185.5   &0.035             &12.855 \\
  b154456.5  & \des &62.92   &0.2081            &21.739 \\
  154724.531 &    \ecys   &65.42   &5.75              &215.731   \\
  215040.8 &\ecys &     229.5   & 15.9    & 1199.4 \\   
 215088.203 &\ecys &      288.76   & 8.46  &  533.6  \\
 215109.047  &\ecys &     183.46   &  12.49   & 651.1  \\
 215119.203  &\ecys &     135.84   &  12.63   & 369.666 \\
 215126.703  &\ecys &     316.4    &  6.56    &  502.8  \\
 215173.234  &\ecys &     346.23   &  5.91    &  469.4  \\
 215400.797  &\ecys &     156.87   &  11.84   &  340.336 \\
 215427.984 &\ecys &      156.88   &  12.85   & 340.332 \\ 
 227897.625 &   \ecys   &242.47  &6.59              &594.744 \\
  227906.719 &    \ecys   &214.46  &6.89              &609.457 \\
  227918.656 &    \ecys   &274.75  &3.33              &577.798 \\
  227960.234 &    \ecys   &311.27  &3.03              &558.56 \\
  227967.594 &     \ecys  &190.74  &6.87              &621.8 \\
  228017.375 &    \ecys   &352.    &3.44              &537.313 \\
  228087.297 &    \ecys   &396.91  &0.91              &513.431 \\
  228090.531 &    \ecys   &156.23  &5.45              &319.903 \\
  228104.609 &    \ecys   &171.34  &5.45              &315.959 \\
  228160.297 &    \ecys   &171.35  &7.02              &631.917 \\
\hline
\end{longtable} 
\begin{longtable}{ccccc}
\caption
{\label{tab:g342007} Transitions observed towards G34.26+0.15 during
the 1997 observing run.}
\\
\hline
\hline
$\nu$ & Molecule & $E_{\rm l}$ & $\int{T_{MB}\:dv}$ & $\mu^{2}S$\\
(MHz) &             &  (cm$^{-1}$)      &   (K \kms)  &  $(D^2)$\\
\hline
98523.883 & C$_2$H$_5$CN &  44.251 &    1.23 & 113.939\\
98524.664 & C$_2$H$_5$CN &  54.287 &  0.96  & 96.525\\
98532.070  & C$_2$H$_5$CN &  65.860 & 1.16  & 76.412\\
98533.9845   & C$_2$H$_5$CN & 35.754 & 1.64 & 128.685\\
98544.148 & C$_2$H$_5$CN & 78.967 & 0.68 & 53.615\\
98559.867 & C$_2$H$_5$CN & 93.604 & 0.24 & 28.149\\
98564.836 & C$_2$H$_5$CN & 28.801 & 1.10 & 140.761\\
98566.797 & C$_2$H$_5$CN & 28.801 & 1.35 & 140.761\\
98701.109 & C$_2$H$_5$CN & 23.398 & 1.37 & 150.140\\
b104177.000 & \de\ & 102.625  & 0.1033  & 79.56847 \\
104419.312 & C$_2$H$_3$CN &  71.4640 & 0.57  & 112.653\\
104432.797 & C$_2$H$_3$CN &  30.9450 & 0.57  & 148.58\\
104437.516 & C$_2$H$_3$CN &  90.9290 & 0.71  & 95.507\\
104453.930 & C$_2$H$_3$CN &  30.9460 & 0.89  & 148.58\\
104461.5160 &  C$_2$H$_3$CN  & 113.3510 & 0.425  & 151.236 \\
b111783.241 & \de\ & 17.550   & 0.1639  & 27.2197 \\
  111802.93  & \de\ & 42.142   & 0.0192              & 2.53886    \\  
 b111813.417  & \de\ & 118.013  & 0.1493  & 96.5637 \\
 b147205.839 & \de\ & 22.082   & 0.2814  & 19.52177  \\           
 b147456.863 & \de\ & 257.141  & 0.0763  & 145.57157 \\
161261.141 & C$_2$H$_5$CN & 95.144 & 9.19 & 213.\\
161265.734 & C$_2$H$_5$CN & 108.253  & 6.32 & 199.066\\
b161285.603 & \de\ & 75.305   & 0.279   & 12.35331 \\ 
161289.996 & \de\ & 333.668  & 0.0566   & 162.13361  \\                     
161303.484 &  C$_{2}$H$_{5}$CN  & 139.060  & 6.89 & 166.301\\
161304.922 &  C$_{2}$H$_{5}$CN  & 73.536 & 6.89  & 235.930\\
161367.344 & C$_2$H$_5$CN & 175.957  & 2.87  & 126.974\\
161379.844 & C$_2$H$_5$CN & 65.047 & 6.24 & 244.939\\
161381.875 & C$_2$H$_5$CN & 65.047 & 4.05 & 244.939\\
161397.047 & C$_2$H$_3$CN &  97.1140 & 2.31  & 217.164\\
161445.391 & C$_2$H$_3$CN &  67.1130 & 2.21  & 234.339\\
161450.359 & C$_2$H$_3$CN &  67.1130 & 2.21  & 234.339\\
161475.172 & C$_2$H$_5$CN & 52.712  & 7.47   & 258.004\\
161581.203 & C$_2$H$_5$CN & 58.115 & 5.75    & 252.312\\
172998.766 & C$_2$H$_5$CN & 55.578 & 5.92 & 293.998\\
173092.859 & C$_2$H$_3$CN &  54.7720 & 4.69     & 259.424\\
 b173293.064 & \de\ & 34.233   & 0.3016  & 46.73276  \\
                 
b209515.532 & \de\ & 41.221   & 0.7994  & 54.9886   \\
 b209812.329 & \de\ & 168.885  & 0.1865  & 64.85627 \\
224002.109 & C$_2$H$_5$CN & 166.828  & 9.23 & 309.672\\
224003.438 & C$_2$H$_5$CN & 152.185 & 3.69 & 320.879\\
224045.75 & C$_2$H$_5$CN & 200.696 & 3.26 & 283.720\\
224084.281 & C$_2$H$_5$CN & 219.912 & 3.88 & 268.960\\
224088.2035 & C$_2$H$_5$CN & 127.510 & 6.79 & 339.753\\
224186.344 & C$_2$H$_5$CN & 262.876 & 2.43 & 235.930\\
224206.609   & C$_2$H$_5$CN & 117.492 & 4.95 & 347.406\\
224208.078 & C$_2$H$_5$CN & 117.492 & 4.95 & 347.406\\
b224310.9220 &  C$_2$H$_3$CN  &  84.2220  &  5.83  &  2.17 \\
224315.9380  &  C$_2$H$_5$CN &  311.8460  &  2.62  &  396.362 \\
224419.812  & C$_2$H$_5$CN & 109.033 & 5.32 & 353.894\\
224458.859 & C$_2$H$_5$CN & 109.036  & 5.64 & 353.894\\
237396.969 & C$_2$H$_3$CN & 168.5260 & 1.40        & 336.209\\
237405.1880  &  C$_2$H$_5$CN &  101.8870  &  9.42   &  380.894 \\
237411.906 & C$_2$H$_3$CN & 149.0560 & 1.67       & 343.797\\
237485.016 & C$_2$H$_3$CN & 132.5700 & 1.50       & 350.218\\
  237548.541 & \de\ & 447.259  & 0.152      & 34.69413   \\         
b237620.371 & \de\ & 32.291   & 0.443   & 23.17836  \\
237638.016 & C$_2$H$_3$CN & 119.0860 & 1.76  & 355.471\\
\hline
\end{longtable}
\begin{table}[htbp]
\caption []
{\label{tab:g342006} Transitions observed towards G34.26+0.15 during the
1996 observing run.}
\begin{center}
\begin{tabular}{*{5}{c}}
\hline
\hline
$\nu$ & Molecule & $E_{\rm l}$ & $\int{T_{MB}\:dv}$ & $\mu^{2}S$\\
(MHz) &             &  (cm$^{-1}$)      &   (K km/s)     &  $(D^2)$\\
\hline
 215040.8    &\ecys&     229.5   & 15.83    & 1199.4   \\
 215088.203  &\ecys&      288.76  & 7.      & 533.6    \\
 215109.047  &\ecys&      183.46  & 10.73   & 651.1   \\
 215119.203  &\ecys&      135.84  & 11.35   & 369.666  \\
 215126.703  &\ecys&      316.4   & 4.87    & 502.8\\
 215173.234  &\ecys&      346.23  & 4.37    & 469.4\\
 215400.797  &\ecys&     156.87   & 9.2     & 340.336\\
 215427.984  &\ecys&      156.88  & 11.88    & 340.332 \\     
\hline
\end{tabular}
\end{center}
\end{table}
\end{appendix}
%
\begin{appendix}
\section*{Appendix C: Boltzmann plots} 
 
\begin{figure*}[t!]
 \centering
 \resizebox{\textwidth}{!}{\includegraphics[angle=-90]{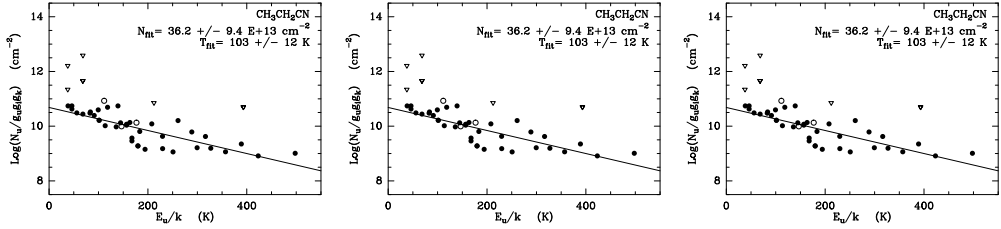} %
 			   \includegraphics[angle=-90]{7485C1a.ps}%
			   \includegraphics[angle=-90]{7485C1a.ps}}
 \caption[Boltzmann plots for G10.47]
 {\label{fig:g10.ps}{Boltzmann plots for G10.47. $T_{\rm fit}$ and 
$N_{\rm fit}$ are the ``best-fit'' values fot the rotational temperature 
and the total column densities (see Tab.~\ref{tab:Results}).
Filled circles indicate ``strong'' transitions; open circles correspond to
``weak'' transitions. Open triangles indicate upper limits for
undetected lines, estimated from the 3 $\sigma$ level in the spectrum.
The error on the quantity on the y-axis is about 20\% , and it
is due to calibration uncertainties.}}
\end{figure*}

\begin{figure*}[t!]
 \begin{center}
 \resizebox{\textwidth}{!}{\includegraphics[angle=-90]{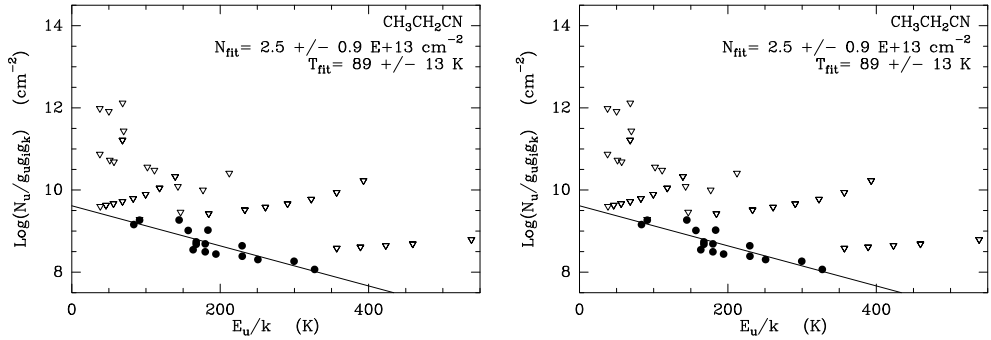}%
			   \includegraphics[angle=-90]{7485C2a.ps}}
 \end{center}
 \caption[Boltzmann plots for G10.62]
 {\label{fig:g106.ps}{Same as Fig.~\ref{fig:g10.ps}, but for G10.62.}}
\end{figure*}

\begin{figure*}[t!]
 \begin{center}
 \resizebox{\textwidth}{!}{\includegraphics[angle=-90]{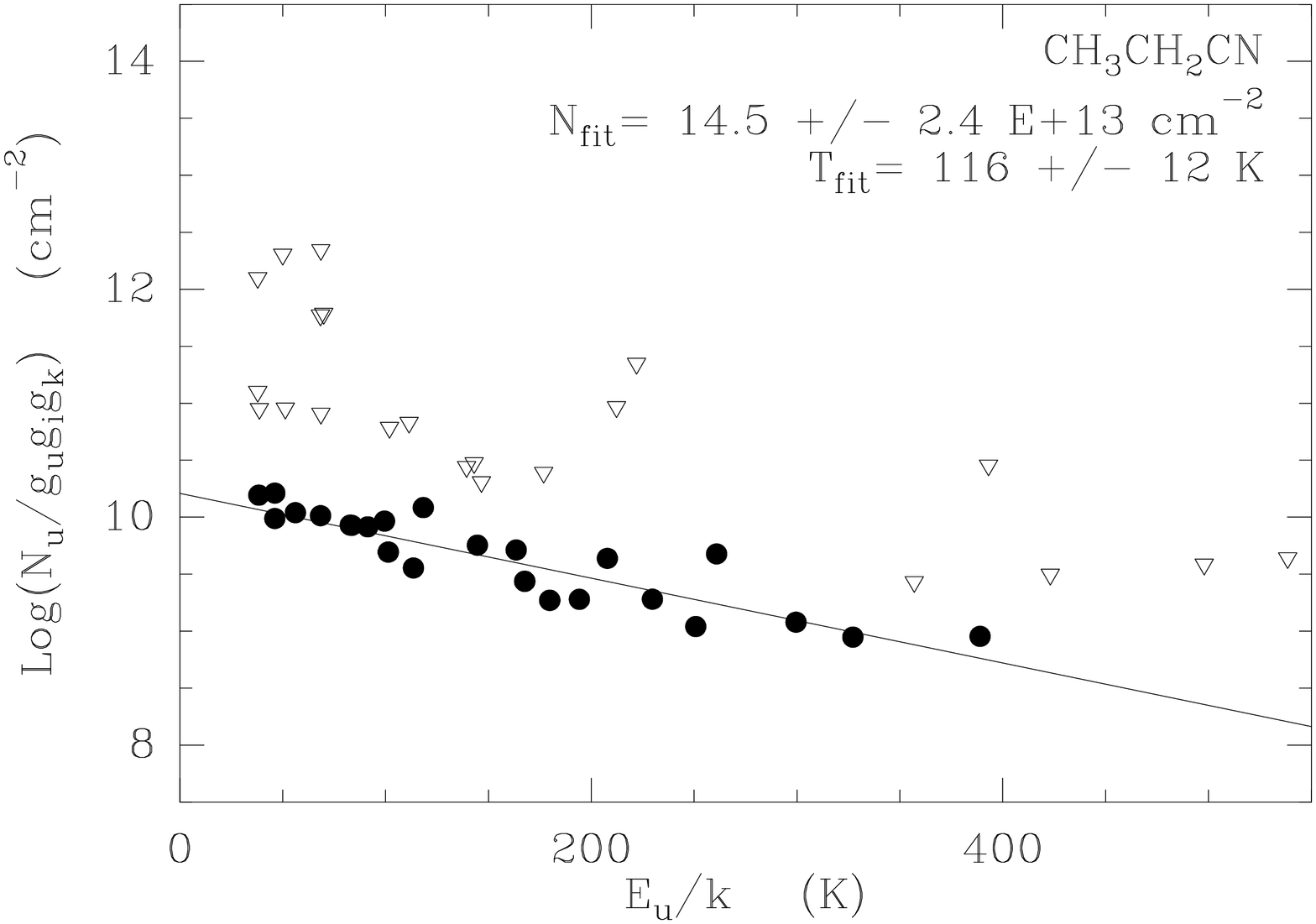} %
 			   \includegraphics[angle=-90]{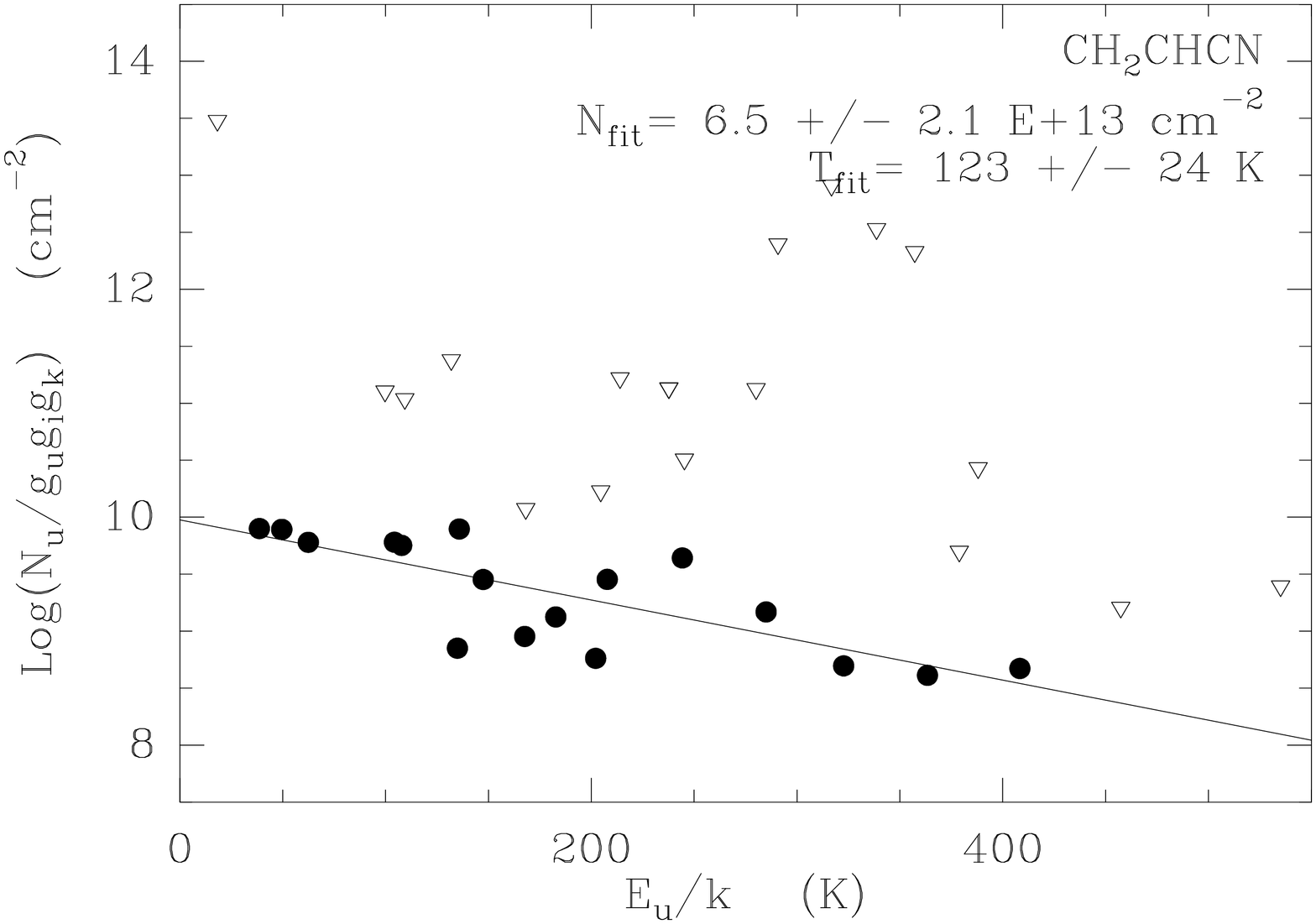} %
			   \includegraphics[angle=-90]{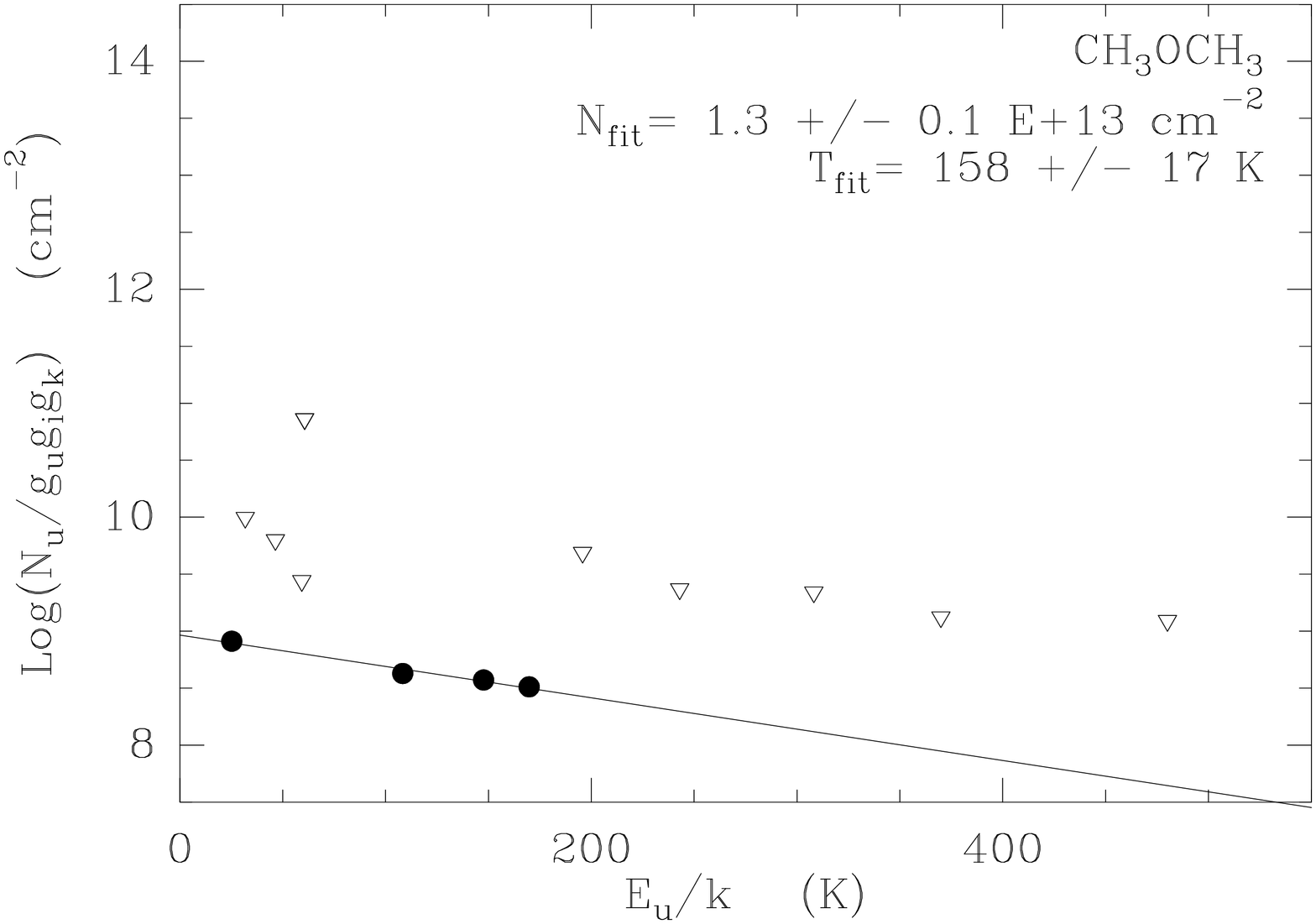}}
 \end{center}
 \caption[Boltzmann plots for G19.96]
 {\label{fig:g19.ps}{Same as Fig.~\ref{fig:g10.ps}, but for G19.61.}}
\end{figure*}

\begin{figure*}[t!]
 \begin{center}
 \resizebox{\textwidth}{!}{\includegraphics[angle=-90]{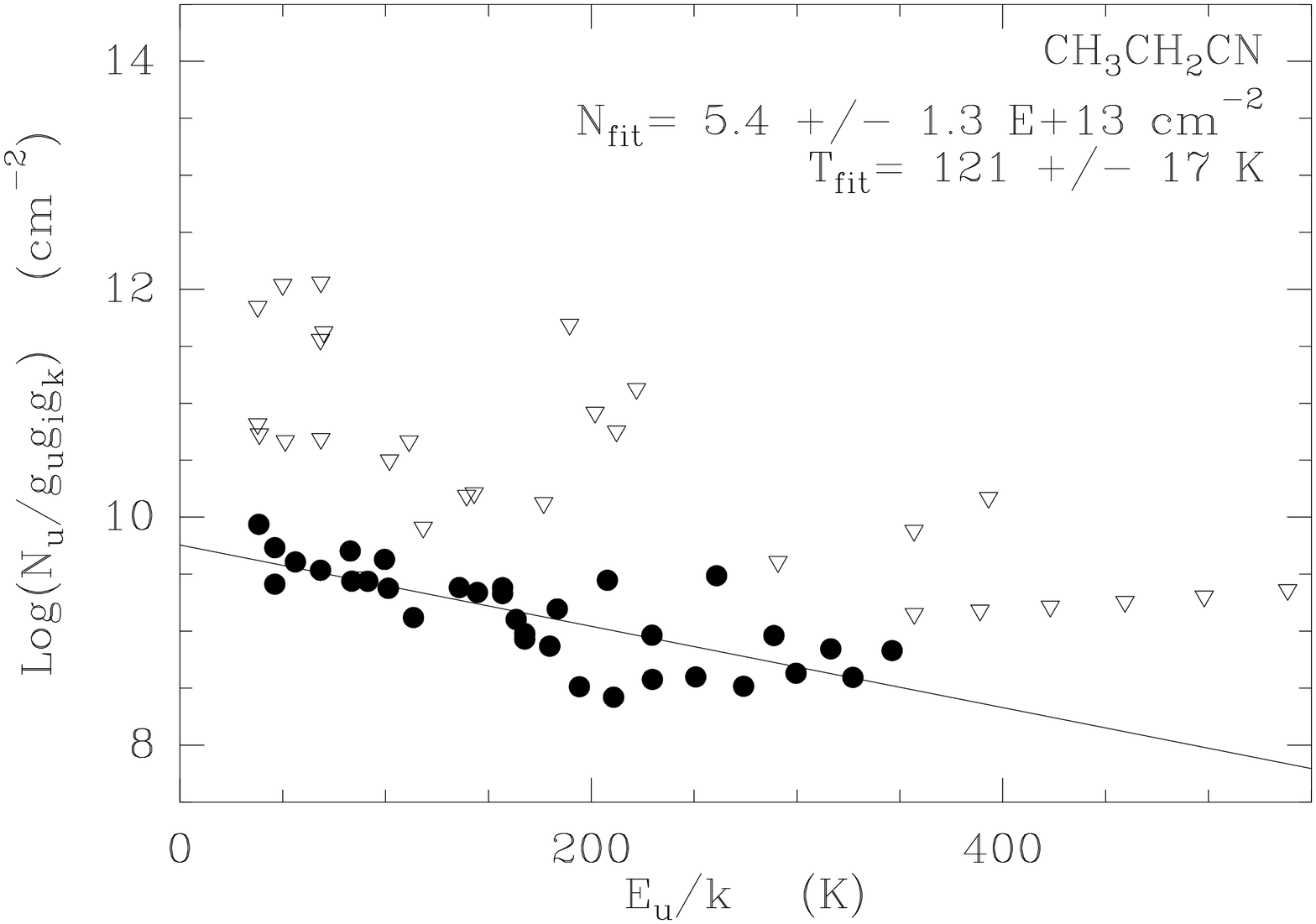}%
 			   \includegraphics[angle=-90]{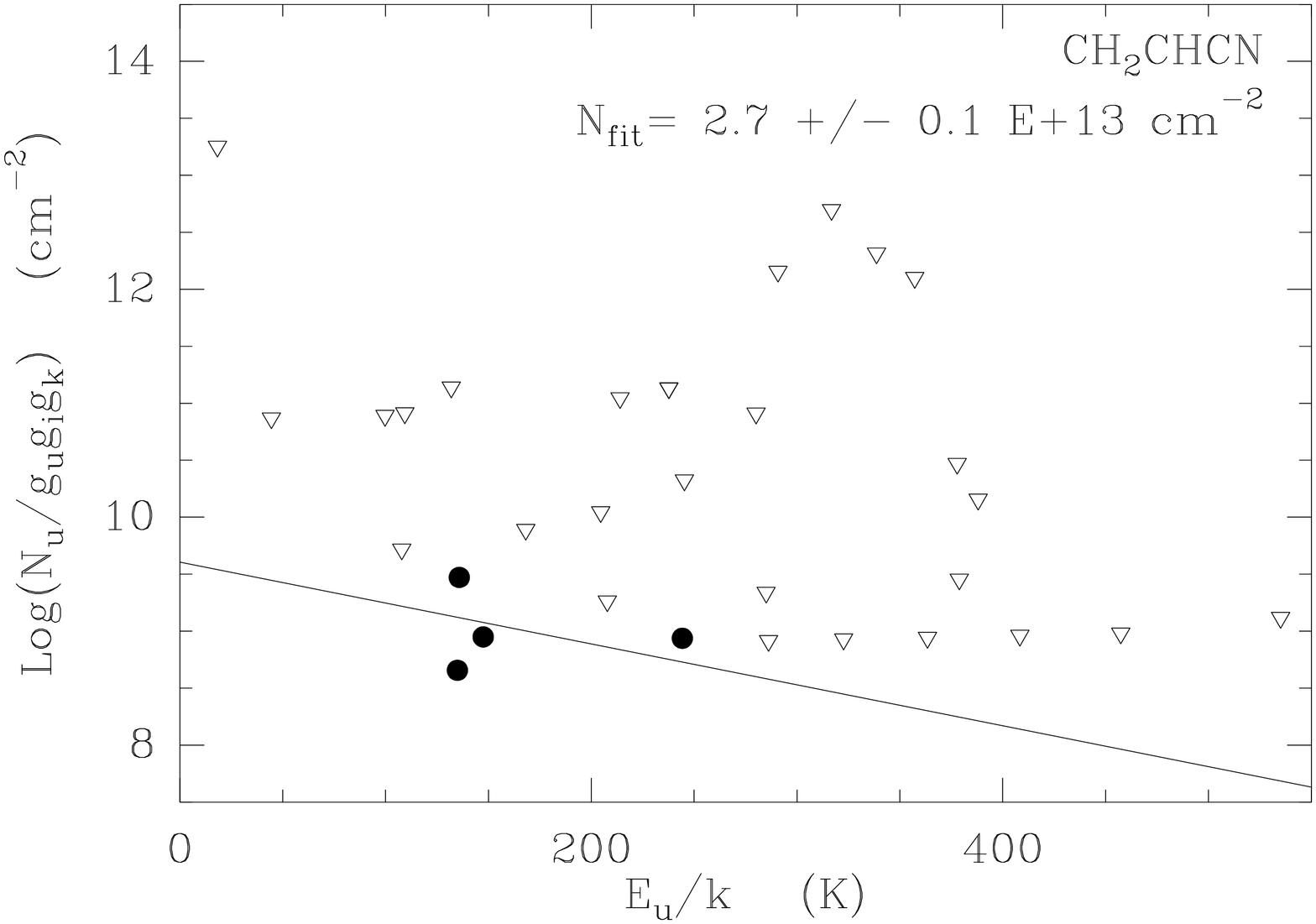}%
			   \includegraphics[angle=-90]{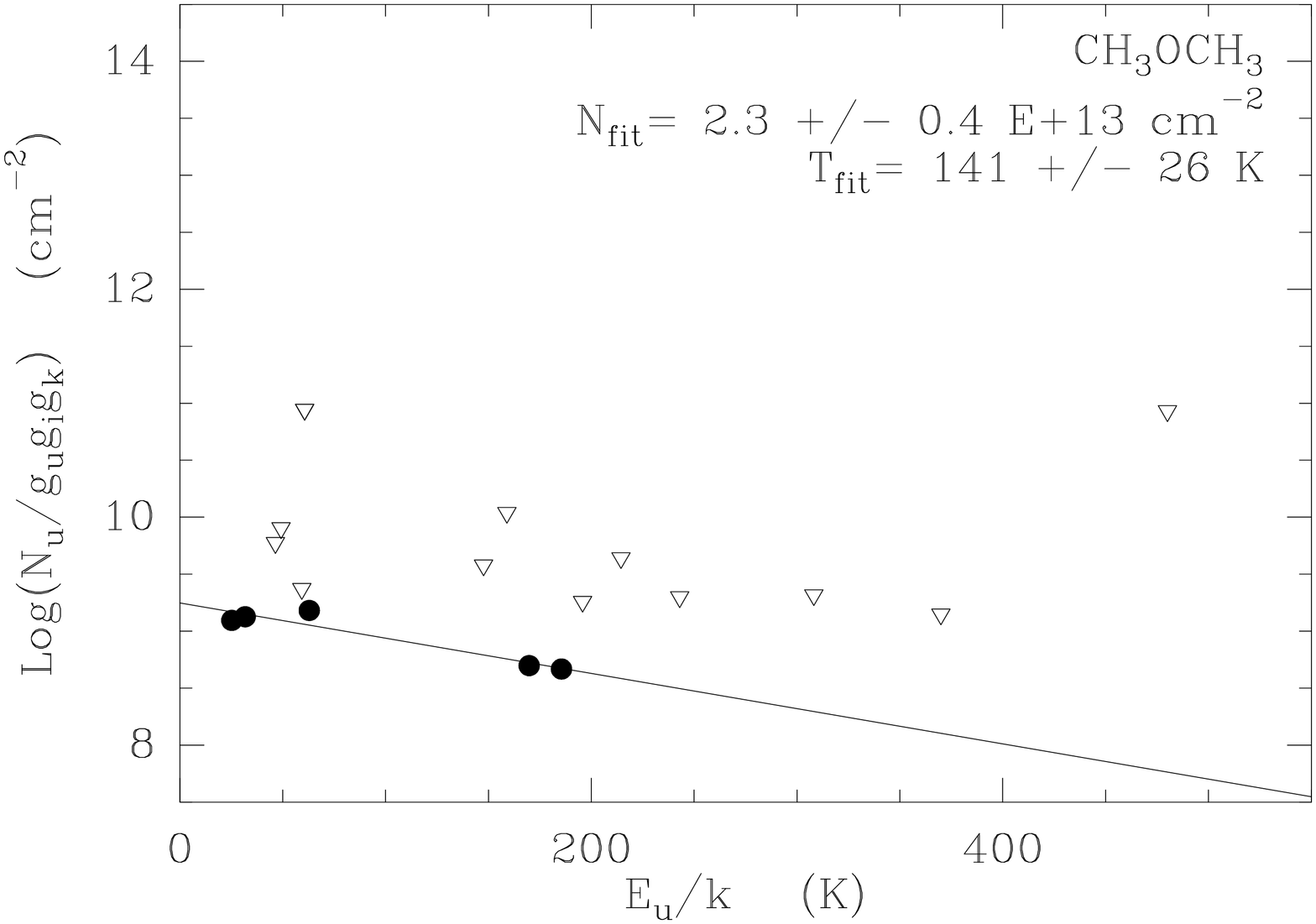}} 
 \end{center}
 \caption[Boltzmann plots for G29.96]
 {\label{fig:g29.ps}{Same as Fig.~\ref{fig:g10.ps}, but for G29.96.}}
\end{figure*}

\begin{figure*}[t!]
 \begin{center}
 \resizebox{\textwidth}{!}{\includegraphics[angle=-90]{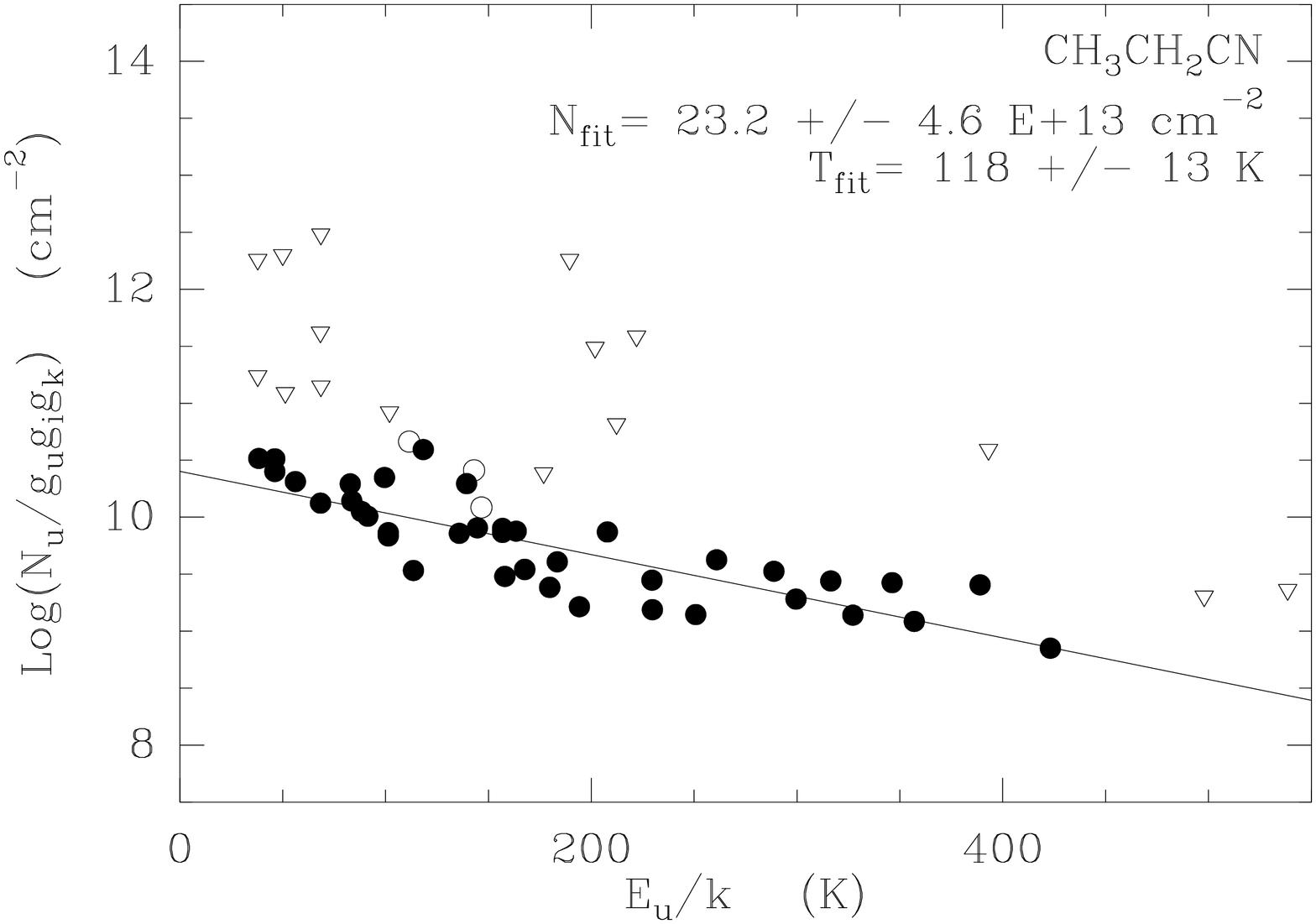} %
 			   \includegraphics[angle=-90]{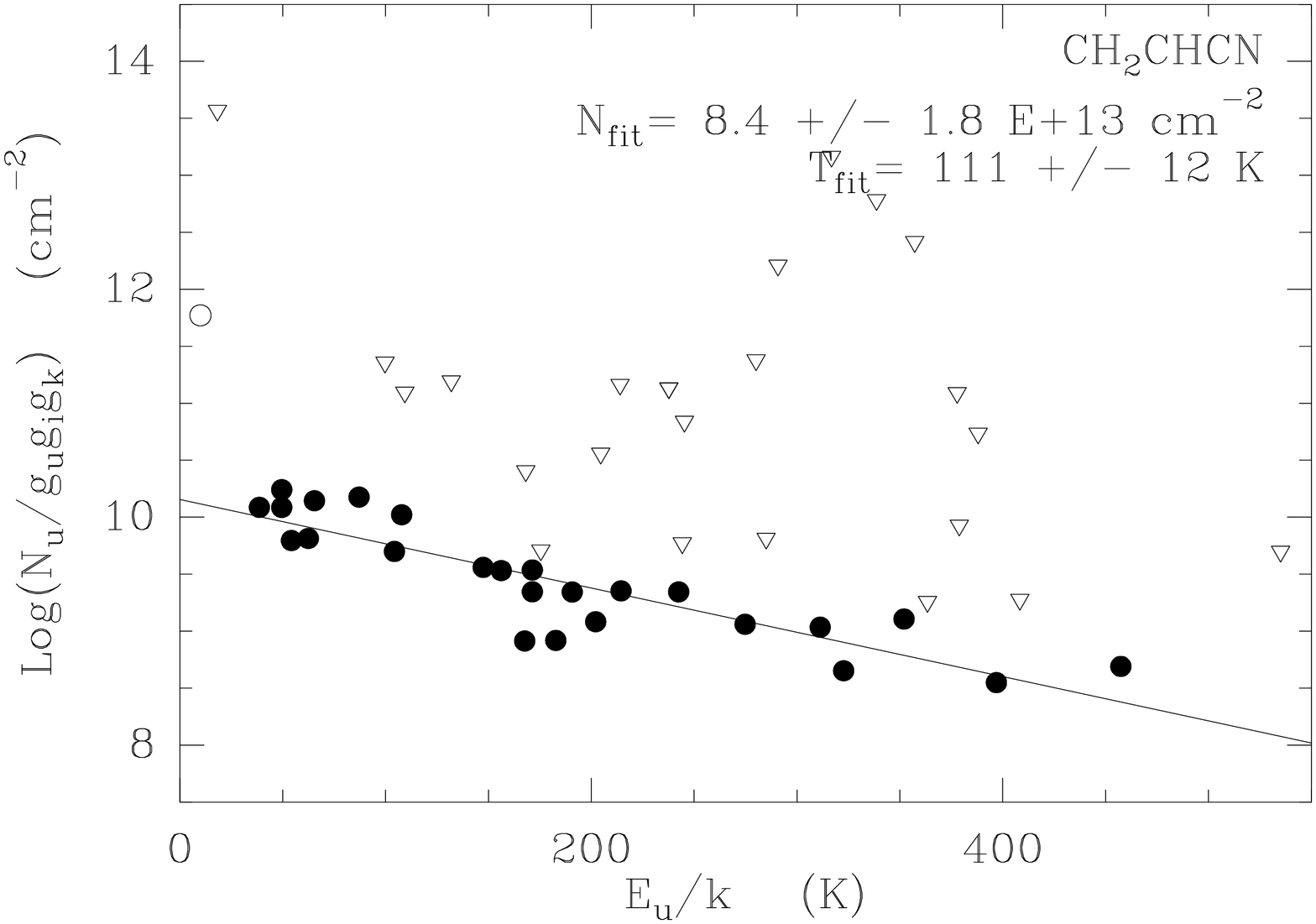}%
			   \includegraphics[angle=-90]{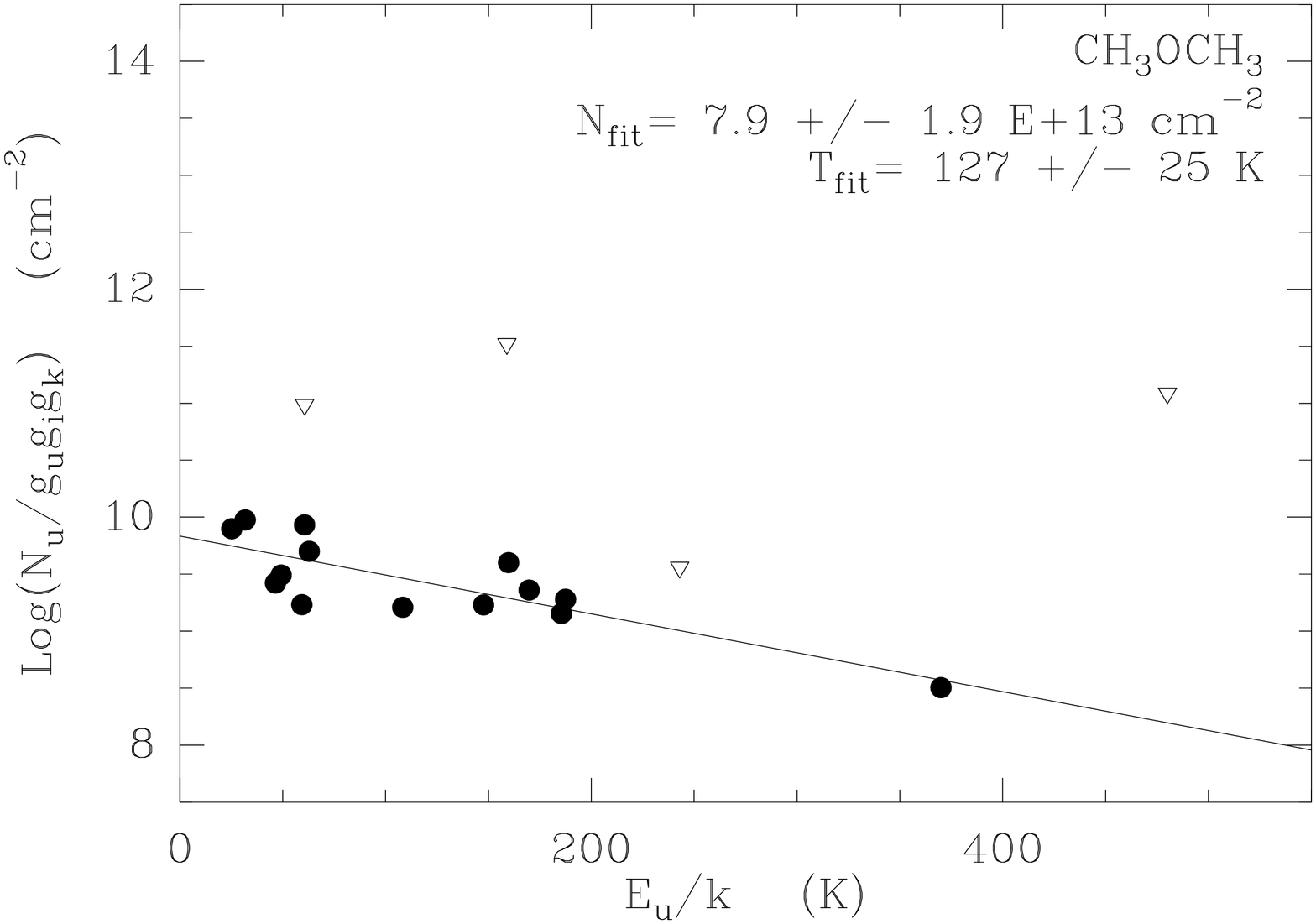}}
 \end{center}
 \caption[Boltzmann plots for G31.41]
 {\label{fig:g31.ps}{Same as Fig.~\ref{fig:g10.ps}, but for G31.41.}}
\end{figure*}

\begin{figure*}[t!]
 \begin{center}
 \resizebox{\textwidth}{!}{\includegraphics[angle=-90]{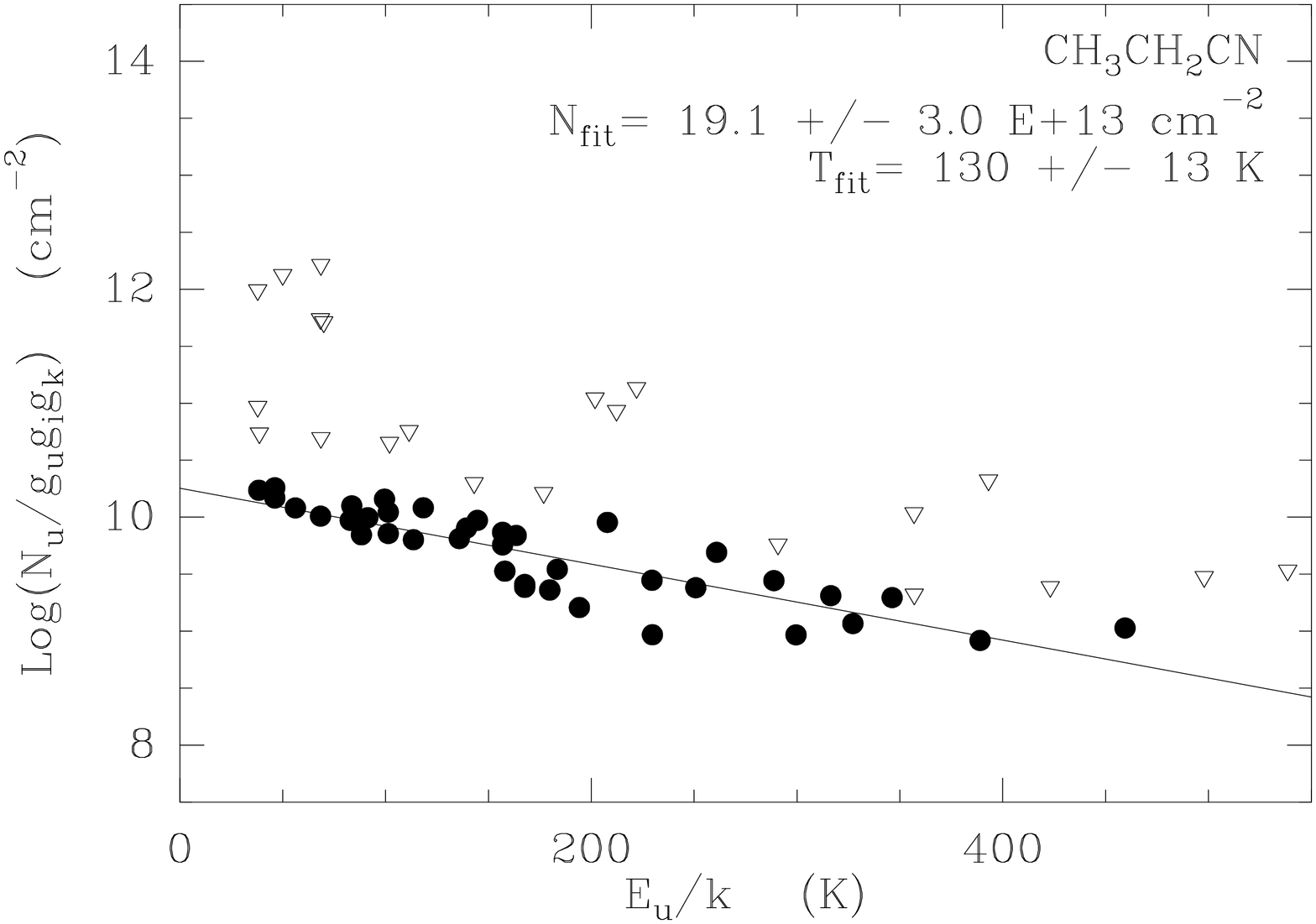} %
 			   \includegraphics[angle=-90]{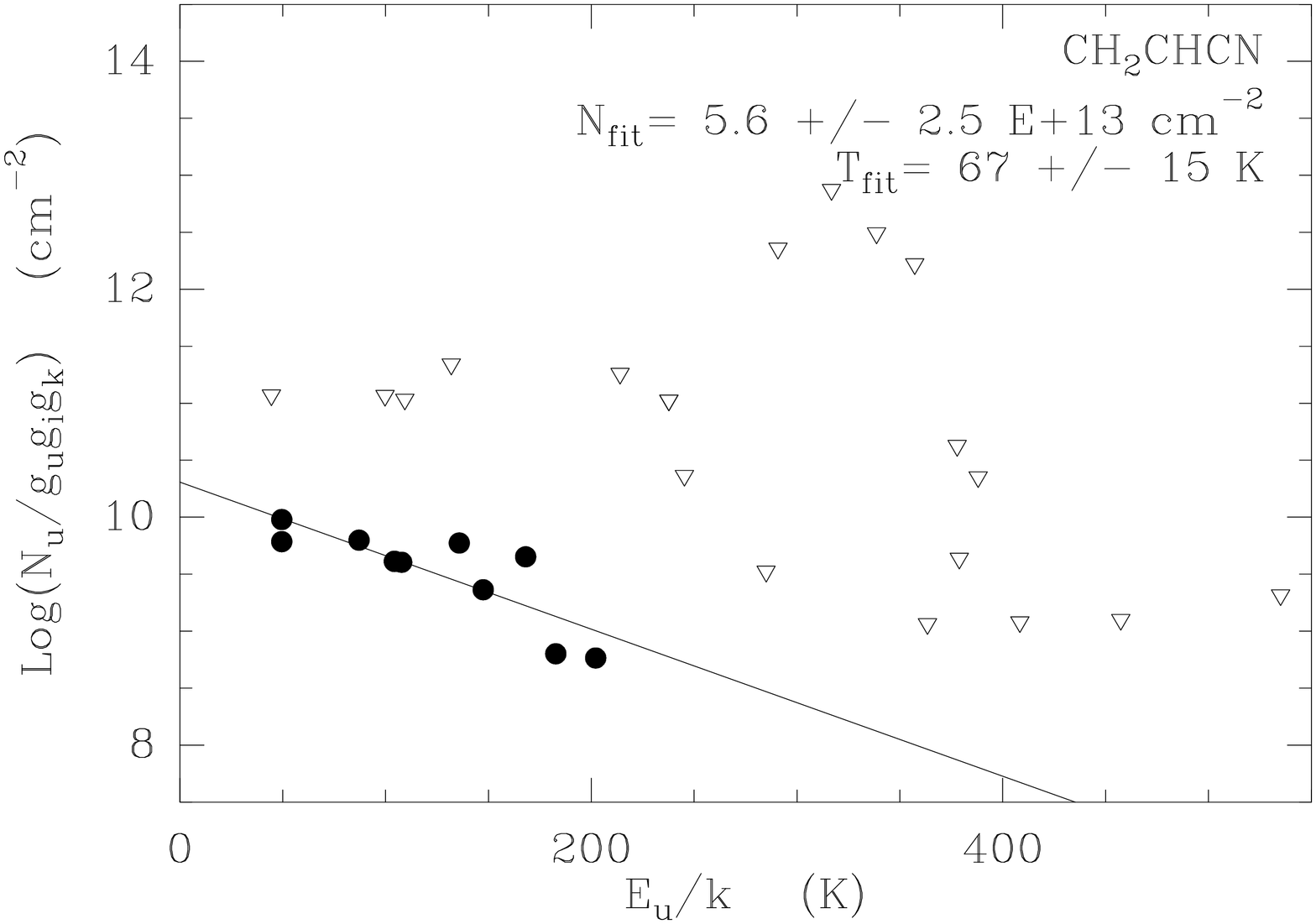} %
			   \includegraphics[angle=-90]{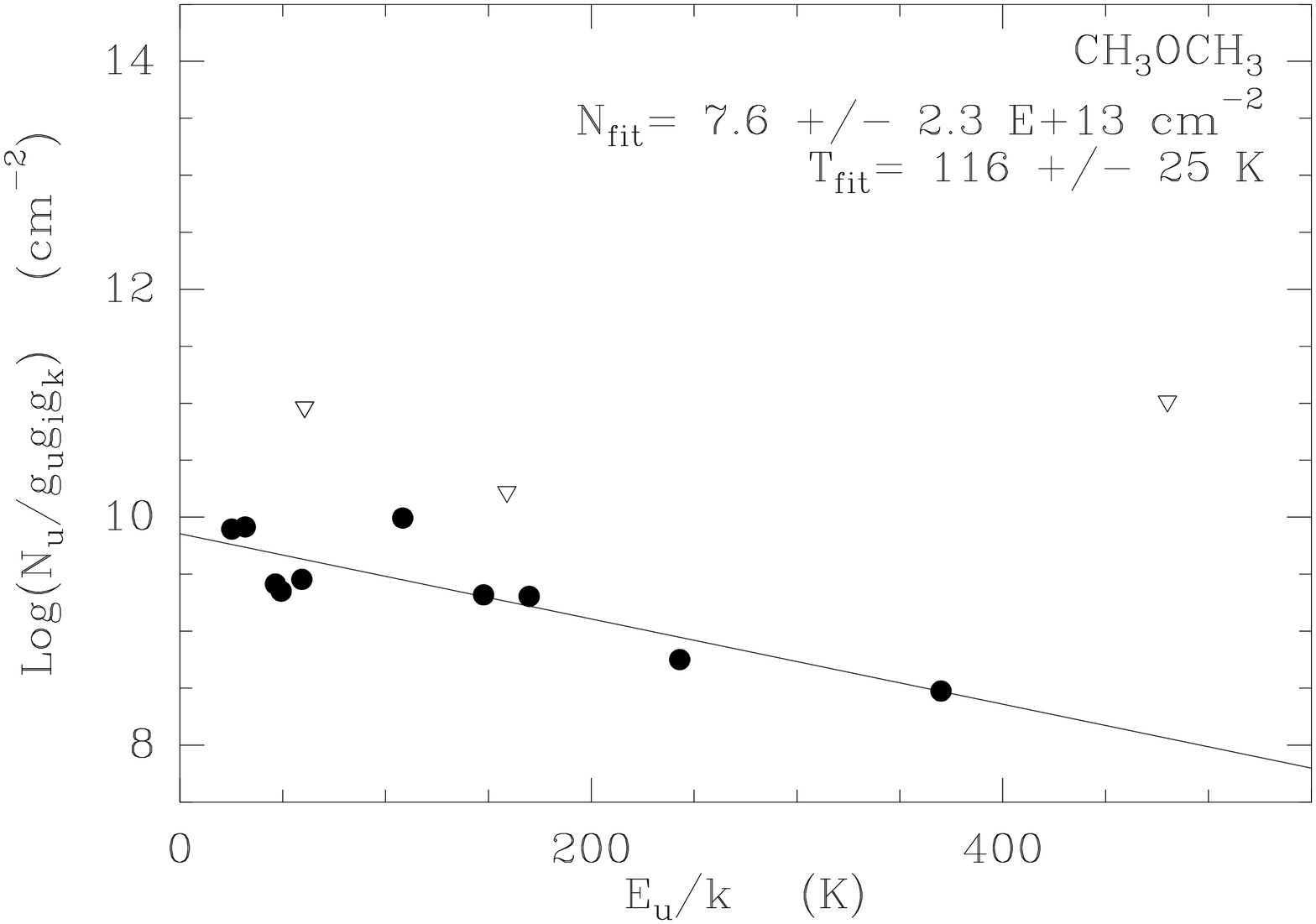}} 
 \end{center}
 \caption[Boltzmann plots for G34.26]
 {\label{fig:g34.ps}{Same as Fig.~\ref{fig:g10.ps}, but for G34.26.}}
\end{figure*}

\end{appendix}


\newpage

\end{document}